\documentclass[final]{jfm}

\usepackage[title]{appendix}
\usepackage{graphicx}
\usepackage{newtxtext}
\usepackage{newtxmath}
\usepackage{natbib}
\usepackage{hyperref}
\usepackage{soul}
\hypersetup{
    colorlinks = true,
    urlcolor   = blue,
    citecolor  = black,
}

\newcommand{\RomanNumeralCaps}[1]
\linenumbers

\usepackage{subcaption}
\usepackage[table]{xcolor}
\usepackage{color,soul}
\usepackage{epstopdf}
\usepackage{amsmath}
\usepackage{mathtools}
\usepackage{amssymb}
\usepackage{upgreek}
\usepackage{float}
\usepackage{contour}

\definecolor{ao(english)}{rgb}{0.0, 0.5, 0.0}

\newcommand{\ALrevise}[1]{\textcolor{black}{#1}}

\title{On the laminar solutions and stability of accelerating and decelerating channel flows}

\author{Alec J. Linot\aff{1},
  \corresp{\email{aleclinot@gmail.com}} 
  Peter J. Schmid\aff{2} \and
  Kunihiko Taira\aff{1}}
\affiliation{\aff{1}Department of Mechanical and Aerospace Engineering, University of California, Los Angeles, CA 90095, USA
\aff{2}Department of Mechanical Engineering, King Abdullah University of Science and Technology (KAUST), Thuwal, Saudi Arabia}

\begin{document}

\maketitle
We study the effect of acceleration and deceleration on the stability of channel flows. To do so, we derive an exact solution for laminar profiles of channel flows with arbitrary, time-varying wall motion and pressure gradient. This solution then allows us to investigate the stability of any unsteady channel flow.
In particular, we \ALrevise{restrict our investigation to} the nonnormal growth of perturbations \ALrevise{about time-varying base} flows with exponentially decaying acceleration and deceleration, with comparisons to growth \ALrevise{about} a constant \ALrevise{base} flow (i.e., the time-invariant simple shear or parabolic profile). We apply this acceleration and deceleration through the velocity of the walls and through the flow rate. For accelerating \ALrevise{base} flows, \ALrevise{perturbations} never grow larger than \ALrevise{perturbations} \ALrevise{about} a constant \ALrevise{base} flow, while decelerating flows show massive amplification of \ALrevise{perturbations} -- at a Reynolds number of $500$,  \ALrevise{properly timed perturbations} \ALrevise{about} the decelerating \ALrevise{base} flow grow $\mathcal{O}(10^5)$ times larger than \ALrevise{perturbations} grow \ALrevise{about} a constant \ALrevise{base} flow. This amplification increases as we raise the rate of deceleration and the Reynolds number. We find that this amplification arises due to a transition from spanwise perturbations leading to the largest amplification, to streamwise perturbations leading to the largest amplification that only occurs in the decelerating \ALrevise{base} flow. By evolving the optimal perturbations through the linearized equations of motion, we reveal that the decelerating \ALrevise{base flow} achieves this massive amplification through \ALrevise{the Orr mechanism, or the} down-gradient Reynolds stress mechanism, which accelerating and constant \ALrevise{base} flows cannot maintain. 

\begin{abstract}

\end{abstract}

\newpage

\section{Introduction}
\label{sec:Intro}

In the study of flow stability, a frequently overlooked topic is the stability of unsteady, aperiodic flows. Despite this \ALrevise{oversight}, these transient flows appear in many systems like airfoil gust encounters \citep{Jones2020}, start-up flow in a pipe \citep{Kataoka1975}, and particle sedimentation \citep{Guazzelli2011}. A common feature of unsteady flows is that under acceleration, the flow tends to stabilize, while under deceleration, the flow tends to destabilize. Such trends have been observed and analyzed for both periodic flows \citep{Gerrard1971,Hino1983} and transient flows \citep{Kurokawa1986,Shuy1996,He2000}. Understanding the mechanism by which this stabilization or destabilization occurs is highly valuable as it could either be used as a tool to actuate steady flows, or it could be used to inform control of already unsteady flows. Two key challenges with systematically investigating the stability of unsteady laminar flows are: (1) the base profile about which to perform the analysis may be unknown, and (2) standard linear stability methods examine the long-time dynamics of a \ALrevise{time-invariant} linear operator, while for unsteady flows we wish to examine transient dynamics associated with a time-varying linear operator.

Although analytical laminar profiles are known for various boundary conditions \citep{Drazin2006}, solutions are not known for arbitrary boundary conditions. Two widely studied geometries with many different analytical laminar solutions are pipe flow and channel flow. The different analytical solutions for pipe flow correspond to different pressure gradients (these pressure gradients may be dictated by a prescribed flow rate), and the different analytical solutions for channel flows correspond to different pressure gradients and wall motion. The simplest cases are either constant pressure gradient or constant, opposite wall motion (in the channel flow case). The former results in a parabolic flow (Poiseuille) and the latter in simple shear flow (Couette) \citep{Bird2015}.

Unlike the constant pressure gradient case, an unsteady pressure gradient yields different solutions between the pipe and channel geometries. First, we discuss some of the unsteady solutions in pipe flow. A widely studied flow type is pulsatile or Womersley flow \citep{Womersley1955}. This flow corresponds to pipe flow driven by a periodic pressure gradient. Although Womersley's derivation is widely cited, earlier derivations of this profile exist \citep{Szymanski1932}, as noted in \citet{Urbanowicz2023}. Other common solutions for pipe flow are startup flow by either a discontinuous change in the pressure gradient \citep{Szymanski1932} or a linear ramp change in the pressure gradient \citep{Ito1952}. \citet{Kannaiyan2021} later extended these startup solutions for prescribing the flow rates instead of pressure gradients. \citet{Fan1964} also found solutions for general pressure gradients and for rectangular ducts. For a more complete review of solutions in pipe flow, we refer the reader to \citet{Urbanowicz2023}.

Next, we survey analytical solutions for channel flows. Two classical solutions in this domain are Stokes' first and second problems \citep{Schlichting2017,Batchelor2000,Liu2008}. These solutions correspond to the cases of instantaneously moving a wall from rest and periodic wall motion. In addition to these solutions, there also exists a solution for a periodic pressure gradient in channel flows \citep{Majdalani2008} -- this differs from Womersley flow in a pipe. \citet{Majdalani2008} also provided solutions for arbitrary periodic pressure gradients. This work was extended by \citet{Lee2017} to include pressure gradients that are not periodic in both pipes and channel flow, for motionless initial conditions. Finally, \citet{Daidzic2022} derived an analytical solution for arbitrary periodic pressure gradients or wall motion. We emphasize that the analytical solutions presented in this work are for arbitrary wall motion and pressure gradients (not necessarily periodic) and for arbitrary initial conditions. Furthermore, we will show how to compute the pressure gradient for a prescribed flow rate. This is an important extension for investigating accelerating and decelerating pressure-driven flow.

Once a laminar \ALrevise{flow} profile is known, the next challenge consists of performing stability analysis about this profile. When the laminar profile exhibits periodic time-varying dynamics, traditional methods of linear stability analysis about a fixed point can be extended with Floquet analysis \citep{Davis1976}. \citet{Kerczek1974} applied Floquet analysis to Stokes' second problem and found that the flow was even more stable than a motionless fluid. Later, \citet{kerczek1982} applied Floquet analysis to pulsatile channel flow and again showed that the motion had a stabilizing effect, and \citet{Tozzi1986} later performed Floquet analysis for pulsatile pipe flow. More recently, \citet{Pier2017} provided a comprehensive Floquet analysis of pulsatile channel flow in which they found destabilization of the flow at low frequencies. In this linearly unstable regime, they found a `cruising' regime where nonlinearity is sustained over a period and a `ballistic' regime where trajectories exhibit large growth to a nonlinear phase before returning to small amplitudes within a cycle. 

While Floquet analysis is appropriate for periodic flows it does not apply to aperiodic flows. One approach to investigating the stability of time-varying flows is to consider the stability of the instantaneous profile as if it were `frozen' (i.e., the quasi-steady state approximation). For example, linear stability analysis has been performed using this quasi-steady state approximation in start-up pipe flow \citep{Kannaiyan2022}. However, this approach breaks down when the laminar profile changes faster than perturbations grow or decay in the linear stability analysis, and the linear stability analysis does not provide this timescale. \citet{Shen1961} discusses further \ALrevise{challenges with} this approach.

An alternative to linear stability analysis is the energy method \citep{Serrin1959,Joseph1976}. Whereas linear stability indicates long-time growth, the energy method reveals when a perturbation will lead to immediate growth in energy \ALrevise{$E$ (i.e., this method finds perturbations where $dE/dt>0$ at the instant the perturbation is applied)}. To apply this method to unsteady base flows, the quasi-steady state approximation must again be taken. Because this method quantifies the instantaneous behavior, the assumption is less detrimental than the frozen stability analysis\ALrevise{, which reveals the asymptotic stability}. Additionally, an advantage of this method is that it can be formulated in terms of the relative energy of the perturbation in relationship to the base flow \citep{Shen1961}.
\citet{Conrad1965} applied this method to accelerating and decelerating channel profiles with wall boundary conditions of $1-e^{-\kappa t}$ and $e^{-\kappa t}$ (among other profiles). Their results showed that acceleration increased the critical Reynolds number, while deceleration decreased it. However, we note that the energy method dramatically underestimates the critical Reynolds number at which flows go through transition\ALrevise{. Moreover, it does not provide the shape of the perturbation or the amount of growth that perturbations exhibit}.   

We overcome these problems associated with both linear stability analysis and the energy method by instead investigating stability using optimal perturbation theory of the time-varying linearized equations. In this method, we find the perturbation energy growth over a finite time window \citep{Schmid2001}. This differs from the energy method in that it restricts the perturbations to physically realizable fields, it does not account for nonlinearity, and it amounts to the growth over a finite window. The optimal perturbation method captures the effects of nonnormal growth missed by linear methods \citep{Trefethen1993}. \citet{Butler1992} computed the optimal perturbations for constant wall motion in a channel flow and found that pairs of streamwise vortices produce the largest growth. 
\citet{Reddy1993} also found the optimal perturbations for constant wall motion and for constant pressure-gradient in channel flows. They again showed that the largest growing perturbations only vary in the spanwise direction. Similarly, \citet{Schmid1994} found that azimuthal-dependent perturbations lead to the largest growth when computing the optimal perturbations for constant pressure gradient pipe flow. In both \citet{Reddy1993} and \citet{Schmid1994}, the energy growth was shown to scale as the Reynolds number squared.

Optimal \ALrevise{perturbations} have also been determined for some unsteady flows. \citet{Biau2016} computed the optimal perturbations for Stokes' second problem and found that streamwise perturbations resulted in the largest growth, which scales exponentially with the Reynolds number. \citet{Xu2021} investigated growth in pulsatile pipe flows and found that \ALrevise{, at certain Womersley numbers and amplitudes, helical perturbations} dominated with an exponential scaling \ALrevise{at high Reynolds numbers and quadratic scaling at low Reynolds numbers}. Finally, one of the few studies of an unsteady, aperiodic flow was performed by \citet{Nayak2017}. They computed the optimal \ALrevise{perturbation} for channel flow impulsively stopped from a constant pressure gradient. The optimal perturbations in this case are again streamwise structures. Our investigation of accelerating and decelerating flows will link together the differences in $\Rey$ scaling observed between the constant and unsteady flows described here.




In the present work, we investigate the transient growth of perturbations in unsteady channel flows that exhibit exponentially decaying acceleration and deceleration. Section \ref{sec:Laminar} derives the analytical solutions for arbitrary wall motion (Sec.\ \ref{sec:LaminarWall}) and pressure gradients (Sec.\ \ref{sec:LaminarPres}) for channel flows. In Section \ref{sec:Stability}, we investigate the transient growth of perturbations to laminar solutions associated with acceleration and deceleration of the wall velocity and flow rate. Section \ref{sec:Stability1} presents the approach taken to compute this transient growth through the linearized equations of motion and examples of this growth at a specific wavenumber. Following this, in Section~\ref{sec:Stability2}, we compute the maximum growth as we vary Reynolds numbers and acceleration or deceleration. Notably, as we increase deceleration, \ALrevise{perturbations} become far more amplified, and the most amplified \ALrevise{perturbations} shift from spanwise structures to streamwise structures. Acceleration shows less amplification and the most amplified \ALrevise{perturbations} maintain a spanwise structure. \ALrevise{In Section~\ref{sec:Stability3}, we study the evolution of these perturbations and find that energy in the decelerating case grows via the Orr mechanism at high Reynolds number and deceleration rates. We then validate the growth of these perturbations in direct numerical simulations in Section~\ref{sec:Stability4}. We find the optimal timing of these perturbations in Section~\ref{sec:Stability5}.}
Finally, in Section~\ref{sec:Conclusions}, we summarize our results and discuss future prospects.

\section{Exact solutions for time-varying wall-driven and pressure-driven channel flow}
\label{sec:Laminar}

\begin{figure}
    \includegraphics[width=\textwidth]{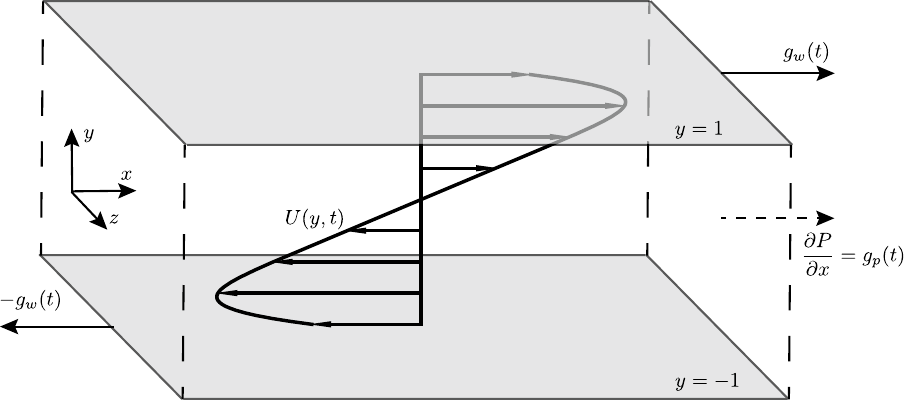}
  \caption{Diagram of mixed wall and pressure-driven channel flow, with an example snapshot of the laminar flow for exponentially decaying wall motion.}
\label{fig:Diagram}
\end{figure}

We first need to determine the underlying laminar flow solutions to investigate the stability of unsteady laminar flows. 
Figure \ref{fig:Diagram} illustrates the configuration of interest -- an incompressible fluid confined between two plates moving with arbitrary speed in opposite directions and with an arbitrary pressure gradient. 

We seek laminar profiles in this domain that satisfy the \ALrevise{incompressible} Navier-Stokes equations (NSE)
\begin{equation} \label{eq:NSE}
	\dfrac{\partial \mathbf{u}}{\partial t}+\mathbf{u}\cdot \nabla \mathbf{u}=-\nabla p+\dfrac{1}{\Rey}\nabla^2 \mathbf{u}, \quad \nabla \cdot \mathbf{u}=0,
\end{equation} 
which have been nondimensionalized by some characteristic velocity $U_b$, the channel half-height $h$, and the kinematic viscosity $\nu$, defining the Reynolds number as $\Rey=U_b h/\nu$. For arbitrary wall motion and pressure gradient, the characteristic velocity $U_b$ varies, and we will mention natural choices for specific examples. In Eq.\ \ref{eq:NSE} we define spatial coordinates in the streamwise $x\in[-\infty,\infty]$, wall-normal $y\in[-1,1]$, and spanwise $z\in[-\infty,\infty]$ directions with the velocity vector $\mathbf{u}=[u,v,w]$, and pressure $p$. 

For finding laminar solutions to Eq.\ \ref{eq:NSE}, we restrict our search to streamwise velocity profiles that only depend on the wall-normal location and time $U(y,t)$. Inserting functions of this form into Eq.\ \ref{eq:NSE} yields
\begin{equation} \label{eq:NSELam}
    \dfrac{\partial U}{\partial t}=-g_p(t)+\dfrac{1}{\Rey}\dfrac{\partial^2 U}{\partial y^2},
\end{equation}
where $g_p(t)$ is the pressure gradient. Here, we prescribe boundary conditions 
\begin{equation} \label{eq:NSELamBC}
    U(y=\pm 1,t)=\pm g_w(t)+g_e(t).
\end{equation}
\ALrevise{It is important to prescribe the boundary conditions as in Eq.\ \ref{eq:NSELamBC}, as it allows for arbitrary top and bottom wall motion. Furthermore, this formulation allows us to seek even and odd solutions to satisfy these boundary conditions.}
We also prescribe the initial condition
\begin{equation} \label{eq:NSELamIC}
    U(y,t=0)=h_o(y)+h_e(y),
\end{equation}
where function $h_o(y)$ is an odd function with boundary conditions $h_o(\pm 1)=\pm g_w(0)$ and the function $h_e(y)$ is an even function with boundary conditions $h_e(\pm 1)=g_e(0)$. The superposition of these terms allows for all possible $y$-dependent initial conditions. For example, if we wish to start from uniform shear then $h_o(y)=y$ and $h_e(y)=0$, or if we want to start with a parabolic profile then $h_o(y)=0$ and $h_e(y)=1-y^2$. Detailed reasons for this split in symmetries will be presented in Sec.\ \ref{sec:LaminarWall} and Sec.\ \ref{sec:LaminarPres}. 

Equation~\ref{eq:NSELam} is simply a heat equation with a forcing due to the pressure gradient $g_p(t)$. The linear nature of this equation allows us to add solutions together via linear superposition \citep{Deen2012}. Thus, solving this equation requires adding solutions together in order to recast the problem into a canonical form we can subsequently solve with a sum over basis functions, e.g., Fourier modes. Specifically, this involves adding solutions such that the resulting partial differential equation (PDE) can be exactly reconstructed by our choice of basis functions. To this end, we first find a solution for odd wall motion ($g_w\neq0$, $g_e=0$, $g_p=0$), after which we seek a solution for arbitrary pressure gradients and even wall motion ($g_w=0$, $g_e\neq0$, $g_p\neq0$). These solutions can be summed to drive the flow with both wall motion and a pressure gradient. We will refer to odd wall motion cases as wall-driven flow (WDF) and to pressure gradient cases as pressure-driven flow (PDF).

\ALrevise{Finally, although we only consider streamwise wall motion, this formulation is valid for arbitrary streamwise and spanwise wall motion. This is straightforward to show if we consider laminar solutions $\mathbf{u}=[U(y,t),0,W(y,t)]$. Inserting this solution into Eq.\ \ref{eq:NSE}, we obtain Eq.\ \ref{eq:NSELam} and
\begin{equation} \label{eq:NSELam2}
    \dfrac{\partial W}{\partial t}=-g_{p,z}(t)+\dfrac{1}{\Rey}\dfrac{\partial^2 W}{\partial y^2},
\end{equation}
where $g_{p,z}(t)$ denotes the pressure gradient in the spanwise direction. The boundary conditions and initial condition would match those above (except in the spanwise direction), thus the solution we present in the streamwise direction is also valid in the spanwise direction. Combining these two solutions then allows us to find the laminar solution for arbitrary in-plane wall motion and pressure gradients.}
\subsection {Wall-driven flow} \label{sec:LaminarWall}

In the case of time-varying wall-driven flow, we seek odd functions $U(y,t)=-U(-y,t)$, which is a natural choice because the boundary conditions satisfy this behavior (note $U_C=0$ here). In Cartesian coordinates, this implies that we solve the heat equation using a sine basis. We achieve this by seeking solutions of the form
\begin{equation} \label{eq:NSELamWall}
    U(y,t)=f_w(y,t)+\dfrac{\Rey}{6}\dfrac{dg_w}{dt}(y^3-y)+g_w(t)y.
\end{equation}
Inserting this expression into Eqs.\ \ref{eq:NSELam}, \ref{eq:NSELamBC}, and \ref{eq:NSELamIC} results in an equation for $f_c$
\begin{equation} \label{eq:NSELamWallPDE}
    \dfrac{\partial f_w}{\partial t}+\dfrac{\Rey}{6}\dfrac{d^2g_w}{dt^2}(y^3-y)=\dfrac{1}{\Rey}\dfrac{\partial^2 f_w}{\partial y^2}
\end{equation}
with boundary conditions 
\begin{equation}
    f_w(y=\pm 1,t)=0
\end{equation}
and initial condition
\begin{equation} \label{eq:NSELamWallIC}
    f_w(y,0)=h_o-\dfrac{\Rey}{6}\left. \dfrac{dg_w}{dt}\right|_{t=0}(y^3-y)-g_w(0)y.
\end{equation}

Through this linear superposition of solutions, $f_w$ is now in a suitable form to be represented as
\begin{equation} \label{eq:Wallexp}
    f_w(y,t)=\sum_{n=1}^\infty \hat{f}_{w,n}(t)\sin(n\pi y).
\end{equation}
Notably, by including the additional terms in Eq.\ \ref{eq:NSELamWall} both the boundary condition for $f_w$, the initial condition for $f_w$, and all terms in Eq.\ \ref{eq:NSELamWallPDE} go to zero at the boundary, just like the sine basis. Had we omitted $(\Rey/6)(dg_w/dt)(y^3-y)$ in Eq.\ \ref{eq:NSELamWall}, Eq.\ \ref{eq:NSELamWallPDE} would contain $y$, which has different boundary conditions at $y=1$ and $y=-1$. Thus, if we were to recreate $y$ with a periodic function there would be a discontinuity at the boundary, resulting in Gibbs phenomena \citep{Graham2013}. In Appendix \ref{sec:AppendixA} we elaborate on this alternative approach, and show that the error is larger than using Eq.\ \ref{eq:NSELamWall}.

Next, we find the coefficients $\hat{f}_{w,n}(t)$ of the sine expansion. By combining Eq.\ \ref{eq:Wallexp} with Eq.\ \ref{eq:NSELamWallPDE} and taking the inner product with $\sin(m\pi y)$, we obtain
\begin{equation} \label{eq:PCFODE}
    \dfrac{d \hat{f}_{w,m}}{dt}=-\dfrac{2 \Rey (-1)^m}{(\pi n)^3}\dfrac{d^2g_w}{dt^2}-a_m\hat{f}_{w,m}
\end{equation}
where $a_m=(\pi m)^2/\Rey$. Solving for this equation, we arrive at 
\begin{equation} \label{eq:PCFODESol}
    \hat{f}_{w,m}= e^{-a_nt}\left( -\dfrac{2 \Rey (-1)^n}{(\pi n)^3} \int_0^t e^{a_nt'} \left. \dfrac{d^2g_w}{dt^2}\right|_{t=t'} dt' +C_{1,m} \right),
\end{equation}
where $C_{1,m}$ can be determined by taking the inner product of the initial condition (Eq.\ \ref{eq:NSELamWallIC}) with $\sin(m\pi y)$
\begin{equation}
    C_{1,m}=-\dfrac{2 \Rey (-1)^m}{(\pi m)^3}\left. \dfrac{dg_w}{dt}\right|_{t=0}+\int_{-1}^1(h_o-g_w(0)y)\sin(m\pi y) dy.
\end{equation}
If $C_{1,m}=0$ the initial condition is the simple shear profile.
Finally, substituting $\hat{f}_{w,n}$ into Eq.\ \ref{eq:NSELamWall}, we find the laminar flow solution
\begin{multline} \label{eq:CouetteLam}
    U(y,t)= \sum_{n=1}^\infty e^{-a_nt} \left[-\dfrac{2 \Rey (-1)^n}{(\pi n)^3} \left( \int_0^t e^{a_nt'}\left.\dfrac{d^2g_w}{dt^2} \right|_{t=t'} dt'+\left. \dfrac{dg_w}{dt} \right|_{t=0}\right) +C_{1,n}\right]\sin(n\pi y) \\
    +\dfrac{\Rey}{6} \dfrac{dg_w}{dt}(y^3-y)+g_w(t) y.
\end{multline}


\begin{figure}
    \includegraphics[width=\textwidth]{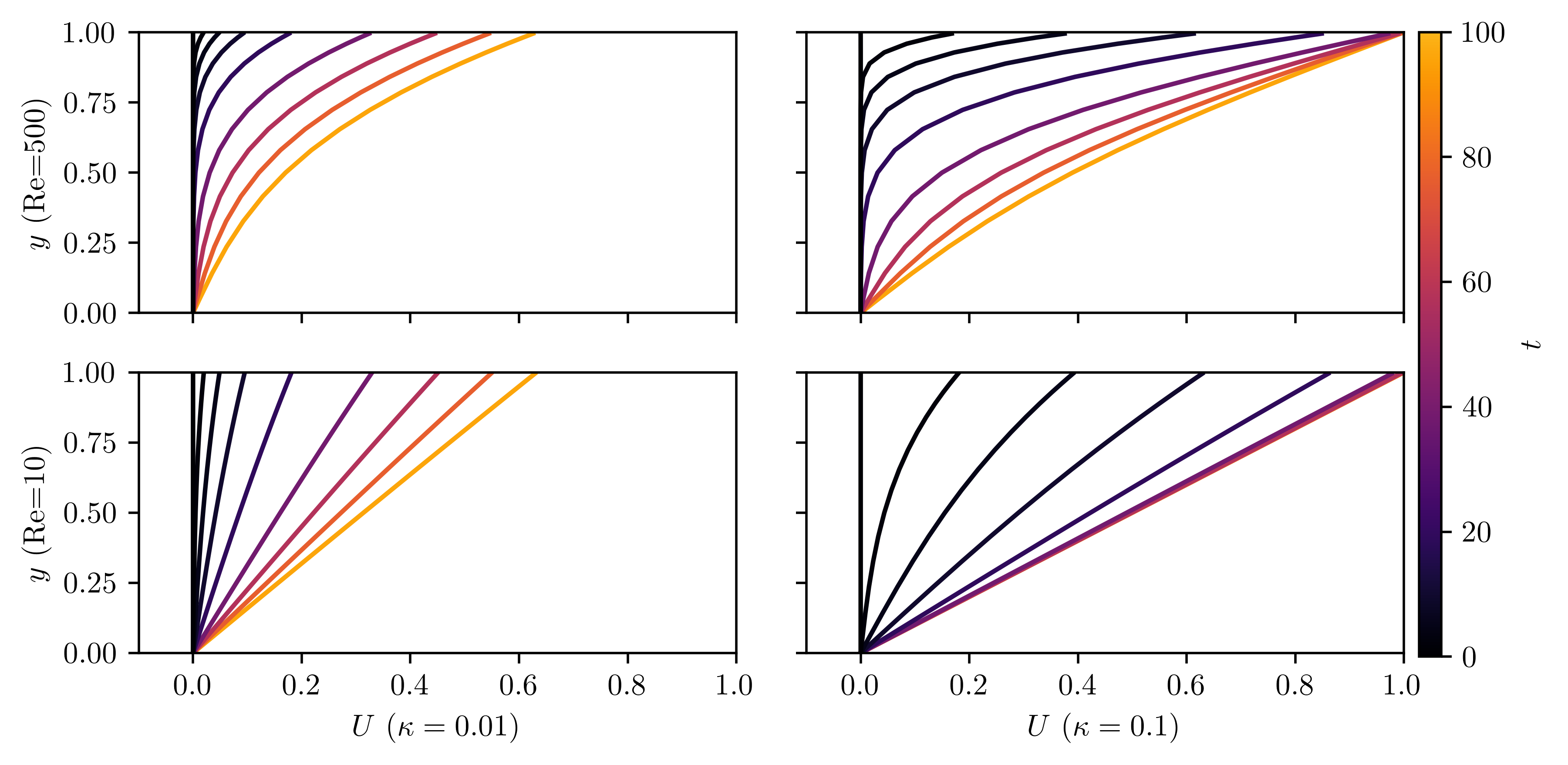}
  \caption{Laminar accelerating WDF. The Reynolds number and acceleration parameter for each flow are noted in the figure. The solution is shown at times $t=[0,2,5,10,20,40,60,80,100]$.}
\label{fig:CBaseflowAcc}
\end{figure}

\begin{figure}
    \includegraphics[width=\textwidth]{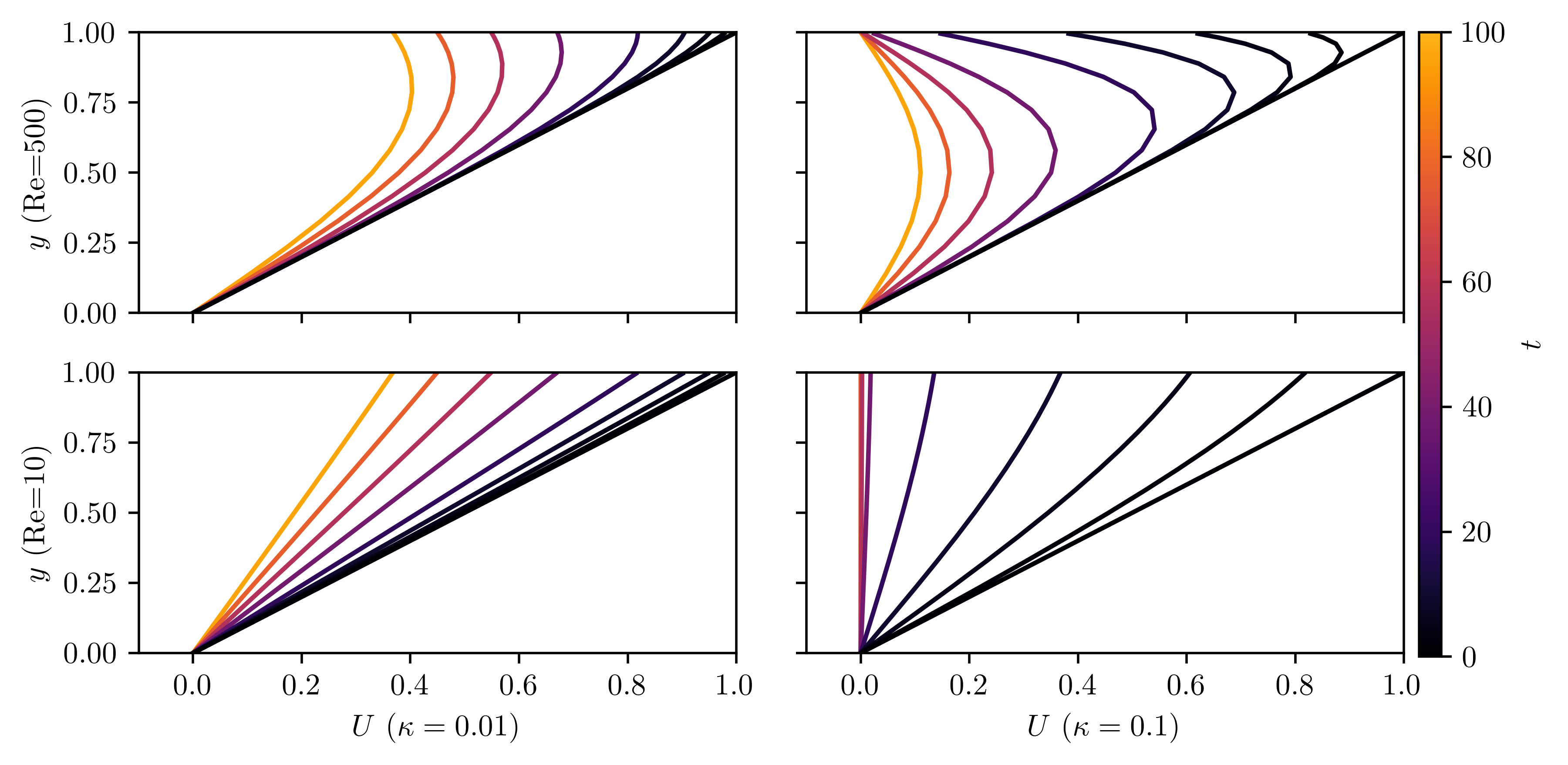}
  \caption{Laminar decelerating WDF. The Reynolds number and deceleration parameter for each flow are noted in the figure. The solution is shown at times $t=[0,2,5,10,20,40,60,80,100]$.}
\label{fig:CBaseflowDec}
\end{figure}

For many $g_w(t)$ profiles of interest, the integral in \ref{eq:CouetteLam} may be evaluated directly. In Table \ref{Table}, we provide some specific profiles of interest, and in Appendix \ref{sec:AppendixA} we validate the solution against other known laminar solutions. Here, we intend to study the effect of acceleration and deceleration on stability characteristics. Hence, the two profiles we concentrate on are exponentially decaying acceleration and deceleration from simple shear. These profiles are given by
\begin{equation}
    g_w(t)=1-e^{-\kappa t}   
\end{equation}
for acceleration and 
\begin{equation}
    g_w(t)=e^{-\kappa t} 
\end{equation}
for deceleration. In both cases, we set the characteristic velocity for the Reynolds number as the maximum wall velocity over an infinite time horizon. We further discuss the nuances of this choice of nondimensionalization in Sec.\ \ref{sec:Nondim}.

\ALrevise{The analytical laminar solution derived here, in combination with the derivation in the subsequent section enables us to consider arbitrary in-plane wall motion and pressure gradients. Thus, the approach we take with the prescribed exponentially decaying acceleration and deceleration can be applied to any in-plane flow, opening the possibility of investigating a wide range of unsteady flows. In particular, we hope this approach will inspire further investigation into aperiodic flows, which have largely been ignored compared to periodic flows. We choose to focus on the exponentially decaying acceleration and deceleration since it is one of the simplest forms of acceleration and deceleration that is bounded. Had we constantly accelerated or decelerated the wall, the flow would grow unbounded. Furthermore, constant deceleration would eventually turn into acceleration in the opposite direction. }

\ALrevise{This form of acceleration and deceleration is also of practical interest. This type of flow can be experienced in the startup and shutdown of internal flows, including pipe flow \citep{GREENBLATT2004}, and flow around moving bodies, such as accelerating and decelerating airfoils \citep{SENGUPTA2007}. Also, unsteady separation bubbles can impose rapid acceleration or deceleration on the flow near the body, for example during dynamic stall. \citet{McCroskey1981} directly compares the vortex structures of an impulsively started plate to vortex structures during dynamic stall. 
Furthermore, exponentially decaying acceleration and deceleration are relevant problems when a controller is used to target a set point. If the system is first-order then exponential decay occurs when we apply a step change to the system \citep{Seborg2010}. Additionally, exponentially decaying acceleration and deceleration are a reasonable proxy for the type of profile exhibited by an overdamped PID controller.}



\ALrevise{Use of an analytical solution, as opposed to a numerical simulation, also offers many advantages. First, our analytical solution depends upon both $\Rey$ and $\kappa$, so we would have to perform a numerical simulation any time we changed these parameters. This would be far slower than evaluating the analytical solution, especially if we considered a flow with even more non-dimensional parameters. Second, at sufficiently high Reynolds numbers the numerical simulations could become unstable and trigger turbulence due to transient growth prevalent in these flows, which we will further discuss in Sec.\ \ref{sec:Stability}. Third, even if the solution does not diverge, numerical errors will accumulate in time, while we would not see these errors in our analytical laminar solution (instead, errors in the analytical solution will stem from truncating the summation). This last point is especially important since the stability analysis requires both first and second derivative information on the laminar profile. Finally, just knowing the form of the analytical solution is useful. For example, we can use Eq.\ \ref{eq:CouetteLam} to prescribe wall motion that achieves desired laminar profiles. Perhaps, this fact could be utilized to achieve the fastest transition between flow profiles or to find a transition that minimizes the transient growth of perturbations. Regardless, knowledge of the analytical solution can be a powerful tool in understanding and controlling flows.}

Figure \ref{fig:CBaseflowAcc} shows the accelerating laminar profiles at $\Rey=\{10,500 \}$ and $\kappa=\{0.01,0.1\}$. Similarly, Fig.\ \ref{fig:CBaseflowDec} shows the deceleration laminar profiles for the same parameters. At low Reynolds numbers $\Rey$ and low values of $\kappa$ the flow closely resembles simple shear at different shear rates (i.e., $U=g_w(t)y$), whereas with increased $\Rey$ and $\kappa$ the profiles become more curved. This curvature stems from a delay in the transfer of momentum from the wall to the middle of the channel. At higher $\Rey$ this transfer is slower and at higher $\kappa$ the rate of change of velocity at the wall increases. At the largest values of $\Rey$ and $\kappa$ the difference between acceleration and deceleration is exemplified. In this case, we observe that the accelerating profile maintains a positive gradient throughout the domain, while the decelerating profile shows a negative gradient near the wall. In Sec.\ \ref{sec:Stability}, we investigate the influence of these profiles on the stability of the flow.

\subsection {Pressure-driven flow}  \label{sec:LaminarPres}

We follow a similar approach to find the laminar flow solution for time-varying pressure gradients. In the case of pressure-driven flow, we expect the solution to be even $U(y,t)=U(-y,t)$. In Cartesian coordinates, this suggests a solution of the heat equation using a cosine basis. We thus seek solutions of the form
\begin{equation} \label{eq:NSELamPres}
    U(y,t)=f_p(y,t)+
    \frac{\Rey}{2} g_p(t)(y^2-1) + g_e(t).
\end{equation}
Inserting this expression into Eqs.\ \ref{eq:NSELam}, \ref{eq:NSELamBC}, and \ref{eq:NSELamIC} results in an equation for $f_p$ of the form 
\begin{equation} \label{eq:NSELamPresPDE}
    \dfrac{\partial f_p}{ \partial t}+\dfrac{\Rey}{2}\dfrac{dg_p}{dt}(y^2-1)+\dfrac{dg_e}{dt}=\dfrac{1}{\Rey}\dfrac{\partial^2 f_p}{\partial y^2},
\end{equation}
with boundary conditions 
\begin{equation}
    f_p(y=\pm 1,t)=0,
\end{equation}
and initial condition
\begin{equation} \label{eq:NSELamPresIC}
    f_p(y,0)=h_e-\dfrac{\Rey}{2}g_p(0)(y^2-1)-g_e(0).
\end{equation}
To solve Eq.\ \ref{eq:NSELamPresPDE}, we represent $f_p$ as
\begin{equation} \label{eq:Presexp}
    f_p(y,t)=\sum_{n=0}^\infty \hat{f}_{p,n}(t)\cos\left[\left(n+\dfrac{1}{2}\right)\pi y\right].
\end{equation}
The boundary conditions associated with all terms in Eqs.\ \ref{eq:NSELamPresPDE} and \ref{eq:NSELamPresIC} are satisfied by our cosine basis. To solve for the coefficients $\hat{f}_{p,n}(t)$, we repeat the procedure used for WDF. First, we combine Eq.\ \ref{eq:Presexp} with Eq.\ \ref{eq:NSELamPresPDE} and take the inner product of the result with $\cos((m+1/2)\pi y)$ leading to 
\begin{equation} \label{eq:PPFODE}
    \dfrac{d \hat{f}_{p,m}}{dt}=\dfrac{16 \Rey (-1)^m}{(2 \pi m+\pi)^3}\dfrac{dg_p}{dt}- \dfrac{4 (-1)^m}{2 \pi m+\pi} \dfrac{dg_e}{dt}-b_m\hat{f}_{p,m}
\end{equation}
with $b_m=(2 \pi m+\pi)^2/(4\Rey)$. Solving Eq.\ \ref{eq:PPFODE} produces 
\begin{equation} \label{eq:PPFODESol}
    \hat{f}_{p,m}= e^{-b_mt} \left[\dfrac{16 \Rey (-1)^m}{(2 \pi m+\pi)^3} \int_0^t e^{b_mt'}\left.\dfrac{dg_p}{dt}\right|_{t=t'} dt'-\dfrac{4 (-1)^m}{2 \pi m+\pi} \int_0^t e^{b_mt'}\left.\dfrac{dg_e}{dt}\right|_{t=t'} dt' +C_{2,m} \right].
\end{equation}
Then, to solve for the constant $C_{2,m}$, we take the inner product of the initial condition (Eq.\ \ref{eq:NSELamPresIC}) with $\cos((m+1/2)\pi y)$ to obtain
\begin{equation}
    C_{2,m}=\int_{-1}^1\left[h_e-\dfrac{\Rey}{2}g_p(0)(y^2-1)-g_e(0)\right]\cos\left[\left(m+\dfrac{1}{2}\right)\pi y\right] dy.
\end{equation}
If the initial profile is parabolic and there is no wall motion, then $C_2$=0. Finally, by inserting $\hat{f}_{p,n}$ into Eq.\ \ref{eq:NSELamPres}, we find the laminar profile
\begin{multline} \label{eq:NSELamPresfin}
    U(y,t)= \\
    \sum_{n=0}^\infty e^{-b_nt} \left[\dfrac{16 \Rey (-1)^n}{(2 \pi n+\pi)^3} \int_0^t e^{b_nt'}\left.\dfrac{dg_p}{dt}\right|_{t=t'} dt'-\dfrac{4 (-1)^n}{2 \pi n+\pi} \int_0^t e^{b_nt'}\left.\dfrac{dg_e}{dt}\right|_{t=t'} dt' +C_{2,n} \right] \\
    \cos\left[\left(n+\dfrac{1}{2}\right)\pi y\right]+\dfrac{\Rey}{2}g_p(t)(y^2-1) + g_e(t) .
\end{multline}
In Table \ref{Table}, we provide analytical expressions for the integral in Eq.\ \ref{eq:NSELamPresfin} for representative flows. In Appendix \ref{sec:AppendixA}, we validate the solution against Womersley flow.

As with the wall-driven flow, we study the effect of acceleration and deceleration in the pressure-driven case \ALrevise{(here we let $g_e(t)=0$)}. A natural first choice for examining the impact of acceleration and deceleration might be to set the pressure gradient to the same profiles used for the wall velocities (i.e., $g_p(t)=1-e^{-\kappa t}$ and $g_p(t)=e^{-\kappa t}$). In the case of pressure-driven flow, we nondimensionalize the velocity by the maximum centerline velocity. However, the pressure gradient required to satisfy this nondimensionalization is $g_p=-2/\Rey$ either at $t=0$ or as $t\rightarrow \infty$. This means that $g_p(t)=1-e^{-\kappa t}$ and $g_p(t)=e^{-\kappa t}$ do not properly satisfy the correct profiles, and multiplying these quantities by $-2/\Rey$ would result in a different pressure gradient profile for different Reynolds numbers. 

Instead of prescribing the same pressure gradient across all cases, we enforce an exponentially decaying accelerating or decelerating flow rate according to 
\begin{equation} \label{Qacc}
    Q(t)=\dfrac{2}{3}(1-e^{-\kappa t})
\end{equation}
and
\begin{equation} \label{Qdec}
    Q(t)=\dfrac{2}{3}(e^{-\kappa t}),
\end{equation} 
\ALrevise{respectively.}
These flow rates correspond to a unit centerline streamwise velocity for a parabolic profile. Note that flow rate and mean velocity are synonymous here. As with wall motion, we delve into the details of this nondimensionalization in Sec.\ \ref{sec:Nondim}.

Prescribing a flow rate requires a corresponding pressure gradient to achieve this flow rate. We first show that accounting for all continuous flow rates requires pressure gradients that can undergo a step change at $t=0$, after which we compute the pressure gradients $g_p$ and use it to approximate $U$. We calculate the flow rate by integrating Eq.\ \ref{eq:NSELamPresfin} in the wall-normal direction and dividing by twice the channel height to result in 
\begin{equation} \label{eq:Q}
    Q(t)=\sum_{n=0}^\infty e^{-b_nt} \left[ \dfrac{32 \Rey }{(2\pi n+\pi)^4} \int_0^t e^{b_nt'}\left.\dfrac{dg_p}{dt}\right|_{t=t'} dt' +\dfrac{2 (-1)^n C_{2,n}}{2\pi n+\pi}\right] -\dfrac{\Rey}{3} g_p(t).
\end{equation}
Evaluating Eq.\ \ref{eq:Q} at $t=0$ and simplifying the resulting expression leads to   
\begin{equation}
    Q(0)=\sum_{n=0}^\infty \dfrac{2 (-1)^n}{2\pi n+\pi} \int_{-1}^1 h_e(y)\cos\left[\left(n+\dfrac{1}{2}\right)\pi y\right]dy,
\end{equation}
which shows that the initial flow rate only depends on the initial velocity profile, as expected. Taking the derivative of Eq.\ \ref{eq:Q}, we get after simplifications 
\begin{equation} \label{eq:dQ}
    \dfrac{dQ}{dt}=-\sum_{n=0}^\infty \dfrac{8}{(2\pi n+\pi)^2} e^{-b_nt} \int_0^t e^{b_nt'}\dfrac{dg_p(t')}{dt} dt'.
\end{equation}
From this expression, we note that a jump discontinuity is required for $g_p$ as $t\rightarrow 0$. If $g_p$ were a continuous function, Eq.\ \ref{eq:dQ} would indicate that $\lim_{t\rightarrow 0}dQ/dt=0$. However, our desired flow rate results in $\lim_{t\rightarrow 0} dQ/dt=\pm 2k/3$. Thus, satisfying this derivative constraint necessitates a step change in the pressure gradient at $t=0$ which can be formulated by
\begin{equation} \label{eq:gp2}
    g_p(t)=H(t)\hat{g}_0+\hat{g}_p(t),
\end{equation}
with $H(t)$ as the Heaviside function, $\hat{g}_0$ as a constant, and $\hat{g}_p$ as a continuous temporal function. Furthermore, we may use $dQ/dt$ to compute the constant $\hat{g}_0$ by taking the derivative of Eq.\ \ref{eq:gp2} and  combining it with \ref{eq:dQ} to produce 
\begin{equation} \label{eq:dQ2}
    \lim_{t \rightarrow 0}  \dfrac{dQ}{dt}=-\hat{g}_0.
\end{equation}
The above expression indicates that only flow rates with  $\lim_{t\rightarrow 0}dQ/dt=0$ can be constructed with a continuous pressure gradient.

Now that we know the form that the pressure gradient must take, we can numerically solve for $g_p(t)$ to satisfy the flow rates given by Eq.\ \ref{Qacc} and Eq.\ \ref{Qdec}. In short, we use the trapezoidal rule on all the integrals and then solve for $g_p(t)$. We include a detailed description of this procedure in Appendix \ref{sec:AppendixC}.
In Fig.\ \ref{fig:Pres}, we present examples of the numerically computed pressure gradients and the flow rates of the velocity profiles computed using these gradients at $\Rey=500$ and $\kappa=0.1$. Both pressure gradient profiles start with a constant pressure gradient for achieving zero flow (in the case of acceleration) or a parabolic profile (in the case of deceleration). At startup, there is a step change in the pressure gradient that subsequently grows or decays toward the pressure gradient needed to maintain the long-time solution. The excellent match of the prescribed flow rates, and the computed flow rates in Figs.\ \ref{fig:Presc} and \ref{fig:Presd} validate the computed pressure gradients.

\begin{figure}
    \includegraphics[width=\textwidth]{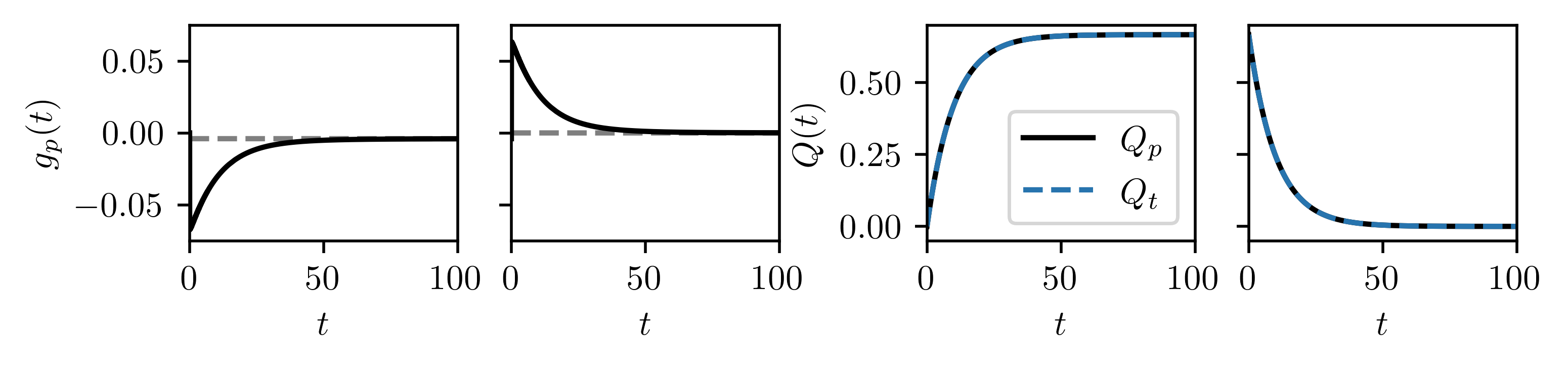}
    
    \captionsetup[subfigure]{labelformat=empty}
    \begin{picture}(0,0)
    \put(10,87){\contour{white}{ \textcolor{black}{a)}}}
    \put(112,87){\contour{white}{ \textcolor{black}{b)}}}
    \put(215,87){\contour{white}{ \textcolor{black}{c)}}}
    \put(293,87){\contour{white}{ \textcolor{black}{d)}}}
    \end{picture} 
    \begin{subfigure}[b]{0\textwidth}\caption{}\vspace{-10mm}\label{fig:Presa}\end{subfigure}
    \begin{subfigure}[b]{0\textwidth}\caption{}\vspace{-10mm}\label{fig:Presb}\end{subfigure}
    \begin{subfigure}[b]{0\textwidth}\caption{}\vspace{-10mm}\label{fig:Presc}\end{subfigure}
    \begin{subfigure}[b]{0\textwidth}\caption{}\vspace{-10mm}\label{fig:Presd}\end{subfigure}
    \vspace{-5mm}

    \caption{(a) and (b) are the pressure gradient for flow rates of $Q_t=(2/3)(1-e^{-\kappa t})$ and $Q_t=(2/3)e^{-\kappa t}$ at $\Rey=500$ and $\kappa=0.1$(the gray dashed line is the expected long-time value). (c) and (d) are the flow rates $Q_p$ computed from applying the pressure gradients in (a) and (b).}
\label{fig:Pres}
\end{figure}
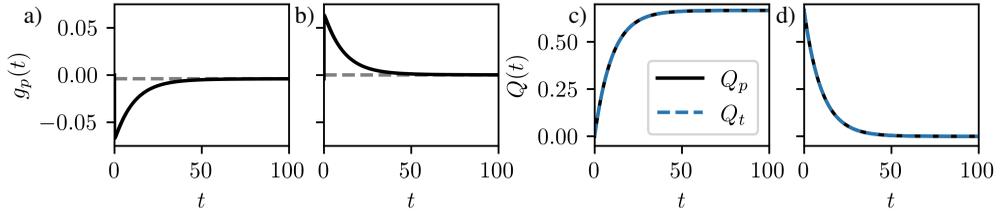

\begin{figure}
    \includegraphics[width=\textwidth]{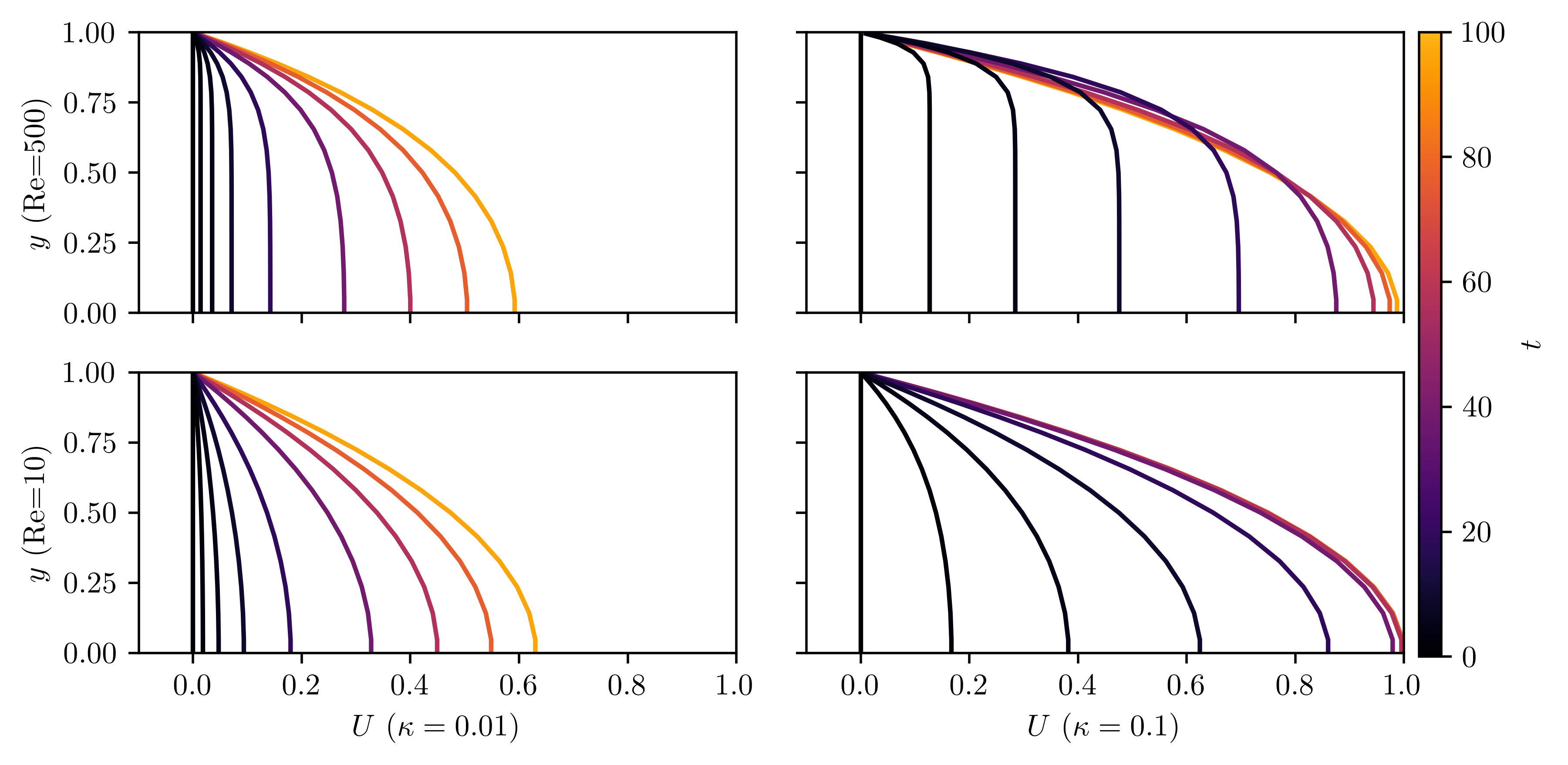}
  \caption{Laminar accelerating PDF. The Reynolds number and acceleration parameter for each flow are denoted in the figure. The solution is shown at times $t=[0,2,5,10,20,40,60,80,100]$.}
\label{fig:PBaseflowAcc}
\end{figure}

\begin{figure}
    \includegraphics[width=\textwidth]{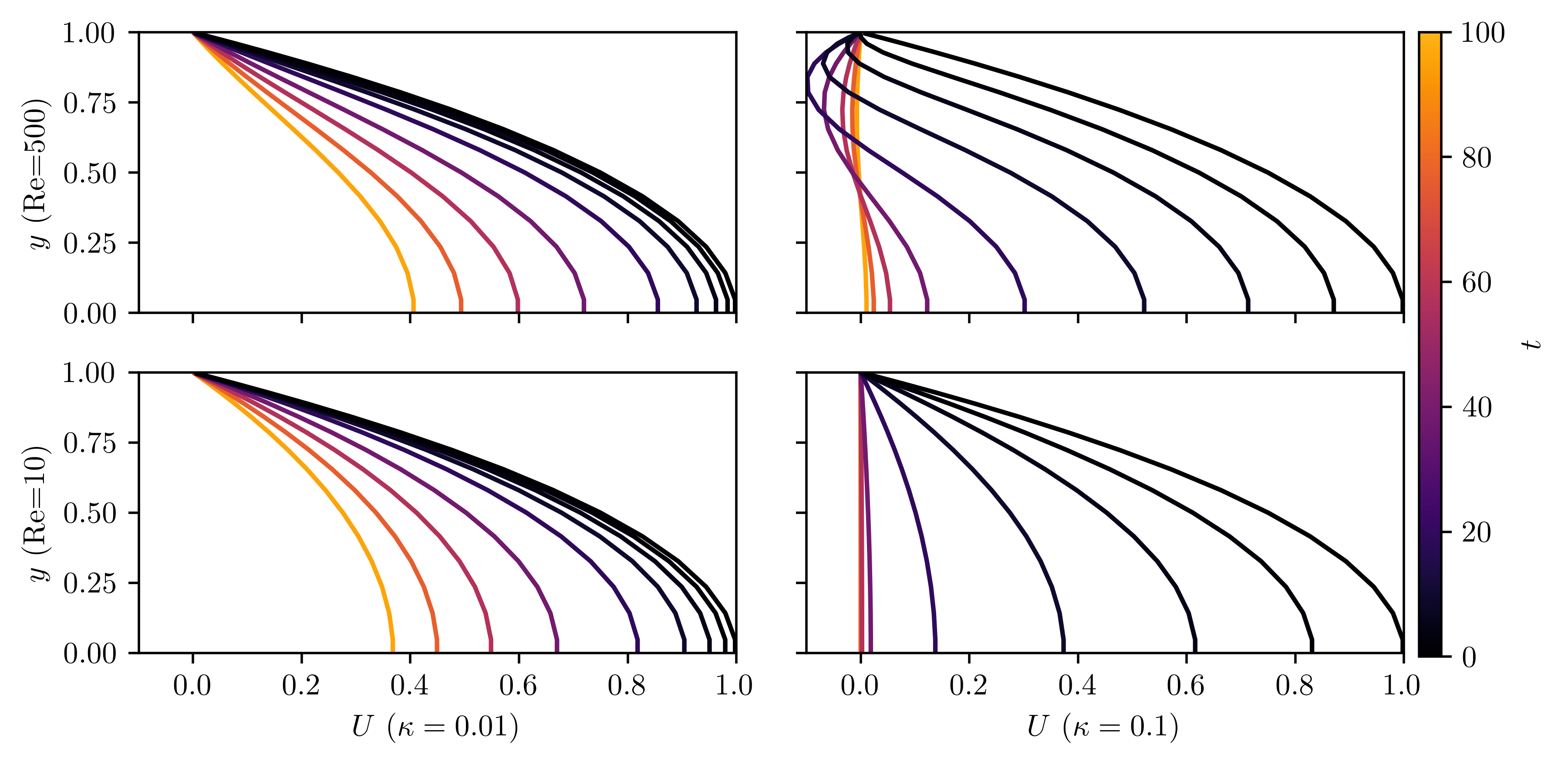}
  \caption{Laminar decelerating PDF. The Reynolds number and deceleration parameter for each flow are denoted in the figure. The solution is shown at times $t=[0,2,5,10,20,40,60,80,100]$.}
\label{fig:PBaseflowDec}
\end{figure}

In Fig.\ \ref{fig:PBaseflowAcc} we display the laminar flow solutions with the numerically computed pressure gradient for accelerating cases at $\Rey=\{10,500\}$ and $\kappa=\{0.01,0.1\}$; in Fig.\ \ref{fig:PBaseflowDec} we show the laminar profiles for decelerating cases with the same parameters. For the accelerating cases, increasing $\Rey$ and $\kappa$ produces transient dynamics that are more plug-like. For the decelerating cases, the gradient is diminished near the wall compared to the accelerating cases, and at sufficiently high $\Rey$ and $\kappa$, the profile exhibits backflow to maintain the appropriate flow rate. Similar to the decelerating WDF, we show in Sec.\ \ref{sec:Stability} that this profile leads to destabilization.

Finally, we emphasize that the solutions provided in Eq.\ \ref{eq:CouetteLam} and Eq.\ \ref{eq:NSELamPresfin} are applicable for odd and even functions, respectively. Owing to linear superposition, \emph{any} function may be represented by simply adding the two solutions together. Thus, the solutions presented in Sec.\ \ref{sec:LaminarWall} and \ref{sec:LaminarPres} are valid for arbitrary \ALrevise{streamwise} wall motion and pressure gradients and can accommodate arbitrary \ALrevise{wall-normal varying} initial conditions.

\subsection{Comments on nondimensionalization} \label{sec:Nondim}

Before investigating the stability of these flows, we discuss the nondimensionalization of the problem setup. Thus far, the equations presented do not depend on the characteristic velocity $U_b$. The choice of $U_b$ depends on the selection of the boundary conditions, or pressure gradient, and should be selected to eliminate one of the dimensional parameters driving the flow. To illustrate this point, let us consider the case of accelerating and decelerating wall motion. In dimensional form, we may write the velocity at the wall as 
\begin{equation}
    g^*_w(t)=u^*_ie^{-\kappa^*t^*}+u^*_f(1-e^{-\kappa^*t^*}),
\end{equation}
which, after nondimensionalization, becomes
\begin{equation}
    g_w(t)=u_ie^{-\kappa t}+u_f(1-e^{-\kappa t})
\end{equation}
with nondimensional scaling of $u_i=u^*_i/U_b$, $u_f=u^*_f/U_b$, and $\kappa=\kappa^*h/U_b$. As such, the most natural choice for nondimensionalization is to select a characteristic velocity that forces $u_i=1$, $u_f=1$, or $\kappa=1$. We consider flows that start from rest ($u_i=0$) or end at rest ($u_f=0$), so we choose the other velocity as our characteristic velocity $U_b$ and vary the nondimensional exponential decay parameter $\kappa$. Notably, this decay parameter is inversely proportional to the characteristic timescale, for example, the half-life $t_{1/2}=\ln(2)/\kappa$. This suggests that, as we vary $\Rey$ and $\kappa$, we vary two timescales: the timescale over which the fluid in the middle of the channel reacts to wall motion due to viscosity and the timescale over which the wall motion reaches the final velocity. While it may also be possible to consider the viscous timescale, we adopt the advective scale due to the fast motion imposed during the acceleration and deceleration process. \ALrevise{The laminar flow depends on both nondimensional parameters $\Rey$ and $\kappa$, so we must vary both when investigating stability.}

In contrast to the wall-driven flow, the selection of the characteristic velocity in the pressure-driven case is less obvious. The pressure gradient drives a flow with a characteristic velocity, but it is difficult to determine this characteristic velocity before prescribing the pressure gradient. To address this issue, we instead use the flow rate to determine the characteristic velocity. Additionally, by prescribing an accelerating or decelerating flow rate, we are directly accelerating or decelerating the mean velocity, whereas accelerating or decelerating the pressure gradient has an unclear effect on the mean velocity. 

As with wall motion, we nondimensionalize around either the long-time or the initial profile, which in the case of the accelerating or decelerating profile corresponds to a parabolic profile. This results in the dimensional flow rate equation
\begin{equation} \label{eq:Flowrate}
    Q^*=\dfrac{1}{2h}\int_{-h}^h\dfrac{U(y=0)}{h^2}(h^2-y^{*2})dy^*.
\end{equation}
After evaluating the integral and using the nondimensionalization of $Q=Q^*/U_b$, Eq.\ \ref{eq:Flowrate} simplifies to
\begin{equation} \label{eq:Flowrate2}
    Q=\dfrac{2U(y=0)}{3U_b}.
\end{equation}
We may then set the characteristic velocity as either the final or initial centerline velocity ($U(y=0)=U_b$), which is satisfied when the flow rate is $Q=2/3$. Thus, we satisfy this nondimensionalization for the 
exponentially decaying acceleration and deceleration cases when either $Q_\infty$ or $Q_0$ equal 2/3 in $Q(t)=Q_\infty(1-e^{-\kappa t})+Q_0e^{-\kappa t}$.
As we investigate accelerating from rest and decelerating to rest in this work, we only need to sweep over the nondimensional exponential decay parameter $\kappa$.



\section{Stability of accelerating and decelerating flows}
\label{sec:Stability}

With the exact laminar solutions for accelerating and decelerating flows established, we proceed to analyze their stability characteristics. As mentioned in Sec.\ \ref{sec:Intro}, studying the stability of these flows is challenged by the fact that many standard methods provide insight into long-time stability properties, but do not account for a general time-dependence of the base flow. To overcome these challenges, we investigate the stability properties of these time-varying base flows via the transient growth of perturbations that evolve according to the linearized Navier-Stokes equations. We start with presenting the linearized equations of motion, which is followed, in Sec.\ \ref{sec:Stability1}, with examples of transient growth in our specific application cases. In Sec.\ \ref{sec:Stability2}, we then sweep over various wavenumbers, Reynolds numbers, and acceleration/deceleration rates to assess the prevalence of transient growth in parameter space. Here we find that higher $\kappa$ and $\Rey$ result in larger energy growth of perturbations, and that decelerating flows exhibit substantially larger growth than constant or accelerating flows. Finally, in Sec.\ \ref{sec:Stability3}, we follow the linear evolution of the optimal perturbations through time, and verify our results, in Sec.\ \ref{sec:Stability4}, against a direct numerical simulation. 

To investigate the evolution of perturbations with respect to a given base flow, we decompose the flow into a laminar base state $\mathbf{U}$ and a perturbation $\mathbf{u}'$ according to 
\begin{equation} \label{eq:ReDecomp}
    \mathbf{u}=\mathbf{U}+\mathbf{u}'
\end{equation}
where $\mathbf{U}=[U,0,0]^T$. We then combine Eq.\ \ref{eq:ReDecomp} with Eq.\ \ref{eq:NSE} and assume that $|\mathbf{u}'|=\mathcal{O}(\epsilon)$ for $0<\epsilon \ll 1$, which results in the linearized Navier-Stokes equations (NSE) 
\begin{equation} \label{eq:NSELin}
	\dfrac{\partial \mathbf{U}}{\partial t}+\mathbf{g}_p-\dfrac{1}{\Rey}\nabla^2 \mathbf{U}+\dfrac{\partial \mathbf{u}'}{\partial t}+\mathbf{U}\cdot \nabla \mathbf{U}+\mathbf{U}\cdot \nabla \mathbf{u}'+\mathbf{u}'\cdot \nabla \mathbf{U}=-\nabla p'+\dfrac{1}{\Rey}\nabla^2 \mathbf{u}',
\end{equation} 
with $\mathbf{g}_p=[g_p,0,0]^T.$ In the above equation, we neglect $\mathcal{O}(\epsilon^2)$ terms.
Because the laminar solutions found in Sec.\ \ref{sec:Laminar} satisfy Eq.\ \ref{eq:NSELam}, the first three terms in Eq.\ \ref{eq:NSELin} cancel out, \ALrevise{and the fifth term is zero by construction}. This simplifies the linearized Navier-Stokes equations such that the only difference between the above equation and the typical form used for time-invariant base flows rests in the time-dependency of $\mathbf{U}$. Taking the divergence of Eq.\ \ref{eq:NSELin} and combining it with the continuity equation, we can find an equation for the pressure perturbation. Inserting this pressure perturbation back into the wall-normal velocity $v'$ equation yields one equation for $v'$ only. The remaining two momentum equations can be combined into an evolution equation for the wall-normal vorticity $\omega_y'=\partial u'/\partial z - \partial w'/\partial x,$ resulting in a system of two equations given as  
\begin{equation} 
	\left( \dfrac{\partial}{\partial t}+U\dfrac{\partial}{\partial x}\right) \nabla^2 v' -\dfrac{\partial^2 U}{\partial y^2}\dfrac{\partial v'}{\partial x}=\dfrac{1}{\Rey}\nabla^4v'
\end{equation} 
and 
\begin{equation} 
	\left( \dfrac{\partial}{\partial t}+U\dfrac{\partial}{\partial x}\right) \omega_y' +\dfrac{\partial U}{\partial y}\dfrac{\partial v'}{\partial z}=\dfrac{1}{\Rey}\nabla^2\omega_y'
\end{equation} 
with boundary conditions conditions
\begin{equation}
    v'(y=\pm1)=\dfrac{\partial v'}{\partial y}(y=\pm1)=\omega_y'(y=\pm1)=0.
\end{equation}

We further simplify these equations by seeking streamwise and spanwise periodic perturbations, which introduces the streamwise wavenumber $\alpha$ and the spanwise wavenumbers $\beta$ such that 
\begin{equation}
    v'(x,y,z,t)=\hat{v}(y,t)e^{i(\alpha x+\beta z)}
\end{equation}
and 
\begin{equation}
    \omega_y'(x,y,z,t)=\hat{\omega}_y(y,t)e^{i(\alpha x+\beta z)}.
\end{equation}
This assumption results in the set of equations
\begin{equation} 
	\dfrac{\partial \hat{v}}{\partial t}=(D^2-k^2)^{-1}\left[ \dfrac{1}{\Rey}(D^2-k^2)^2-i \alpha U(D^2-k^2)+i\alpha \dfrac{\partial^2 U}{\partial y^2} \right]\hat{v}
\end{equation} 
and
\begin{equation} 
	\dfrac{\partial \hat{\omega}_y}{\partial t}=-i\beta \dfrac{\partial U}{\partial y} \hat{v} +\left[-i \alpha U + \dfrac{1}{\Rey}(D^2-k^2) \right] \hat{\omega}_y
\end{equation} 
with boundary conditions
\begin{equation} 
	\hat{v}(y=\pm 1,t)=D \hat{v}(y=\pm 1,t)=\hat{\omega}_y(y=\pm 1,t)=0,
\end{equation} 
where $D=\partial/\partial y$ denotes the wall-normal derivative operator, and $k^2=\alpha^2+\beta^2$ stands for the wavenumber modulus squared. Rearranging and grouping terms in the above equation leads to
\begin{equation} \label{eq:LinearEq}
    \frac{\partial}{\partial t}\left[\begin{array}{l}
    \hat{v} \\
    \hat{\omega}_y
    \end{array}\right]=-\mathrm{i}\left[\begin{array}{cc}
    \mathscr{L}_{\mathrm{os}} & 0 \\
    \mathscr{L}_{\mathrm{c}} & \mathscr{L}_{\mathrm{sq}}
    \end{array}\right]\left[\begin{array}{l}
    \hat{v} \\
    \hat{\omega}_y
    \end{array}\right]
\end{equation}
or
\begin{equation} \label{eq:LinearEq2}
    \frac{\partial \mathbf{q}}{\partial t}=-\mathrm{i}
    \mathscr{L} \mathbf{q},
\end{equation}
with 
\begin{equation}
    \mathscr{L}_{\mathrm{os}}=-(D^2-k^2)^{-1}\left[ \dfrac{1}{i\Rey}(D^2-k^2)^2-\alpha U(D^2-k^2)+\alpha \dfrac{\partial^2 U}{\partial y^2} \right],
\end{equation}
\begin{equation}
    \mathscr{L}_{\mathrm{sq}}=\alpha U - \dfrac{1}{i\Rey}(D^2-k^2),
\end{equation}
and
\begin{equation}
    \mathscr{L}_{\mathrm{c}}=\beta \dfrac{\partial U}{\partial y}.
\end{equation}
The above derivation follows the nomenclature in \citet{Reddy1993}.

Our goal is to investigate the linear evolution of perturbations through Eq.\ \ref{eq:LinearEq2}. This linear equation has solutions of the form 
\begin{equation}
    \mathbf{q}(y,t)=\mathsfbi{A}(t)\mathbf{q}(y,0),
\end{equation}
where $\mathsfbi{A}(t)$ is the fundamental solution operator that satisfies 
\begin{equation}
    \dfrac{\partial}{\partial t} \mathsfbi{A}(t)=e^{-i\mathscr{L}(t)t} \mathsfbi{A}(t) \qquad\qquad \mathsfbi{A}(t=0)=\mathsfbi{I}.
\end{equation}
In the above, we assume a discretization in the wall-normal $y$-direction which results in a finite-dimensional representation of the fundamental solution in terms of a matrix $\mathsfbi{A}(t)$. 
Numerically, we approximate $\mathsfbi{A}(t)$ as the finite product of exponentials given by 
\begin{equation}
    \mathsfbi{A}(n \Delta t)=\Pi_{j=1}^n e^{-i\mathscr{L}(j\Delta t)\Delta t}
\end{equation}
with time step $\Delta t$.
By computing the approximate solution matrix $\mathsfbi{A}(t)$, we can track the linear evolution and stability characteristics of infinitesimal perturbations. In the following section, we use the above formalism to determine the maximum linear growth of perturbations through time. 



\subsection {Maximum linear amplification}
\label{sec:Stability1}

Nonnormal linear operators can support large levels of transient growth, even though the operator's eigenvalues indicate asymptotic stability (all eigenvalues have negative real parts). We hence investigate this transient behavior by computing the maximum possible amplification of initial energy density 
\begin{equation} \label{eq:Growth}
    G(t;\alpha,\beta,t_0,\Rey,\kappa)=\max_{\mathbf{q}(t_0)\neq 0} \dfrac{||\mathbf{q}(t)||^2_E}{||\mathbf{q}(t_0)||^2_E}=\max_{\mathbf{q}(t_0)\neq 0} \dfrac{||\mathsfbi{A}(t)\mathbf{q}(t_0)||^2_E}{||\mathbf{q}(t_0)||^2_E},
\end{equation}
where $||\cdot||_E$ is the energy norm, the details of which we address below. We refer to the quantity $G(t)$ as amplification or growth. In the above expression, we emphasize that this amplification depends upon the wavenumbers of the perturbation, $\alpha$ and $\beta$, the time horizon $[t_0, t]$ over which the perturbation is observed, and the parameters of the base flow $\Rey$ and $\kappa$. For brevity, we omit this explicit parameter dependence in what follows.
The transient amplification is taken as the maximum relative increase (or gain) in perturbation energy that can be experienced by any initial \ALrevise{perturbation} $\mathbf{q}(t_0)$ over a given time frame $[t_0,t].$ Notably, a temporal series of $G(t)$ need not arise from the same perturbation $\mathbf{q}(t_0)$, but stem from a range of initial \ALrevise{perturbations}. Consequently, the curve $G(t)$ can be thought of as an envelope bounding the energy amplification of all initial conditions.

A common approach to solving for $G(t)$ is based on the adjoint method \citep{Andersson1999}. However, when the matrix describing the linearized equations of motion is rather small, it becomes computationally tractable to compute $\mathsfbi{A}(t)$ directly. With $\mathsfbi{A}(t)$ explicitly available, we can capitalize on the fact that Eq.\ \ref{eq:Growth} closely resembles the spectral norm of $\mathsfbi{A}(t)$, which only requires a singular value decomposition. 
However, as noted above, the energy norm, not the $L_2$-norm is used in Eq.\ \ref{eq:Growth}. For this reason, we must recast this problem as an equivalent $L_2$-norm problem to proceed.

We define the energy norm according to 
\begin{equation} \label{eq:q}
    ||\mathbf{q}||_E^2=\int_{-1}^1 \left(|D\hat{v}|^2+k^2|\hat{v}|^2+|\hat{\omega}_y|^2\right) dy.
\end{equation}
As shown in \citet{Gustavsson1986}, dividing this quantity by $2k^2$ and integrating over all wavenumbers produces the kinetic energy of $\mathbf{u}'$. As we consider the relative energy amplification for perturbations at a specific set of wavenumbers, this normalization and integration approach is not necessary for the computation of the growth $G(t)$ \ALrevise{(i.e., the normalization constant cancels out in Eq.\ \ref{eq:Growth}). In Appendix \ref{sec:AppendixD} we show how this energy norm can be recast into an equivalent $L_2$-norm by defining a matrix $\mathbf{V}$ such that $||\mathbf{q}||^2_E=||\mathbf{V}\mathbf{q}||^2$. We then compute the maximum amplification as
\begin{equation} \label{eq:Growth2}
    G(t)=\max_{\mathbf{q}(t_0)\neq 0} \dfrac{||\mathbf{V}\mathsfbi{A}(t)\mathbf{q}(t_0)||^2}{||\mathbf{V}\mathbf{q}(t_0)||^2}=\max_{\mathbf{x}(t_0)\neq 0} \dfrac{||\mathbf{V}\mathsfbi{A}(t)\mathbf{V}^{-1}\mathbf{x}(t_0)||^2}{||\mathbf{x}(t_0)||^2}=||\mathbf{V}\mathsfbi{A}(t)\mathbf{V}^{-1}||^2,
\end{equation}
where $\mathbf{x}=\mathbf{V}\mathbf{q}$ and the spectral norm of the matrix corresponds to the largest singular value.}

\begin{figure}
    \includegraphics[width=\textwidth]{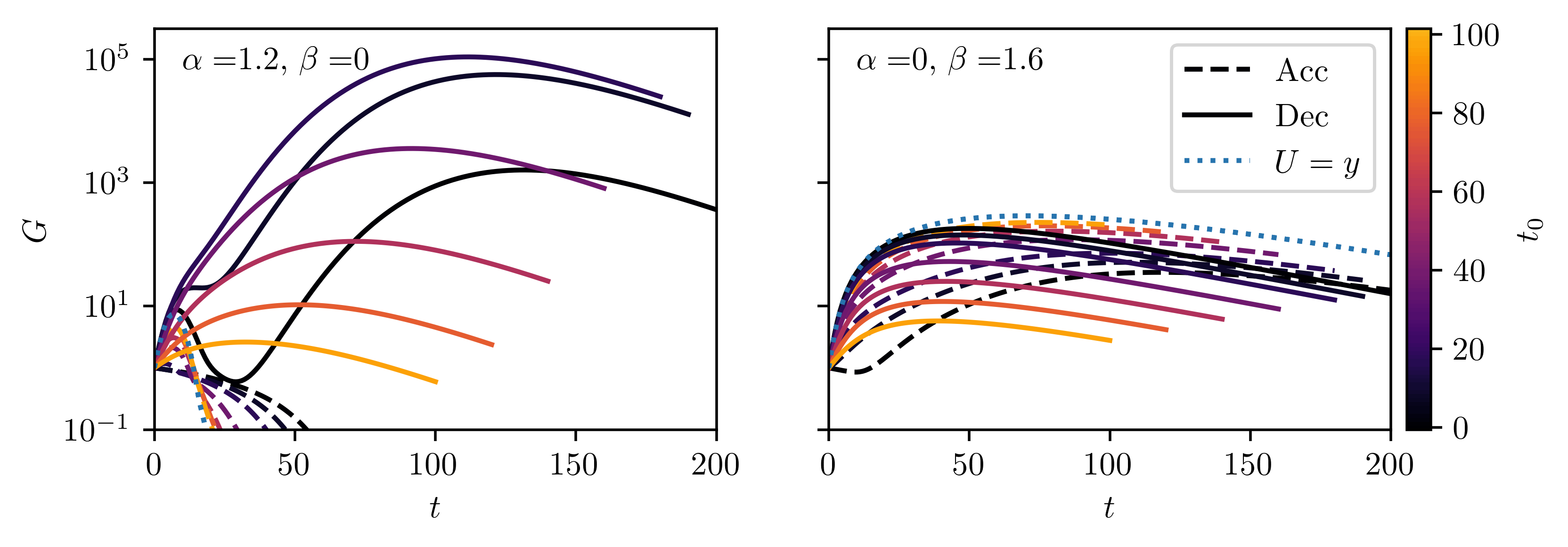}

    \captionsetup[subfigure]{labelformat=empty}
    \begin{picture}(0,0)
    \put(7,128){\contour{white}{ \textcolor{black}{a)}}}
    \put(188,128){\contour{white}{ \textcolor{black}{b)}}}
    \end{picture} 
    \begin{subfigure}[b]{0\textwidth}\caption{}\vspace{-10mm}\label{fig:CGrowtha}\end{subfigure}
    \begin{subfigure}[b]{0\textwidth}\caption{}\vspace{-10mm}\label{fig:CGrowthb}\end{subfigure}
    \vspace{-5mm}
    
    \caption{Amplification $G(t)$ for perturbations applied at different times $t_0$ to accelerating (``Acc''), decelerating (``Dec''), and constant WDF at $\Rey=500$ and $\kappa=0.1$. Amplification shown for (a) $[\alpha,\beta]=[1.2,0]$ and (b) $[\alpha,\beta]=[0,1.6]$. Perturbations at times $t_0=[0,10,20,40,60,80,100]$ are shown.}
\label{fig:CGrowth}
\end{figure}
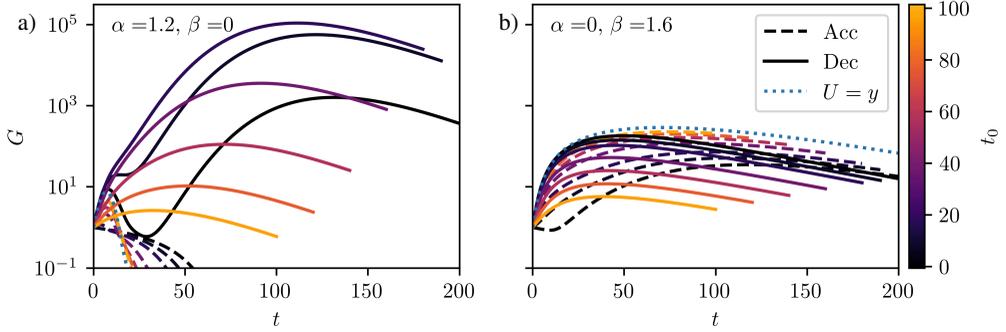

\begin{figure}
    \includegraphics[width=\textwidth]{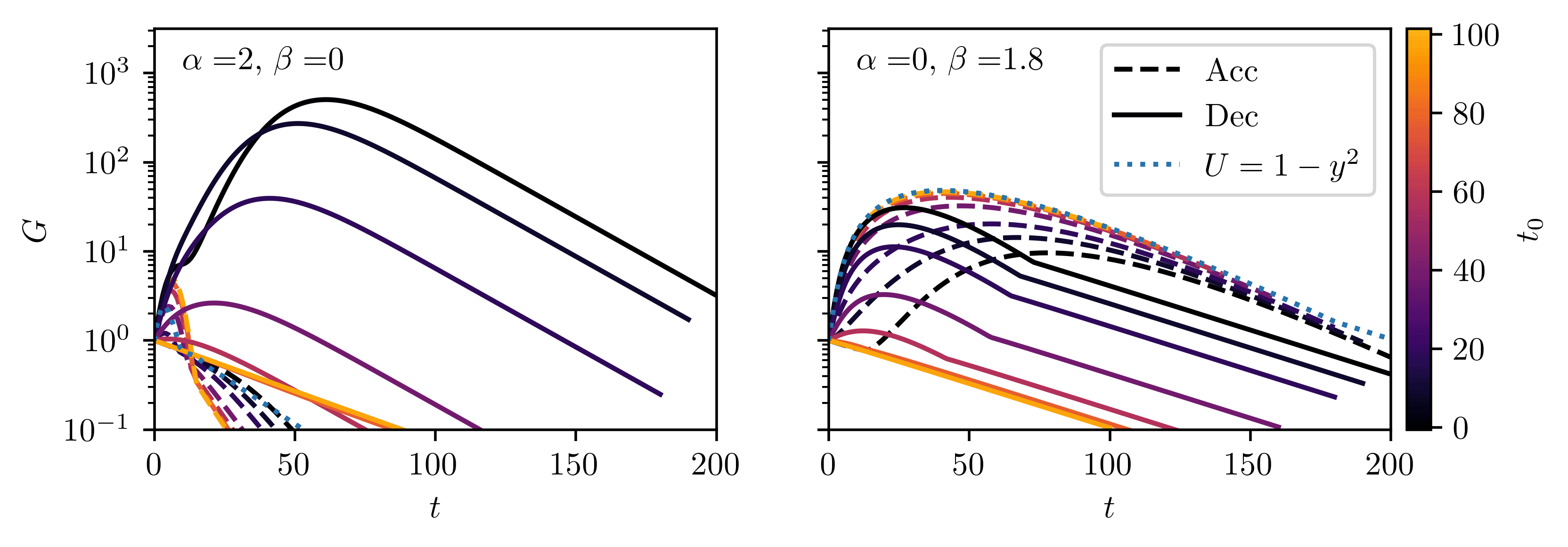}

    \captionsetup[subfigure]{labelformat=empty}
    \begin{picture}(0,0)
    \put(7,128){\contour{white}{ \textcolor{black}{a)}}}
    \put(188,128){\contour{white}{ \textcolor{black}{b)}}}
    \end{picture} 
    \begin{subfigure}[b]{0\textwidth}\caption{}\vspace{-10mm}\label{fig:PGrowtha}\end{subfigure}
    \begin{subfigure}[b]{0\textwidth}\caption{}\vspace{-10mm}\label{fig:PGrowthb}\end{subfigure}
    \vspace{-5mm}
    
    \caption{Amplification $G(t)$ for perturbations applied at different times $t_0$ to accelerating (``Acc''), decelerating (``Dec''), and constant PDF at $\Rey=500$ and $\kappa=0.1$. Amplification shown for (a) $[\alpha,\beta]=[2,0]$ and (b) $[\alpha,\beta]=[0,1.8]$. Perturbations at times $t_0=[0,10,20,40,60,80,100]$ are visualized.}
\label{fig:PGrowth}
\end{figure}
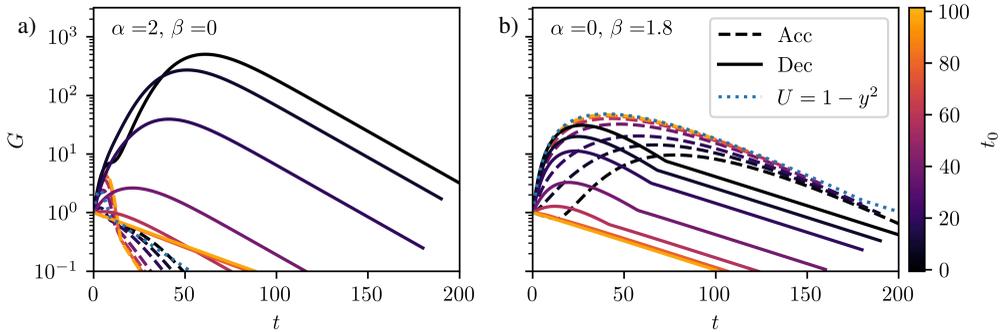

In what follows, we compute $G(t)$ by approximating $\mathbf{V}$ and $\mathsfbi{A}(t)$ on a grid of $M=64$ Chebyshev collocation points using a time step of $\Delta t=0.01$ for both WDF and PDF. We again approximate the base flow with $100$ basis functions as in Sec.\ \ref{sec:Laminar}. To illustrate how the amplification varies over time, we consider accelerating and decelerating WDF and PDF at $\Rey=500$ and $\kappa=0.1$ in Figs.\ \ref{fig:CGrowth} and \ref{fig:PGrowth}. Note that these laminar profiles were shown in Figs.\ \ref{fig:CBaseflowAcc}, \ref{fig:CBaseflowDec}, \ref{fig:PBaseflowAcc}, and \ref{fig:PBaseflowDec}.

In Fig.~\ref{fig:CGrowth}, we plot the amplification $G(t)$ with different parameters for constant, accelerating, and decelerating WDF considering strictly streamwise perturbations (Fig.\ \ref{fig:CGrowtha}) and strictly spanwise perturbations (Fig.\ \ref{fig:CGrowthb}). Most notably, streamwise perturbations subjected to the decelerating base flow result in orders of magnitude larger growth than any perturbation of the constant or accelerating flows, both of which only show small growth at early times. Additionally, the largest amplification in the decelerating cases comes from perturbing the flow at early times, but not at $t_0=0$. In contrast, the accelerating case exhibits the largest growth for perturbations at later times, when the flow behaves more like a steady flow with a constant profile. 

When only spanwise \ALrevise{perturbations} are considered (Fig.~\ref{fig:CGrowthb}) the constant profile exhibits the largest amplification of all cases. The accelerating and decelerating cases gradually transition between the amplification of the constant profile and the amplification in the case of no flow, as the time of the initial perturbation $t_0$ varies. Naturally, the accelerating case transitions from lower amplification to higher amplification, and the decelerating case transitions from higher amplification to lower amplification, as $t_0$ increases.
This suggests that the shapes of the accelerating and decelerating base flows are less important to the growth of spanwise perturbations.

In Fig.~\ref{fig:PGrowth} we show the maximum amplification for constant, accelerating, and decelerating PDF. Once again Fig.~\ref{fig:PGrowtha} illustrates that streamwise perturbations about the decelerating base flow exhibit orders of magnitude larger amplification than the other cases, and streamwise perturbations in the constant and accelerating flows experience little growth. In Fig.~\ref{fig:PGrowthb}, we again see that spanwise perturbations exhibit the largest amplification for the constant profile, and the accelerating and decelerating cases gradually move between the constant profile and the case of no flow.


Both results for WDF and PDF indicate that perturbations about decelerating laminar base flows may exhibit orders of magnitude greater amplification than perturbations about accelerating or constant laminar profiles. This amplification about decelerating flows occurs predominantly for perturbations with streamwise variations. In contrast, spanwise perturbations lead to the largest energy amplification in constant and accelerating flows.

\subsection {Maximum growth for accelerating and decelerating flows}
\label{sec:Stability2}

\begin{figure}
    \includegraphics[width=\textwidth]{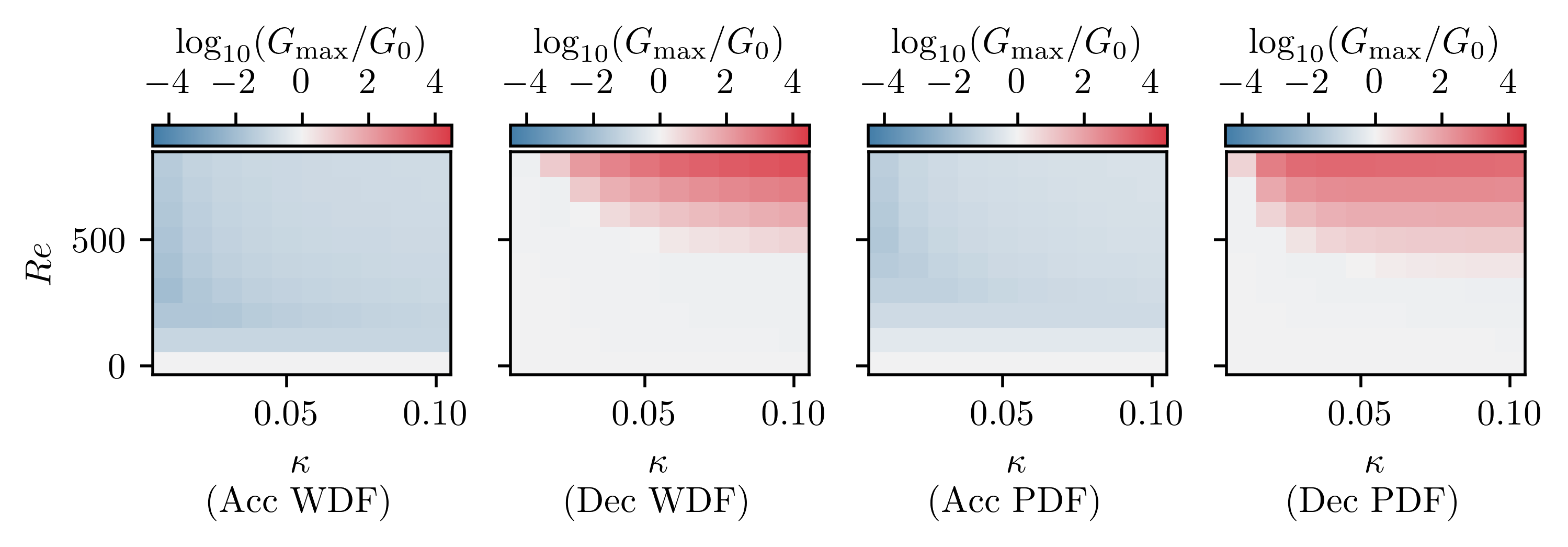}

    \captionsetup[subfigure]{labelformat=empty}
    \begin{picture}(0,0)
    \put(5,98){\contour{white}{ \textcolor{black}{a)}}}
    \put(110,98){\contour{white}{ \textcolor{black}{b)}}}
    \put(198,98){\contour{white}{ \textcolor{black}{c)}}}
    \put(285,98){\contour{white}{ \textcolor{black}{d)}}}
    \end{picture} 
    \begin{subfigure}[b]{0\textwidth}\caption{}\vspace{-10mm}\label{fig:GkRea}\end{subfigure}
    \begin{subfigure}[b]{0\textwidth}\caption{}\vspace{-10mm}\label{fig:GkReb}\end{subfigure}
    \begin{subfigure}[b]{0\textwidth}\caption{}\vspace{-10mm}\label{fig:GkRec}\end{subfigure}
    \begin{subfigure}[b]{0\textwidth}\caption{}\vspace{-10mm}\label{fig:GkRed}\end{subfigure}
    \vspace{-5mm}
    
    \caption{Maximum growth normalized by the maximum growth of perturbations in the constant flow varied over different $\Rey$ and $\kappa$: (a) for accelerating WDF, (b) for decelerating WDF, (c) for accelerating PDF, and (d) for decelerating PDF.}
\label{fig:GkRe}
\end{figure}
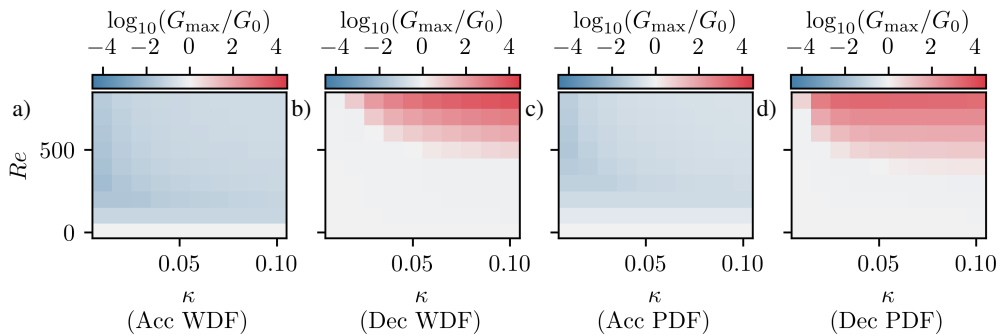

In the previous section, we computed $G(t)$ at specific values of $\Rey, \alpha, \beta,$ and $\kappa$. Here, we \ALrevise{perform a detailed examination of} the maximum growth when sweeping over these parameters, for perturbations to the laminar flows at the beginning of the acceleration or deceleration phase at $t_0=0$.
\ALrevise{Although perturbations at later $t_0$ can exhibit larger growth, the mechanisms of growth at $t_0=0$ and at the optimal $t_0$ tend to be similar in this work. In section \ref{sec:Stability5} we further investigate the effect of optimizing over $t_0$ to illustrate this point.}
In Fig.~\ref{fig:GkRe} we present the maximum growth $G_\text{max}=\max_{\alpha,\beta,t} G(t)$ over a set of $\Rey$ and $\kappa$ for accelerating and decelerating WDF and PDF. As a point of reference, we normalize the growth by the maximum value \ALrevise{$G_0$} obtained from the constant WDF or PDF case. 
Both accelerating WDF and PDF exhibit less growth than the constant laminar flow. The growth relative to the constant laminar profile is lowest at the lowest value of $\kappa$ and moderately low Reynolds number $\Rey$. As $\kappa$ and $\Rey$ increase, the growth appears to level out at around a tenth of the constant flow.

In contrast to the accelerating laminar cases, the decelerating laminar cases exhibit far larger amplification of perturbations than the constant laminar case. For WDF at $\Rey=800$, we see $10^4$ times larger amplification over the constant profile. Upon further increasing $\kappa$ and $\Rey$, the relative amplification continues to grow. Decelerating PDF also exhibits this large amplification at high $\Rey$ and $\kappa$.

The two competing factors are the rate $\kappa$ at which the walls move and the rate at which the flow can react to this motion, i.e., $1/\Rey$. At large values of $\kappa$, the wall motion is fast and the laminar profile is sensitive to changes in $\Rey$ while insensitive to changes in $\kappa$. At the other extreme of small $\kappa$, the laminar flows change so slowly that the growth of perturbations behaves similarly to the growth of perturbations in the constant laminar case or the no-flow case.


\begin{figure}
    \includegraphics[width=\textwidth]{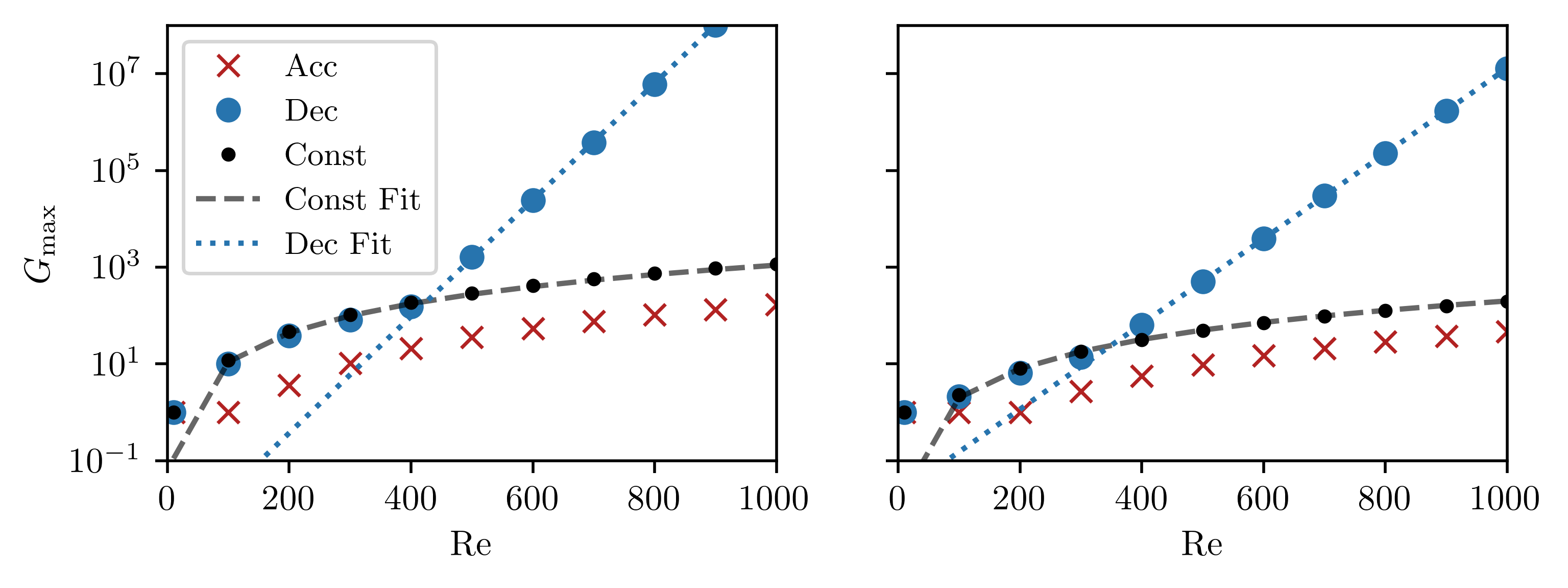}

    \captionsetup[subfigure]{labelformat=empty}
    \begin{picture}(0,0)
    \put(10,140){\contour{white}{ \textcolor{black}{a)}}}
    \put(202,140){\contour{white}{ \textcolor{black}{b)}}}
    \end{picture} 
    \begin{subfigure}[b]{0\textwidth}\caption{}\vspace{-10mm}\label{fig:Gscalea}\end{subfigure}
    \begin{subfigure}[b]{0\textwidth}\caption{}\vspace{-10mm}\label{fig:Gscaleb}\end{subfigure}
    \vspace{-5mm}
    
    \caption{$G_{\text{max}}$ as a function of $\Rey$ for accelerating flow (``Acc''), decelerating flow  (``Dec''), and constant flow (``Const'') at fixed $\kappa=0.1$ for (a) WDF and (b) PDF. The fitting lines are described in the text.}
\label{fig:Gscale}
\end{figure}
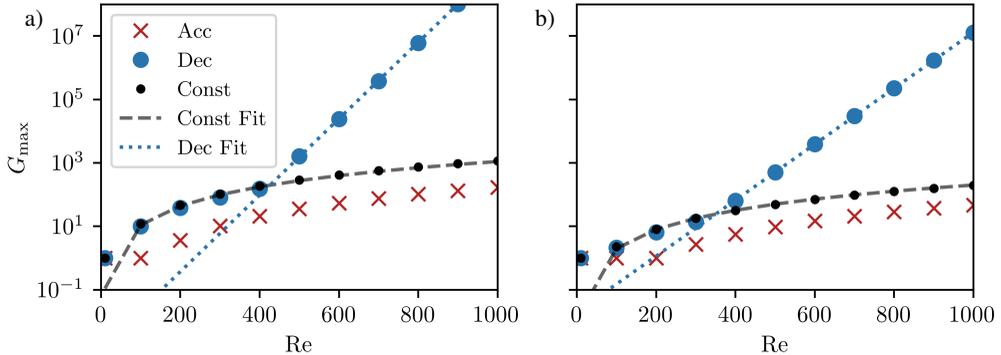

\ALrevise{We focus on this range of $\Rey$ and $\kappa$ values because these values show where a transition in growth is exhibited in the decelerating cases. As $\kappa$ continues to increase, the results converge toward impulsive wall motion (i.e., Stokes' first problem). Similarly, if we continue to increase the Reynolds number, we will find that, at moderate $\kappa$ values, there is a consistent scaling in the maximum amplification as the Reynolds number changes.}
\ALrevise{Figure \ref{fig:Gscale} shows how $G_\text{max}$ varies with $\Rey$ at $\kappa=0.1$ for constant, accelerating, and decelerating WDF and PDF.}
We also show the $\Rey^2$ scaling discussed in \citet{Reddy1993} as ``Const Fit'' (using the same values) and the best exponential fit for the decelerating cases at $\Rey>400$ as ``Dec Fit'' ($G_\text{max}=0.0015\times 10^{0.012\Rey}$ for WDF and $G_\text{max}=0.0199\times 10^{0.009\Rey}$ for PDF). Both the constant and accelerating cases exhibit similar trends over all Reynolds numbers $\Rey$, while the decelerating case exhibits two distinct behaviors. At low $\Rey$, the decelerating case shows the same $\Rey^2$ scaling as the constant case, and at high $\Rey$ the decelerating case shows a $10^{\Rey}$ scaling. \ALrevise{As mentioned in Sec.\ \ref{sec:Intro}, this behavior was also observed in oscillatory flows shown in \citet{Xu2021}. However, the optimal perturbations in their work did not exhibit a shift wavenumber as $\Rey$ increased, which, as we will show, is not the case for these decelerating flows.} 
For low values of $\Rey,$ viscous forces quickly modify the flow in response to the wall velocity, and the growth behaves similarly to a constant flow. When the Reynolds number is high, the flow reacts more slowly to the wall motion, allowing the laminar base flow to exhibit high curvature, leading to the $10^{\Rey}$ scaling. However, unlike the decelerating case, the accelerating case only exhibits the $\Rey ^2$ scaling. Thus, we must further examine the shape of the perturbations to explain this difference in scaling.

\begin{figure}
    \includegraphics[width=\textwidth]{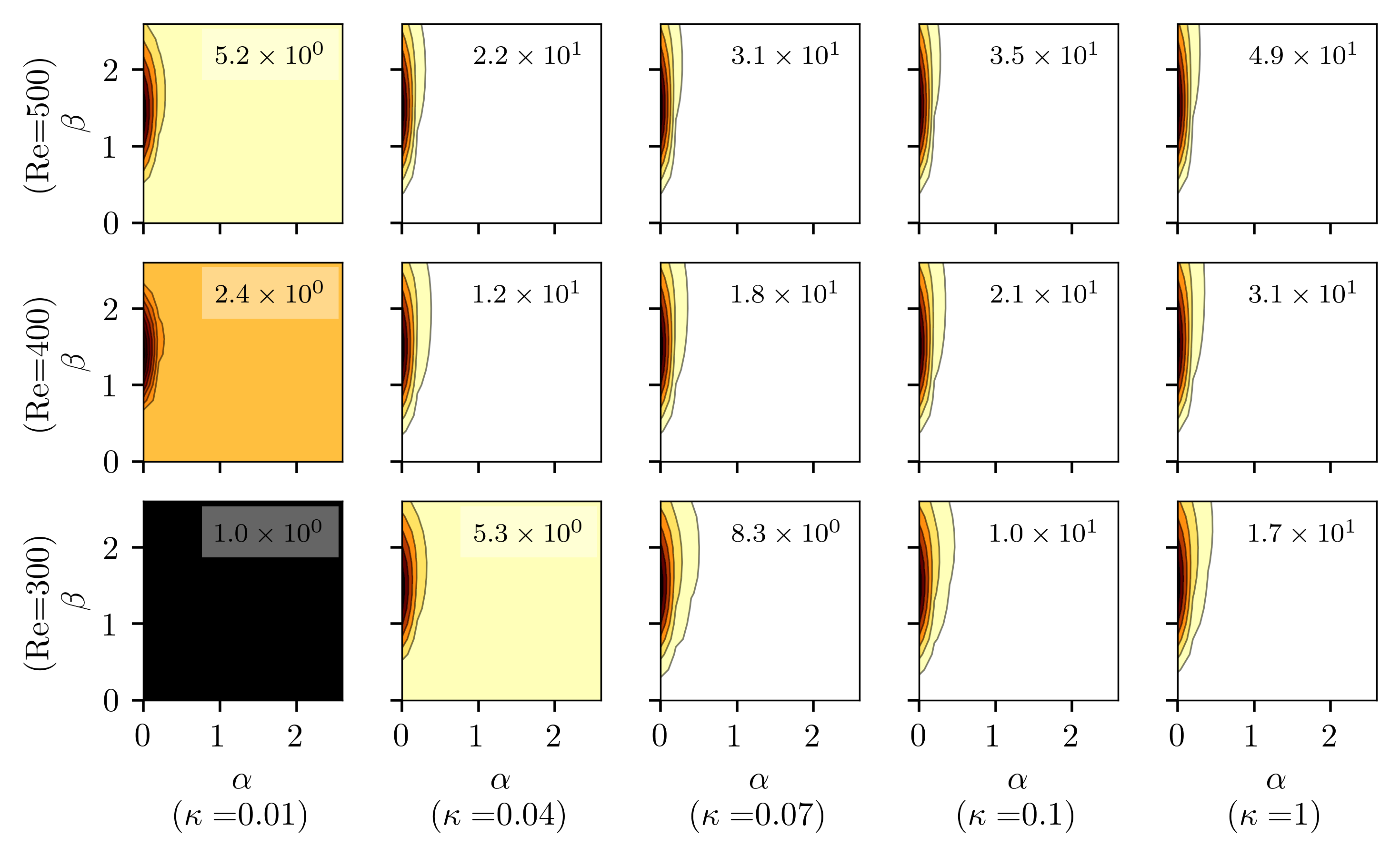}
  \caption{Maximum amplification of perturbations $\max_tG(t)$ for accelerating WDF at different wavenumbers and acceleration rates (denoted in the figure).}
\label{fig:PCFaccsweep}
\end{figure}

\begin{figure}
    \includegraphics[width=\textwidth]{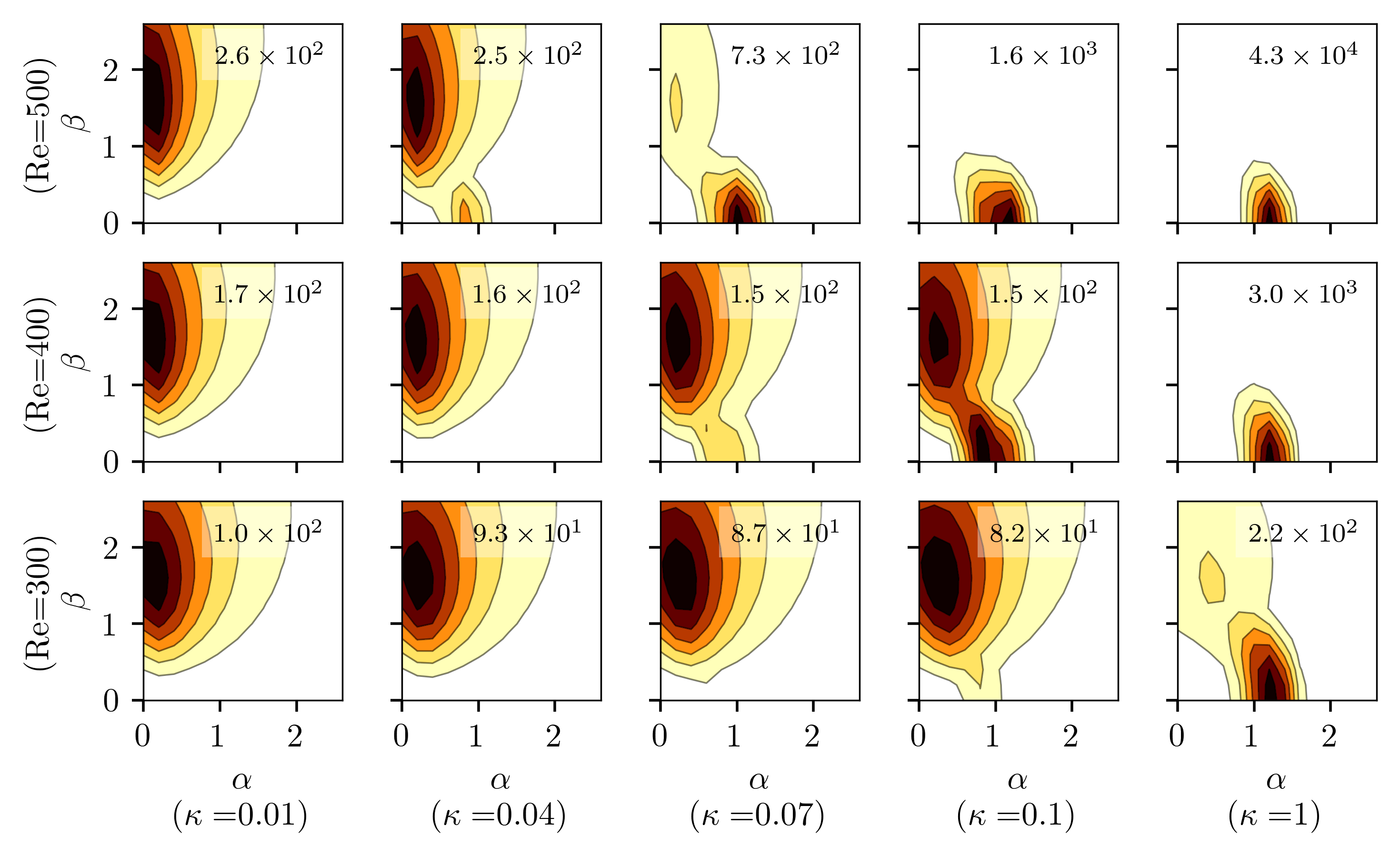}
  \caption{Maximum amplification of perturbations $\max_tG(t)$ for decelerating WDF at different wavenumbers and deceleration rates (denoted in the figure).}
\label{fig:PCFdecsweep}
\end{figure}

To better understand the cause for decreased amplification in the accelerating case and increased amplification in the decelerating case, we consider the wavenumbers at which the amplification occurs. Figure~\ref{fig:PCFaccsweep} shows the maximum amplification $\max_tG(t)$ at multiple values of $\Rey$, $\kappa$, $\alpha$, and $\beta$ for accelerating WDF. Analogously, Fig.~\ref{fig:PCFdecsweep} shows the same results for decelerating WDF. Each figure is normalized by the maximum amplification over all wavenumbers, which is indicated near the top right corner of each plot. 

For accelerating WDF, the most amplified perturbations are ones that vary in the spanwise direction at $[\alpha,\beta]\approx [0,1.6]$. Likewise, the constant laminar profile exhibits maximal amplification for perturbations that vary in the spanwise direction at the same spanwise wavenumber \citep{Reddy1993}. 
Referring back to Eq.\ \ref{eq:LinearEq}, we see that for $\alpha=0$, $\mathscr{L}_\text{os}$ and $\mathscr{L}_\text{sq}$ no longer depend on the laminar base flow. Thus, the influence of the laminar profile only enters through the derivative of the laminar profile in the coupling term $\mathscr{L}_c$. This observation likely explains the weaker dependence on shape exhibited in the amplification shown in Figs.~\ref{fig:CGrowthb} and \ref{fig:PGrowthb}.

\begin{figure}
    \includegraphics[width=\textwidth]{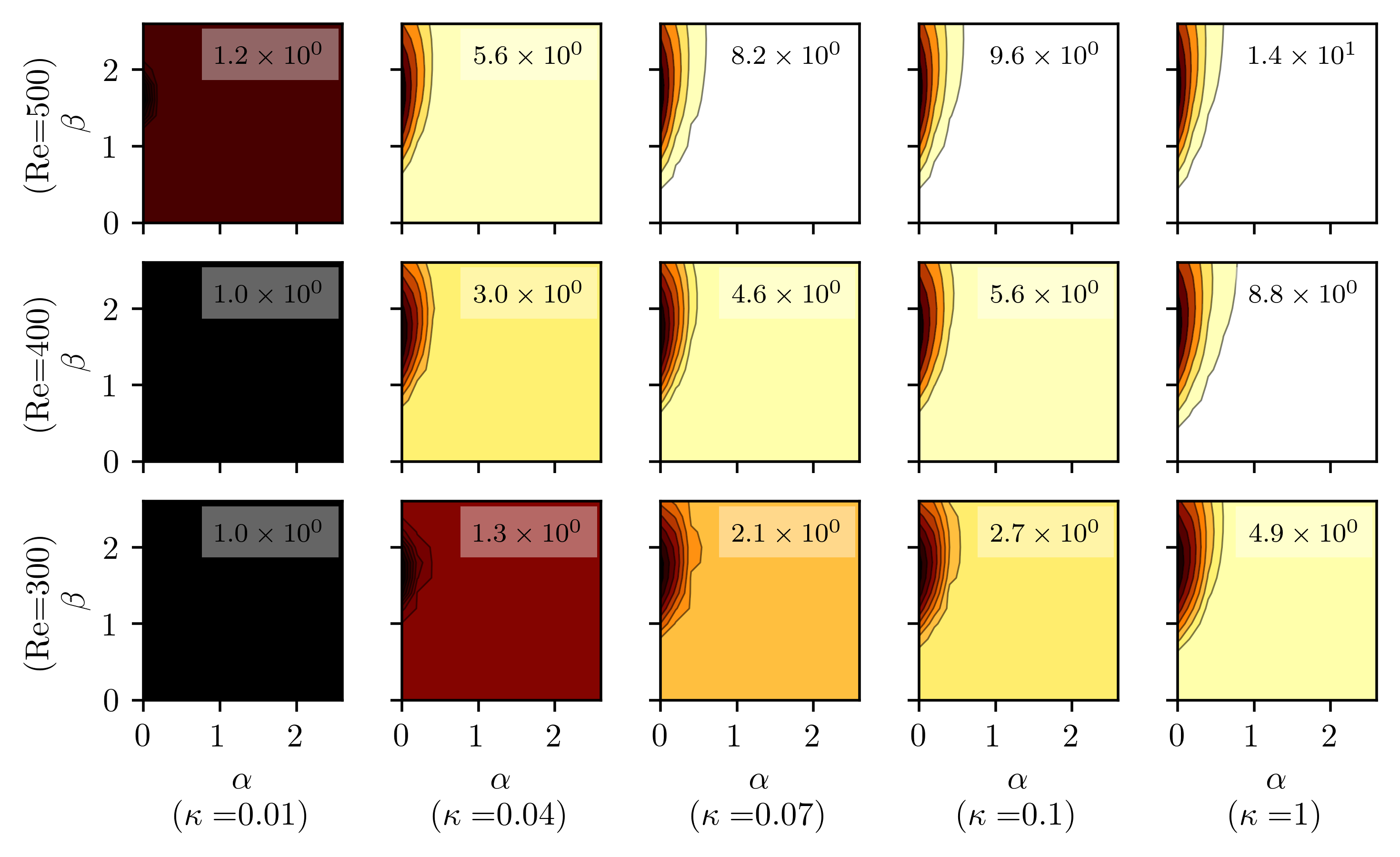}
  \caption{Maximum amplification of perturbations $\max_tG(t)$ for accelerating PDF at different wavenumbers and acceleration rates (denoted in the figure).}
\label{fig:PPFaccsweep}
\end{figure}

Perturbations in decelerating laminar flows exhibit a much different dependency on the wavenumbers as $\Rey$ and $\kappa$ vary. At a low rate of deceleration $\kappa$, the largest maximum amplification arises from spanwise perturbations. Upon increasing $\kappa,$ the largest maximum amplification moves towards perturbations with streamwise variations. A similar trend holds for the Reynolds number, such that for $\Rey=500$ and $\kappa=1$ the maximum amplification is due to a streamwise varying perturbation with wavenumbers $[\alpha,\beta]\approx [1.2,0]$. Referring back to Eq.\ \ref{eq:LinearEq}, we notice that for $\beta=0$ the operator $\mathscr{L}_\text{c}$ vanishes, hence decoupling $\hat{v}$ and $\hat{\omega}_y$. Furthermore, spanwise wavenumbers cause $\mathscr{L}_\text{os}$ and $\mathscr{L}_\text{sq}$ to both depend on the laminar profile. This indicates that the high levels of amplification to perturbations in the decelerating profiles are directly due to the influence of the velocity and second derivative of the velocity for the decelerating laminar flow, and not due to the interactions between wall-normal velocity and vorticity, as is the case for accelerating and constant profiles. 

\begin{figure}
    \includegraphics[width=\textwidth]{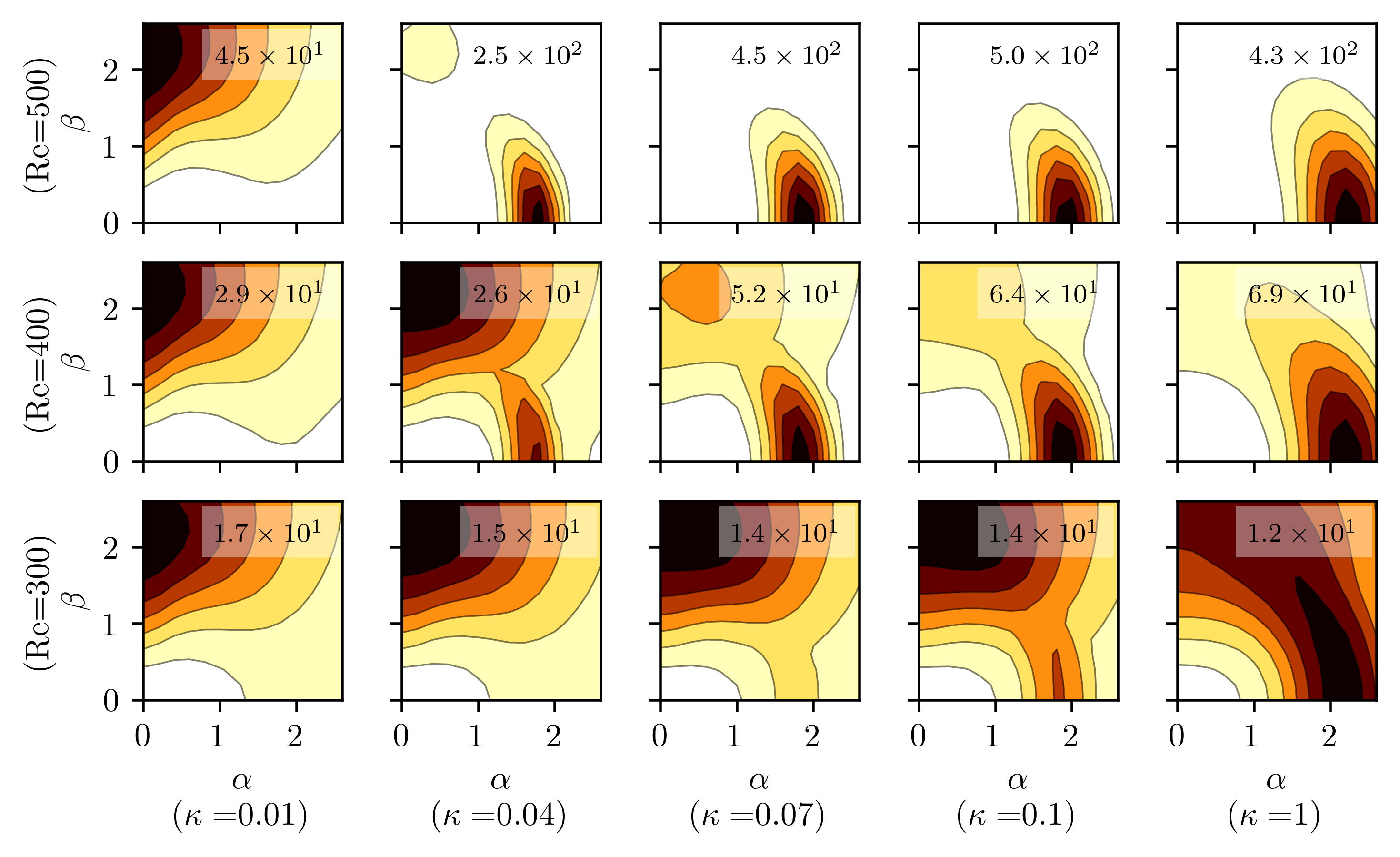}
  \caption{Maximum amplification of perturbations $\max_tG(t)$ for decelerating PDF at different wavenumbers and deceleration rates (denoted in the figure).}
\label{fig:PPFdecsweep}
\end{figure}

Next, we investigate the wavenumbers that lead to the largest amplification for accelerating and decelerating PDF. In Figs.\ \ref{fig:PPFaccsweep} and \ref{fig:PPFdecsweep}, we show the maximum amplification for accelerating and decelerating PDF, respectively. For accelerating flow, there is no growth, or very small growth, at low $\kappa$. As $\kappa$ and $\Rey$ increase, the maximum amplification increases, and the maximum amplification localizes to spanwise wavenumbers.
The maximum amplification occurs at a streamwise wavenumber of $\beta\approx 1.8$, while the constant laminar flow exhibits maximum amplification at a wavenumber of $\beta\approx 2.04$ \citep{Trefethen1993}. 


In the case of decelerating flow, we again see a gradual shift from spanwise dominant to streamwise dominant perturbations. At low values of $\kappa$ and $\Rey$, the maximum amplification is centered at $[\alpha,\beta]\approx[0,2.1]$, similar to the constant profile. On the other extreme, i.e., high values of $\kappa$ and $\Rey$, the maximum amplification is centered on streamwise perturbations with a wavenumber of $[\alpha,\beta] \approx [2.2,0]$. As $\kappa$ and $\Rey$ increase, the maximum amplification shifts between these two types of \ALrevise{perturbations}. 
Moreover, if the Reynolds number is too low, the maximum amplification will never drift towards streamwise \ALrevise{perturbations}, and at higher values of $\Rey,$ the range of $\kappa$ over which this transition occurs becomes much smaller. This transition is reflected in Fig.~\ref{fig:GkRe} where the maximum growth is plotted versus $\Rey$ and $\kappa.$ 

\subsection {Structure of optimal perturbations}
\label{sec:Stability3}

In the previous sections, we explored how the accelerating laminar flow displays maximum amplification for spanwise perturbations, while the decelerating laminar flow shows maximum amplification for streamwise perturbations. These results indicate the shape of the maximally amplified perturbation in the periodic ccordinate directions, but not in the wall-normal direction. In this section, we investigate the structure of the specific perturbations that develop to $\max_t G(t)$. We first outline the calculation of the optimal perturbation and then follow the amplification of that perturbation through time, along with its spatial profile.

For calculating the optimal perturbation, recall that the amplification at some time $t'$ is the energy gain of a perturbation that leads to the largest energy growth over all unit-norm initial perturbations $\mathbf{q}_0$. We compute this quantity by performing the singular value decomposition
\begin{equation}
    \mathbf{V}\mathsfbi{A}(t')\mathbf{V}^{-1}=\mathsfbi{U}_S\Sigma \mathsfbi{V}_S^T,
\end{equation}
where $\mathsfbi{U}_S$ contains the left singular vectors, $\mathsfbi{V}_S$ contains the right singular vectors, and $\Sigma$ contains the singular values, ordered by size $\sigma_1\geq \sigma_2 \geq ... \geq \sigma_N,$ along its diagonal. The maximum amplification is then given by $\sigma_1^2$, and if we consider only the growth along this leading direction we obtain
\begin{equation}
    \mathbf{V}\mathsfbi{A}(t')\mathbf{V}^{-1}\mathbf{v}_{S,1}=\sigma_1 \mathbf{u}_{S,1},
\end{equation}
where $\mathbf{u}_{S,1}$ and $\mathbf{v}_{S,1}$ are the principal left and right singular vectors, respectively.
The right singular vector $\mathbf{v}_{S,1}$ is transformed via the mapping $\mathbf{V}\mathsfbi{A}(t')\mathbf{V}^{-1}$ onto $\mathbf{u}_{S,1}$ and stretched by a factor of $\sigma_1$. 
Owing to this relationship, we conclude that the initial perturbation leading to the maximum amplification at some time $t'$ is given by $\mathbf{v}_{S,1}$. We can subsequently evolve this perturbation in time according to the linearized equations of motion 
\begin{equation}
    \mathbf{v}(t)=\mathbf{V}\mathsfbi{A}(t)\mathbf{V}^{-1}\mathbf{v}_{S,1},
\end{equation}
and, given $\mathbf{v}_{S,1}=1$, we can evaluate the perturbation energy as $G_p(t)=||\mathbf{v}(t)||^2$. In what follows, we compute the perturbation $\mathbf{v}_{S,1}$ from time $t'$ at which $G(t)$ is maximized ($t'=\text{argmax}_t G(t)$). We then display the evolution of the energy and the profile of $\mathbf{v}(t)$ through time. It is important to realize that the amplification that maximizes $G(t')$ at some time $t'$ need not necessarily maximize $G(t)$ for any other time.


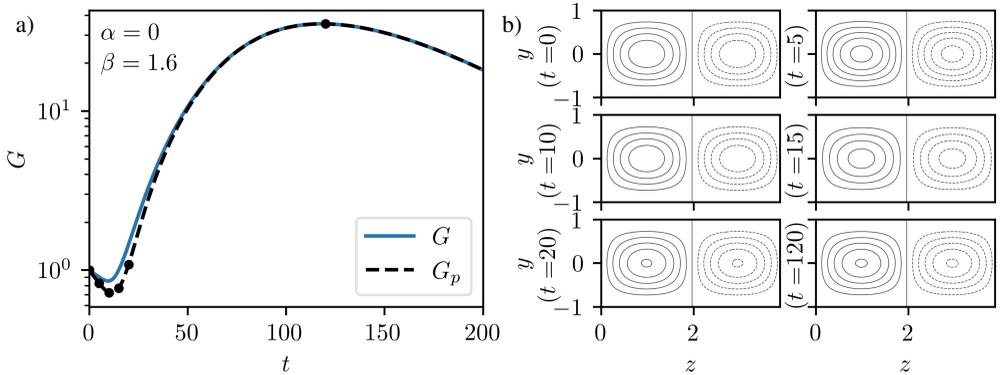
\begin{figure}
    \includegraphics[width=\textwidth]{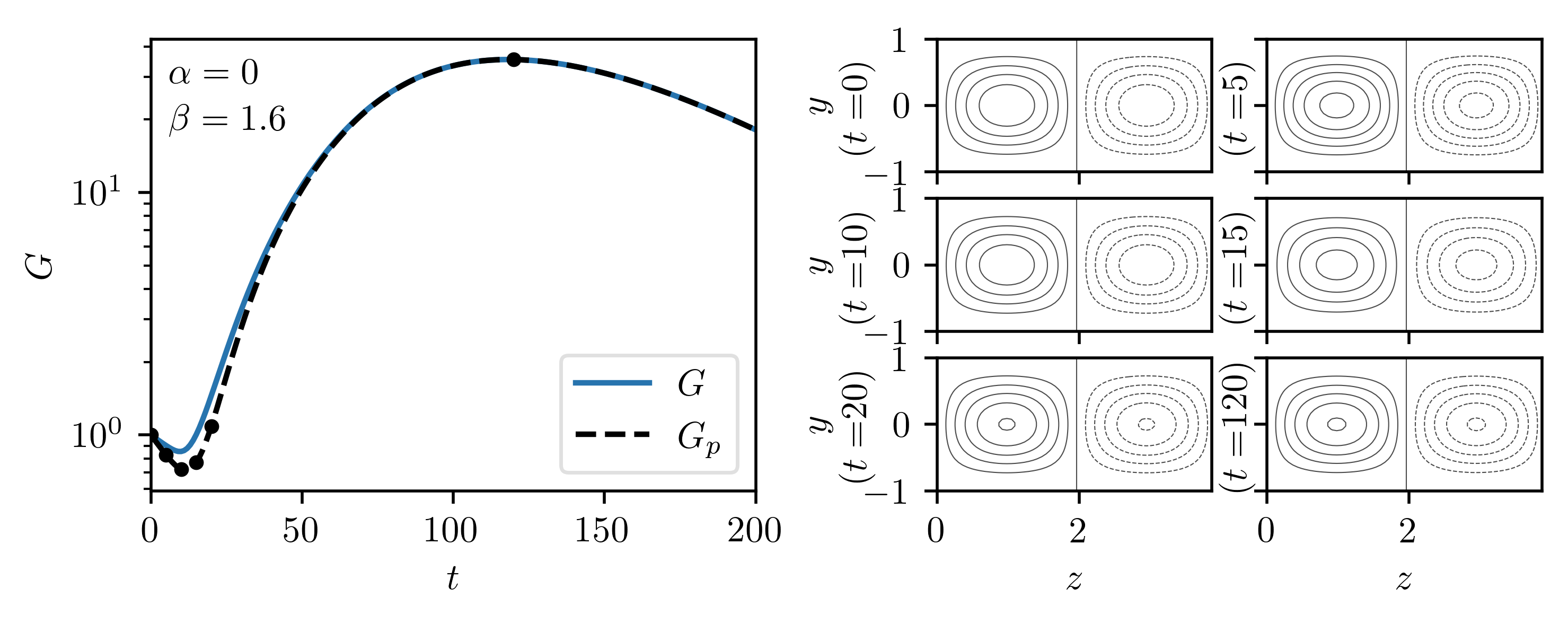}

    \captionsetup[subfigure]{labelformat=empty}
    \begin{picture}(0,0)
    \put(7,145){\contour{white}{ \textcolor{black}{a)}}}
    \put(190,145){\contour{white}{ \textcolor{black}{b)}}}
    \end{picture} 
    \begin{subfigure}[b]{0\textwidth}\caption{}\vspace{-10mm}\label{fig:PCFaccga}\end{subfigure}
    \begin{subfigure}[b]{0\textwidth}\caption{}\vspace{-10mm}\label{fig:PCFaccgb}\end{subfigure}
    \vspace{-5mm}
    
    \caption{(a) Energy of the optimal perturbation and the envelope of growth for accelerating WDF \ALrevise{($\Rey=500$, $\kappa=0.1$)}. (b) Stream function of the perturbation as it evolves in time.}
\label{fig:PCFaccg}
\end{figure}

\begin{figure}
    \includegraphics[width=\textwidth]{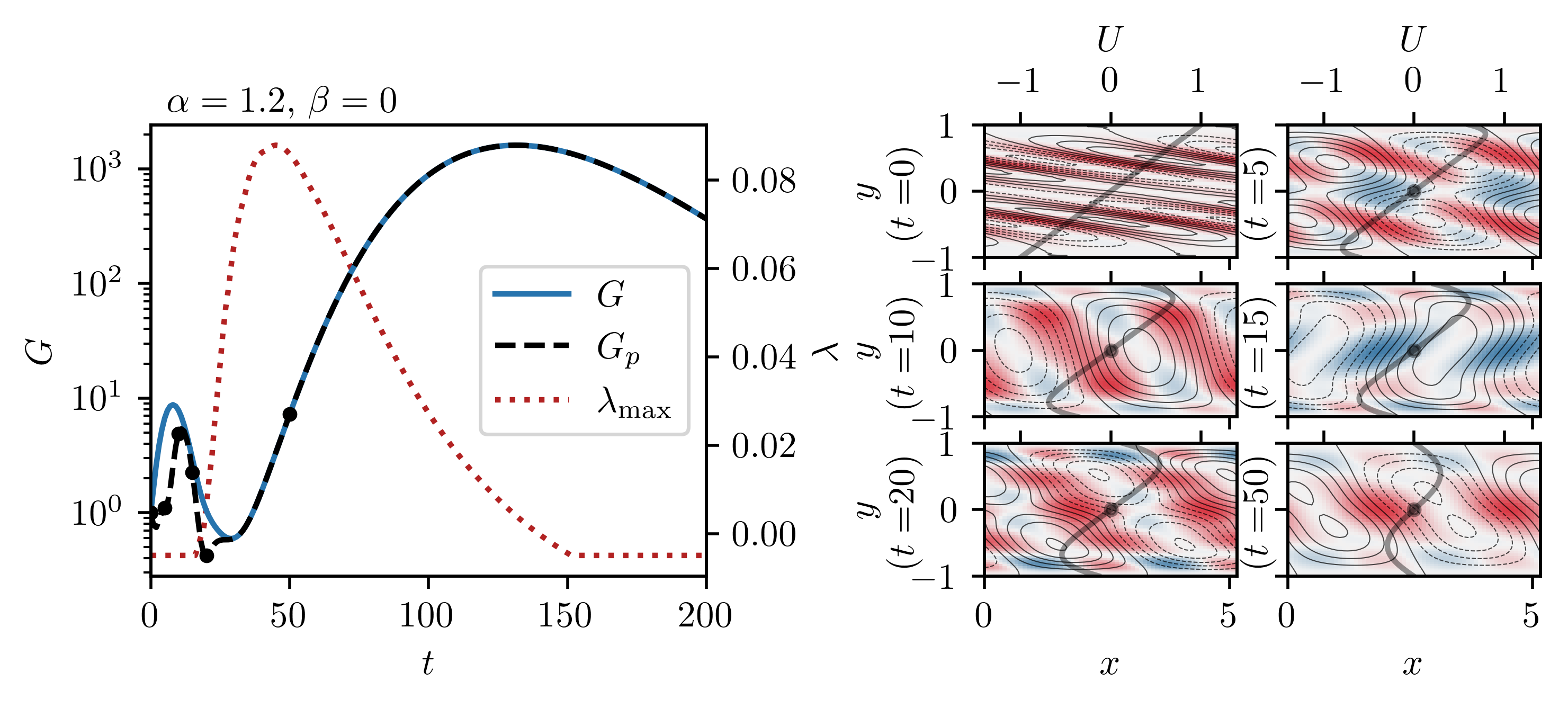}

    \captionsetup[subfigure]{labelformat=empty}
    \begin{picture}(0,0)
    \put(7,145){\contour{white}{ \textcolor{black}{a)}}}
    \put(195,145){\contour{white}{ \textcolor{black}{b)}}}
    \end{picture} 
    \begin{subfigure}[b]{0\textwidth}\caption{}\vspace{-10mm}\label{fig:PCFdecga}\end{subfigure}
    \begin{subfigure}[b]{0\textwidth}\caption{}\vspace{-10mm}\label{fig:PCFdecgb}\end{subfigure}
    \vspace{-5mm}
    
    \caption{\ALrevise{(a) Energy of the optimal perturbation, the envelope of growth, and the instantaneous eigenvalue for decelerating WDF ($\Rey=500$, $\kappa=0.1$). (b) Thin lines are the stream function of the perturbation as it evolves in time. The filled contour plot visualizes the production (Eq.\ \ref{eq:Prod}) normalized by the maximum absolute value (red indicates positive values and blue indicates negative values). The thick black line is the reference laminar profile ($U$), with a dot at the inflection point.}}
\label{fig:PCFdecg}
\end{figure}

\ALrevise{\subsubsection{Optimal perturbations in wall-driven flows}}
First, we consider the accelerating and decelerating WDF cases. In Fig.~\ref{fig:PCFaccga} we show the energy of the optimal perturbation over time for accelerating WDF at $\Rey=500$ and $\kappa=0.1$. For this flow, we compute the optimal perturbation that reaches the maximum amplification at $t'=118$. At short times, the energy of the optimal perturbation falls below the envelope of $G(t)$, and at long times the energy of the optimal perturbation matches $G(t)$. This indicates that other perturbations lead to larger growth at shorter times, but the energy of these perturbations will not reach the long-time energy achieved by the optimal perturbation. In Fig.~\ref{fig:PCFaccgb} we show the evolution of the perturbation at the times indicated in Fig.~\ref{fig:PCFaccga}. We visualize the perturbation with contours of the stream function. The perturbation takes the form of streamwise vortices that initially decay in magnitude, but subsequently grow as the flow accelerates toward the simple shear profile. The shape of this \ALrevise{perturbation} closely resembles that of the optimal \ALrevise{perturbation} in constant WDF and grows via a ``vortex-tilting'' mechanism \citep{Butler1992}. Unlike constant WDF, the perturbation in the accelerating flow exhibits an initial drop in energy due to the transient period in which the accelerating profile develops towards the simple shear profile.

\begin{figure}
    \includegraphics[width=\textwidth]{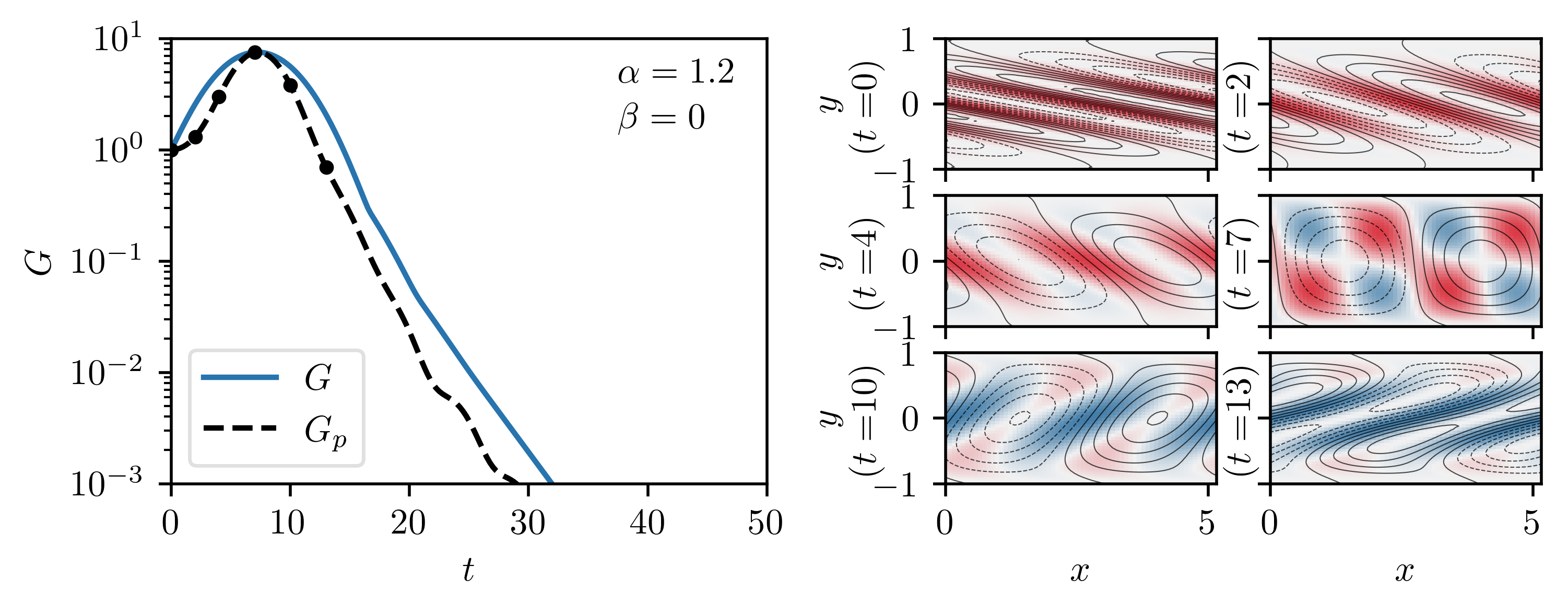}

    \captionsetup[subfigure]{labelformat=empty}
    \begin{picture}(0,0)
    \put(7,145){\contour{white}{ \textcolor{black}{a)}}}
    \put(190,145){\contour{white}{ \textcolor{black}{b)}}}
    \end{picture} 
    \begin{subfigure}[b]{0\textwidth}\caption{}\vspace{-10mm}\label{fig:PCFconstga}\end{subfigure}
    \begin{subfigure}[b]{0\textwidth}\caption{}\vspace{-10mm}\label{fig:PCFconstgb}\end{subfigure}
    \vspace{-5mm}
    \caption{\ALrevise{(a) Energy of the optimal perturbation and the envelope of growth for constant WDF ($\Rey=500$). (b) Thin lines are the stream function of the perturbation as it evolves in time. The filled contour plot visualizes the production (Eq.\ \ref{eq:Prod}) normalized by the maximum absolute value (red is positive and blue is negative).}}
\label{fig:PCFconstg}
\end{figure}
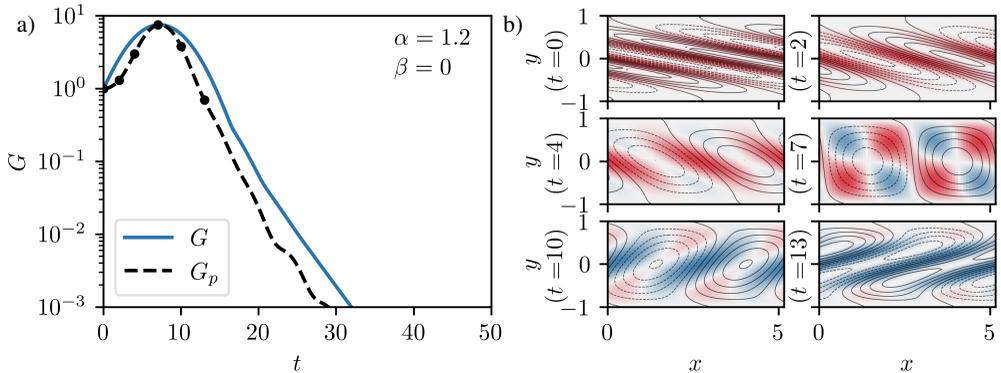

The dynamics of the perturbation in the decelerating case behave much differently from the constant WDF. Figure~\ref{fig:PCFdecga} shows the energy of the optimal perturbation for decelerating WDF \ALrevise{and the largest real part of the eigenvalue of the time-varying linear operator $\lambda_{\text{max}}$} at $\Rey=500$ and $\kappa=0.1$. In this case, we compute the optimal perturbation that reaches its maximum amplitude at $t'=132$. Similar to the accelerating case, the energy of the perturbation falls below $G(t)$ at early times but matches $G(t)$ at long times. Unlike the accelerating case, the perturbation in the decelerating case exhibits initial growth, followed by decay, before growing again to reach much larger energy values. \ALrevise{Notably, the second peak in the growth begins when the instantaneous eigenvalue becomes positive and decays when the eigenvalue becomes negative. }

Along with the energy of the optimal perturbation, in Fig.~\ref{fig:PCFdecgb} we show 
the evolution of the optimal perturbation for decelerating WDF. \ALrevise{We also plot the laminar profile and the spatial energy production
\begin{equation} \label{eq:Prod}
    P=-u'v'\dfrac{\partial U}{\partial y}
\end{equation}
to identify the cause of energy growth. Note that this term lies in the production term of the energy equation \citep{Serrin1959,Conrad1965}
\begin{equation}
    \dfrac{dE}{dt}=\dfrac{d}{dt}\int_V \dfrac{\mathbf{u}'\cdot\mathbf{u}'}{2}dV=-\int_V \dfrac{1}{\Rey} \nabla^2 \mathbf{u}' : \nabla^2 \mathbf{u}'dV-\int_V \mathbf{u}'\cdot\nabla \mathbf{U} \cdot \mathbf{u}'dV.
\end{equation}
In the inviscid case, for a $y$-dependent streamwise baseflow, this equation simplifies to 
\begin{equation}
    \dfrac{dE}{dt}=\int_V -u'v'\dfrac{dU}{dy}dV=\int_V \dfrac{\partial \psi}{\partial y}\dfrac{\partial \psi}{\partial x}\dfrac{dU}{dy}dV=\int_V - \left . \dfrac{dy}{dx}\right |_\psi \dfrac{dU}{dy}dV,
\end{equation} 
where $\psi$ represents the stream function.
Thus, energy increases when the term under the integral is negative, or when the streamlines align opposite the gradient of the laminar flow. This mechanism of growth is referred to as the Orr mechanism or the down-gradient Reynolds stress mechanism in \citet{Butler1992}.}


At $t=0$ the stream function of the initial \ALrevise{perturbation} opposes the laminar shear, causing the initial gain in energy. The streamlines are advected by the laminar profile causing them to rotate and break up at $t=2$ and $t=5$. Over this time window, the energy grows. Then, at $t=10$ the streamlines become vertical and eventually align with the laminar shear at $t=15$ causing the energy to decrease. Finally, the streamlines once again advect and align opposite to the laminar shear, over much of the domain, causing the energy to increase. After $t\approx 50$, there is little change in the shape of the perturbation, which we show at $t=70$.

\ALrevise{Throughout this time series, the energy production follows the anti-aligned streamlines supporting the claim that the Orr mechanism is responsible for this energy growth. In \citet{Moron2022}, the production was also shown to coincide with the inflection point for pulsatile flows. For WDF, the inflection point always remains in the center of the channel. At early times, the location of the inflection point does not match energy production, 
but, at later times, energy is produced at the inflection point. It remains unclear if the inflection point plays an important role in our understanding of the stability of this problem. One approach to further investigate the significance of the inflection point could be to use a master-slave model that eliminates this inflection point, as was performed in \citet{Moron2024}.}

\ALrevise{Next, we show why this Orr mechanism leads to minimal growth in accelerating and constant WDF.}
Figure~\ref{fig:PCFconstg} shows the energy and evolution of the optimal perturbation for constant WDF at the same wavenumbers as in the decelerating case and for $t'=7.25$. The perturbation in constant WDF reaches a similar amplification to the first peak in the decelerating case, but then rapidly dies off. Figure~\ref{fig:PCFconstgb} shows the evolution of this perturbation. The initial \ALrevise{perturbation} closely resembles the perturbation in the decelerating case, and it follows a similar evolution up to $t=7$ wherein the streamlines initially oppose the laminar shear before aligning vertically. However, in contrast to the decelerating case, the constant laminar profile rotates the streamlines and retains this alignment with the laminar shear causing the energy of the perturbation to drop drastically, as shown at $t=10$ and $t=13$. 

We emphasize the key difference in the evolution of the decelerating flow and the accelerating or constant flow is this long-time alignment of the stream function. Streamwise perturbations in decelerating flows exhibit large growth because the stream function aligns opposite the laminar shear. Streamwise perturbations in accelerating or constant flows exhibit small growth because the stream function aligns with the laminar shear. \ALrevise{Thus, a key conclusion of our results is that there is a range of critical Reynolds numbers and deceleration rates where the optimal perturbations switch from spanwise to streamwise perturbations for decelerating flows. This transition occurs when growth due to vortex-tilting is outpaced by growth due to the Orr mechanism. Once the Orr mechanism dominates, the maximum amplification scales exponentially with the Reynolds number resulting in massive amplification not experienced in accelerating or constant flows.}

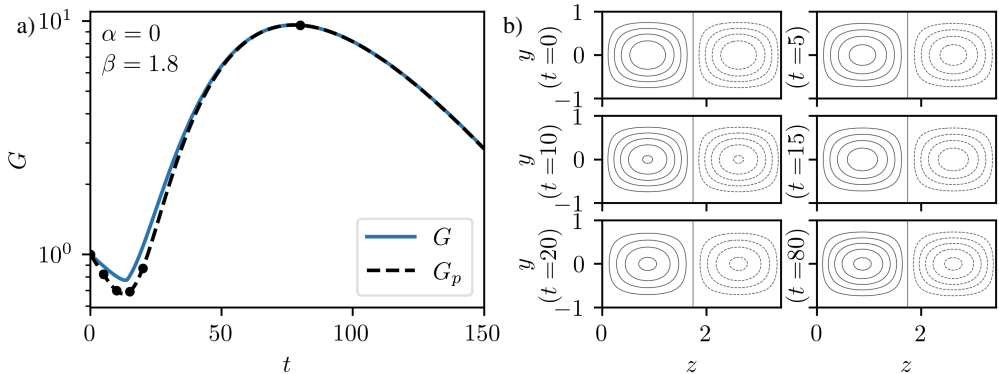
\begin{figure}
    \includegraphics[width=\textwidth]{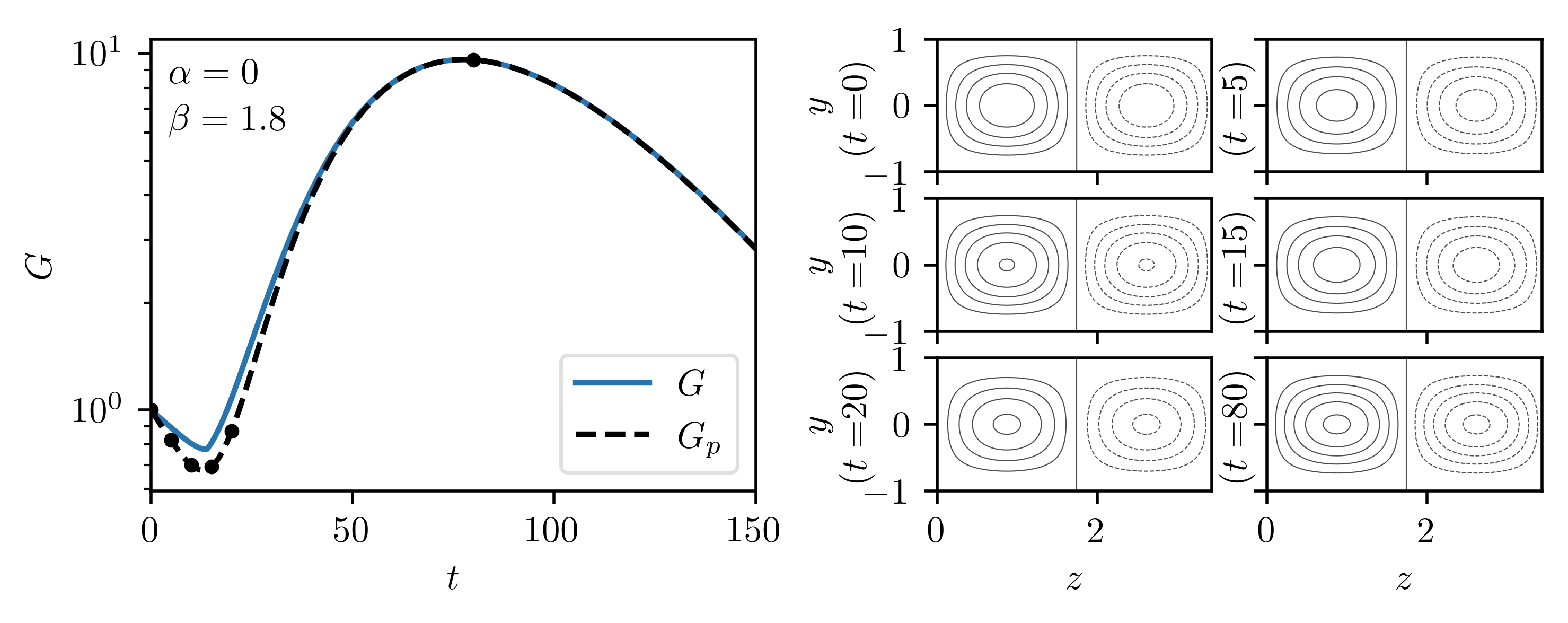}
    \captionsetup[subfigure]{labelformat=empty}
    \begin{picture}(0,0)
    \put(7,145){\contour{white}{ \textcolor{black}{a)}}}
    \put(190,145){\contour{white}{ \textcolor{black}{b)}}}
    \end{picture} 
    \begin{subfigure}[b]{0\textwidth}\caption{}\vspace{-10mm}\label{fig:PPFaccga}\end{subfigure}
    \begin{subfigure}[b]{0\textwidth}\caption{}\vspace{-10mm}\label{fig:PPFaccgb}\end{subfigure}
    \vspace{-5mm}
    
    \caption{(a) Energy of the optimal perturbation and the envelope of growth for accelerating PDF \ALrevise{($\Rey=500$, $\kappa=0.1$)}. (b) Stream function of the perturbation as it evolves in time.}
\label{fig:PPFaccg}
\end{figure}

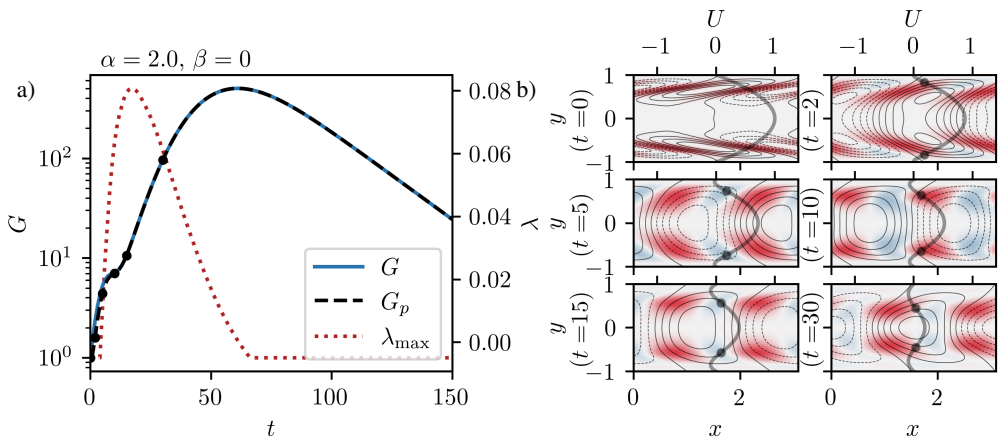
\begin{figure}
    \includegraphics[width=\textwidth]{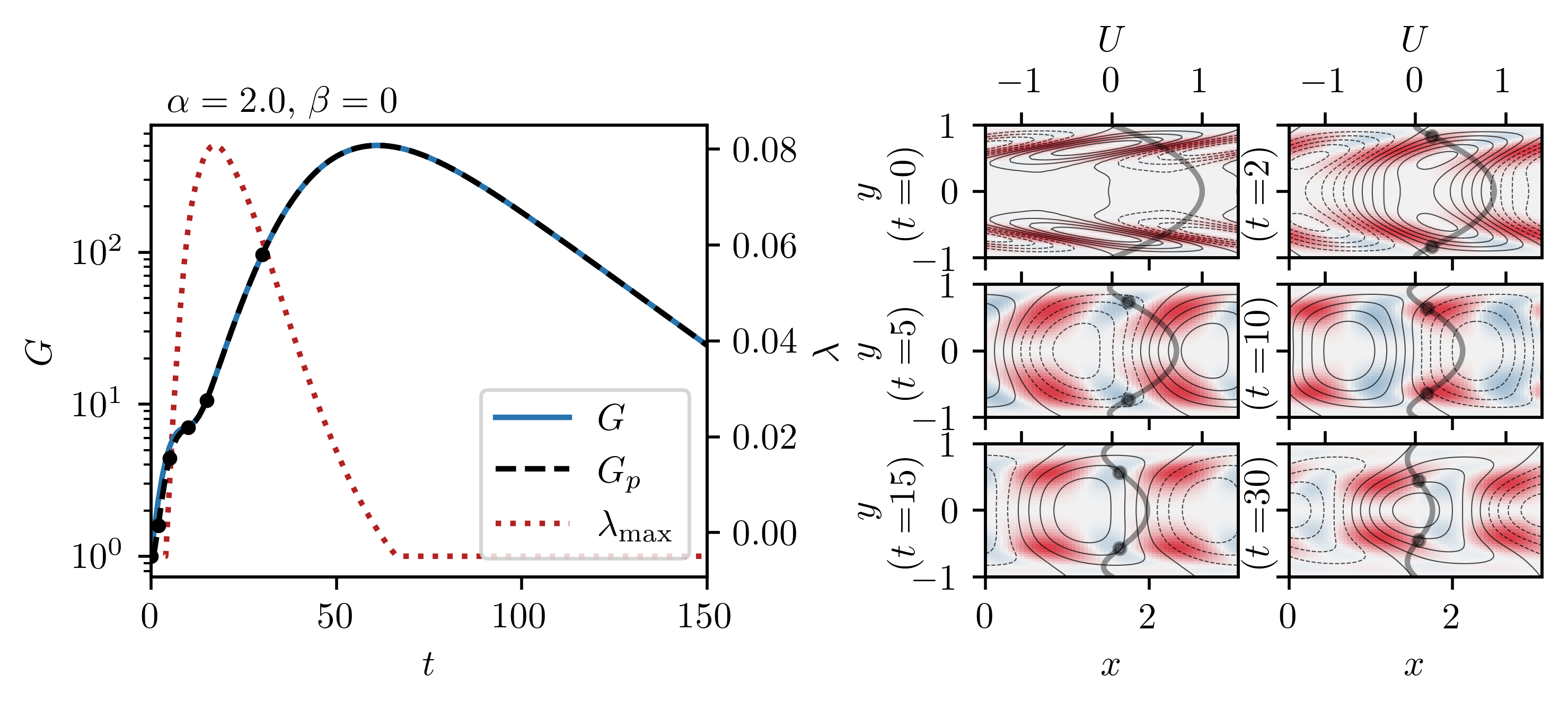}

    \captionsetup[subfigure]{labelformat=empty}
    \begin{picture}(0,0)
    \put(7,145){\contour{white}{ \textcolor{black}{a)}}}
    \put(195,145){\contour{white}{ \textcolor{black}{b)}}}
    \end{picture} 
    \begin{subfigure}[b]{0\textwidth}\caption{}\vspace{-10mm}\label{fig:PPFdecga}\end{subfigure}
    \begin{subfigure}[b]{0\textwidth}\caption{}\vspace{-10mm}\label{fig:PPFdecgb}\end{subfigure}
    \vspace{-5mm}
    
    \caption{\ALrevise{(a) Energy of the optimal perturbation, the envelope of growth, and the instantaneous eigenvalue for decelerating PDF ($\Rey=500$, $\kappa=0.1$). (b) Thin lines are the stream function of the perturbation as it evolves in time. The filled contour plot visualizes the production (Eq.\ \ref{eq:Prod}) normalized by the maximum absolute value (red indicates positive, blue indicates negative values). The thick black line is the reference laminar profile ($U$), with a dot at the inflection point.}}
\label{fig:PPFdecg}
\end{figure}

\ALrevise{\subsubsection{Optimal perturbations in pressure-driven 
flows}}

Next, we turn toward the accelerating and decelerating PDF cases. Figure~\ref{fig:PPFaccga} shows the energy of the optimal perturbation for accelerating PDF at $\Rey=500$ and $\kappa=0.1$. Here, the optimal perturbation reaches the maximum amplitude at $t'=78$. Qualitatively, an accelerating PDF shows the same characteristics as an accelerating WDF. The energy of the perturbation initially drops and then increases as the flow accelerates, and the perturbations again take the form of streamwise vortices. Figure~\ref{fig:PPFaccgb} shows the evolution of the perturbation at the times indicated in Fig.~\ref{fig:PPFaccga}. Again, the shape of this \ALrevise{perturbation} comes in the form of streamwise vortices, which agrees with the constant PDF case in \citet{Butler1992}.

Lastly, we consider the case of decelerating PDF. In Fig.~\ref{fig:PPFdecga} we show the energy of the optimal perturbation that maximizes $G(t')$ at $t'=61$. Unlike the previous cases, even at short times the energy of the perturbation closely follows the envelope over all perturbations $G(t)$. \ALrevise{The decelerating PDF does not exhibit a drop in energy, as seen in the decelerating WDF, because $\lambda_{\text{max}}$ becomes positive sooner.}
Figure~\ref{fig:PPFdecgb} shows the evolution of the optimal perturbation corresponding to the points in Fig.~\ref{fig:PPFdecga}. Again, we show the laminar profile for reference. Similar to the decelerating WDF case, the initial profile opposes the laminar shear leading to growth. As this perturbation evolves, it is advected downstream and the stream function orients vertically which momentarily damps growth. Then, the perturbation again aligns opposite to the laminar shear and stops moving as the base profile tends towards no flow.
For constant PDF, the streamwise perturbations damp out rapidly \citep{Butler1992}, but here the time-varying nature of the base flow allows the perturbation to align opposite to the flow and experience extended periods of growth. 
\ALrevise{Again, we see that the energy is produced at the locations of anti-alignment with the laminar shear. Unlike the WDF case, this production tends to stay near the wall-normal location of the inflection points, similar to \citet{Moron2022}.}




\begin{figure}
    \includegraphics[width=\textwidth]{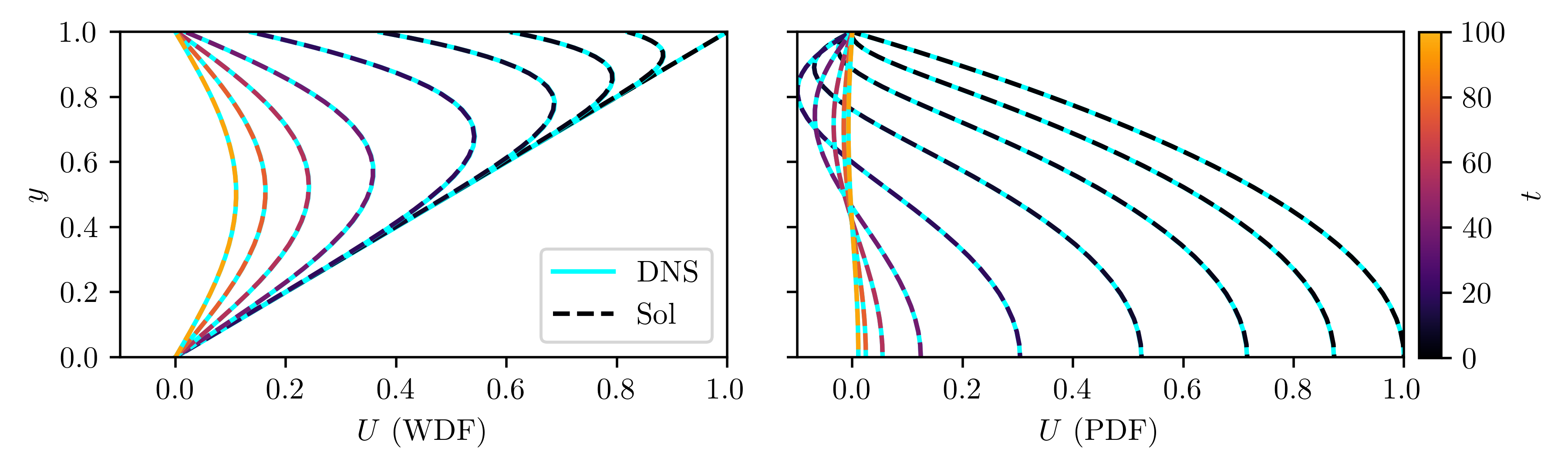}

    \captionsetup[subfigure]{labelformat=empty}
    \begin{picture}(0,0)
    \put(3,110){\contour{white}{ \textcolor{black}{a)}}}
    \put(180,110){\contour{white}{ \textcolor{black}{b)}}}
    \end{picture} 
    \begin{subfigure}[b]{0\textwidth}\caption{}\vspace{-10mm}\label{fig:LaminarDNSa}\end{subfigure}
    \begin{subfigure}[b]{0\textwidth}\caption{}\vspace{-10mm}\label{fig:LaminarDNSb}\end{subfigure}
    \vspace{-5mm}
    
    \caption{\ALrevise{Analytical and DNS solutions for laminar (a) WDF and (b) PDF at $\Rey=500$ and $\kappa=0.1$. }}
\label{fig:LaminarDNS}
\end{figure}
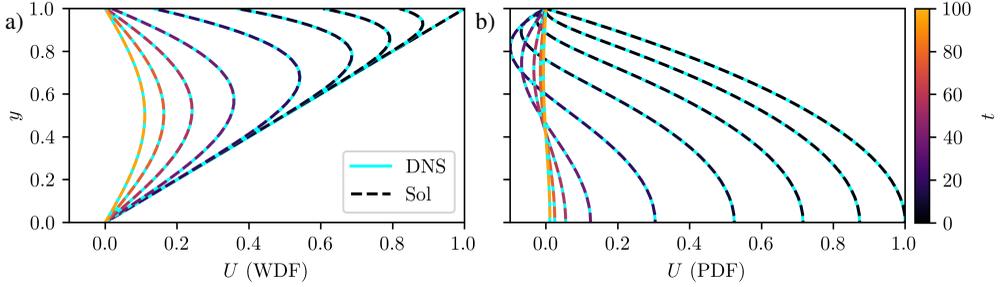

\subsection {Nonlinear evolution of \ALrevise{perturbations}}
\label{sec:Stability4}

Thus far, we have shown the evolution of \ALrevise{perturbations} through the linearized equations of motion. 
\ALrevise{In this section, we investigate the evolution of optimal perturbations in direct numerical simulations (DNS) and the role of these optimal perturbations when perturbing the flow with random noise. We perform DNS of decelerating WDF and PDF at $\Rey=500$ and $\kappa=0.1$}
Our DNS uses a Fourier–Chebyshev pseudo-spectral code implemented in Python \citep{Linot2023_git,Linot2023Control}, which is based on the Channelflow code developed by \citet{Gibson2012,Gibson2021}. We use the Spalart-Moser Runge-Kutta (SMRK2) scheme \citep{Spalart1991} to integrate solutions forward in time. This is a multistage scheme that treats the linear term implicitly and the nonlinear term explicitly. The SMRK2 scheme only requires one flowfield at one instant in time, which we require
to satisfy the decelerating boundary condition at every time step. \ALrevise{In the case of PDF, we directly impose the exponentially decaying flow rate, not the numerical estimate of the pressure gradient shown in Fig.\ \ref{fig:Pres}.}
For more details on the implementation of the DNS, we refer the reader to \citet{Linot2023Control} and \citet{Gibson2012}. 

\ALrevise{First, we validate the results of the DNS by showing that it maintains the analytical laminar profiles. Here, we perform simulations with a timestep of $\Delta t=0.01$ on a grid size of $[N_x,N_y,N_z]=[2,81,2]$ and a domain of $[L_x,L_y,L_z]=[1,2,1]$ with no noise. As the laminar flow only varies in the wall-normal direction the choice of $N_x=2$ and $N_z=2$ was chosen to speed up the computation. Using more grid points in these directions does not influence the results. Figure \ref{fig:LaminarDNS} compares the laminar solutions derived in Sec.\ \ref{sec:Laminar} to the DNS, showing that the DNS and the laminar solution are in excellent agreement.}


\begin{figure}
    \includegraphics[width=\textwidth]{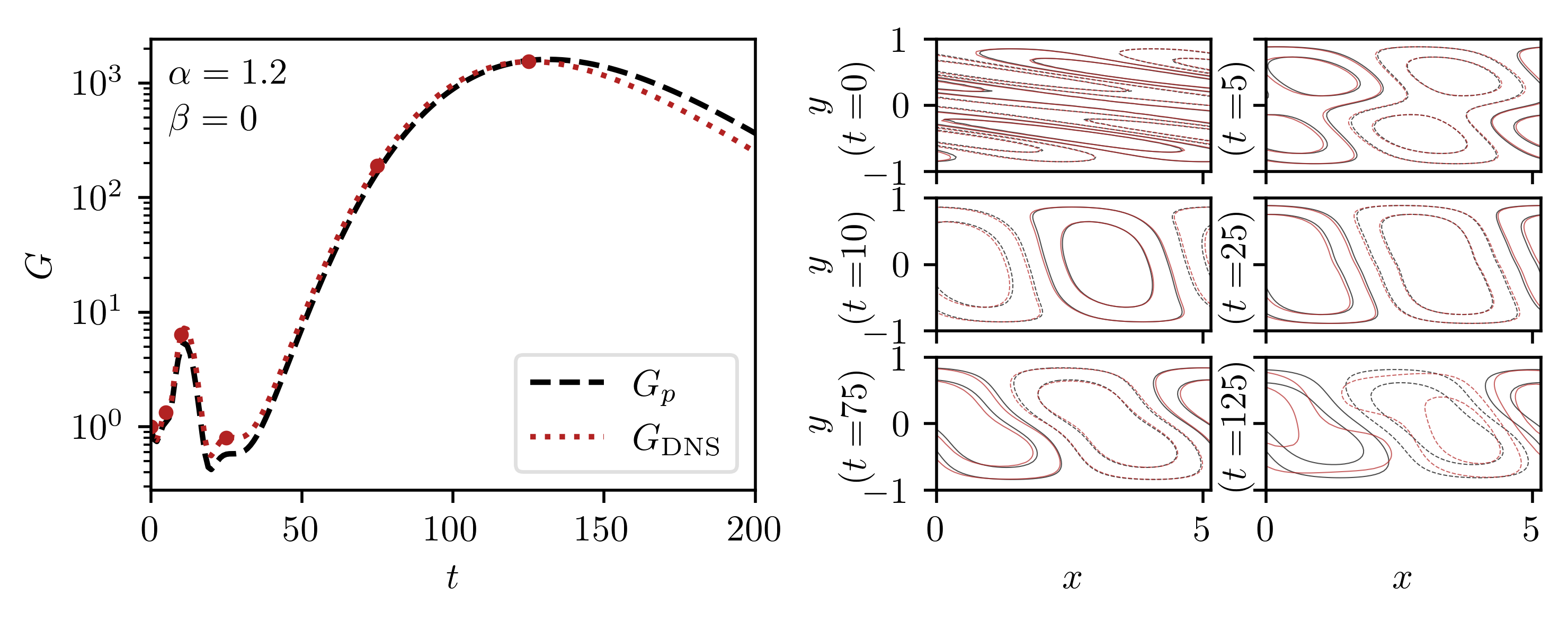}

    \captionsetup[subfigure]{labelformat=empty}
    \begin{picture}(0,0)
    \put(7,145){\contour{white}{ \textcolor{black}{a)}}}
    \put(190,145){\contour{white}{ \textcolor{black}{b)}}}
    \end{picture} 
    \begin{subfigure}[b]{0\textwidth}\caption{}\vspace{-10mm}\label{fig:DNSOptWDFa}\end{subfigure}
    \begin{subfigure}[b]{0\textwidth}\caption{}\vspace{-10mm}\label{fig:DNSOptWDFb}\end{subfigure}
    \vspace{-5mm}
    
    \caption{\ALrevise{(a) Energy of the optimal perturbation for decelerating WDF ($\Rey=500$, $\kappa=0.1$) determined by the linearized equations of motion $G_p$ and by the DNS $G_{\text{DNS}}$. (b) Stream function of the perturbation as it evolves in time (black is from the linearized equations and red is from the DNS).}}
\label{fig:DNSOptWDF}
\end{figure}
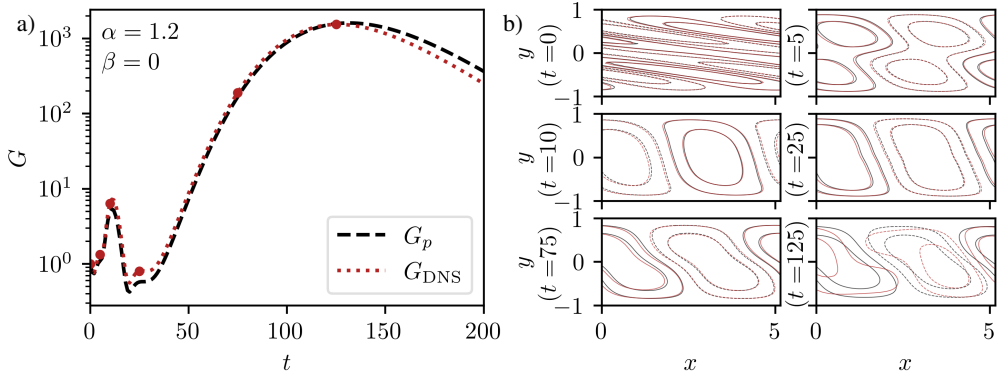

\begin{figure}
    \includegraphics[width=\textwidth]{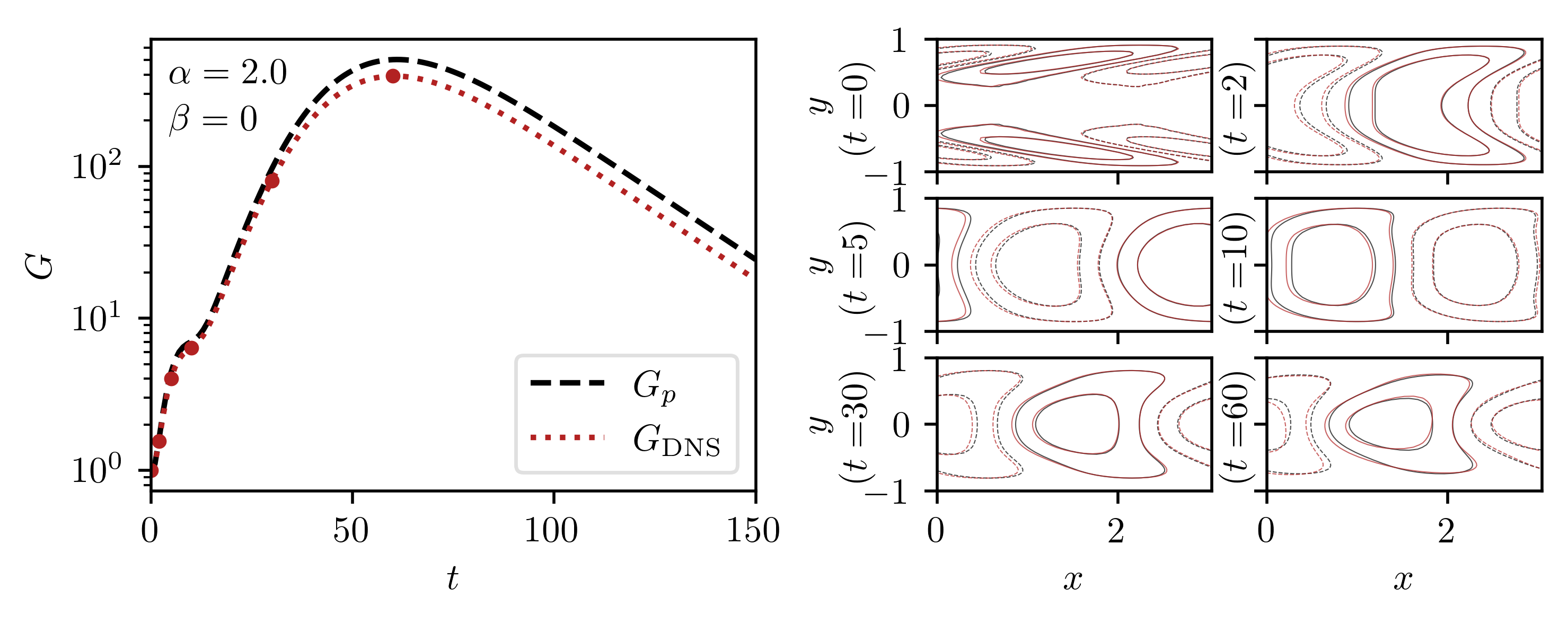}

    \captionsetup[subfigure]{labelformat=empty}
    \begin{picture}(0,0)
    \put(7,145){\contour{white}{ \textcolor{black}{a)}}}
    \put(190,145){\contour{white}{ \textcolor{black}{b)}}}
    \end{picture} 
    \begin{subfigure}[b]{0\textwidth}\caption{}\vspace{-10mm}\label{fig:DNSOptPDFa}\end{subfigure}
    \begin{subfigure}[b]{0\textwidth}\caption{}\vspace{-10mm}\label{fig:DNSOptPDFb}\end{subfigure}
    \vspace{-5mm}
    
    \caption{\ALrevise{(a) Energy of the optimal perturbation for decelerating PDF ($\Rey=500$, $\kappa=0.1$) determined by the linearized equations of motion $G_p$ and by the DNS $G_{\text{DNS}}$. (b) Stream function of the perturbation as it evolves in time (black indicates the linearized equations and red is from the DNS).}}
\label{fig:DNSOptPDF}
\end{figure}
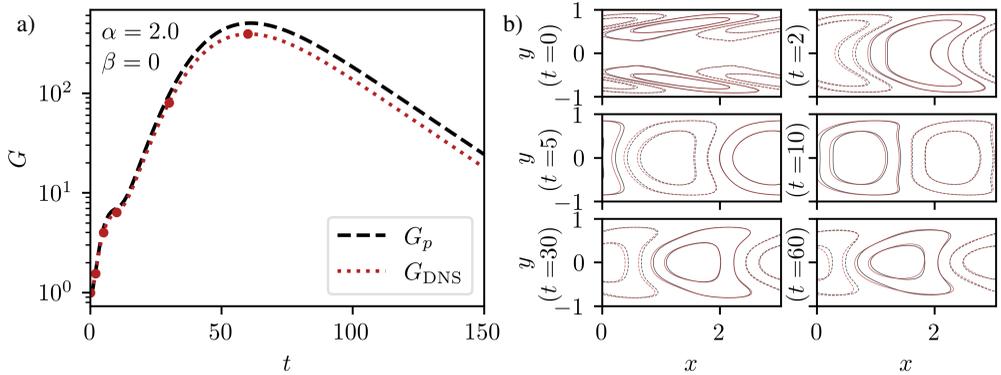

\ALrevise{Next, we consider the effect of applying the optimal perturbations to both of these flows. In both cases, the optimal perturbations are predominantly streamwise structures. Due to this two-dimensionality, we perform the DNS on a grid of size $[N_x,N_y,N_z]=[32,81,2]$ and a domain size of $[L_x,L_y,L_z]=[2\pi/\alpha_{\text{opt}},2,1]$ with a timestep of $\Delta t=0.01$. Simulating on more grid points in the spanwise direction does not change the results, since the initial perturbation does not vary in this direction. We initialize both DNS with an initial condition $\mathbf{u}=\mathbf{U}+\mathbf{u}'$, where we reduce the magnitude of the optimal perturbations shown in Fig.~\ref{fig:PCFdecga} and Fig.~\ref{fig:PPFdecga}, such that $\mathbf{u}'=10^{-3}\mathbf{u}_p'$. With this initialization the energy ratio of the perturbation is $||\mathbf{u}'||_E^2/||\mathbf{U}||_E^2=\mathcal{O}(10^{-7})$ in both cases. Figure \ref{fig:DNSOptWDF} displays, for WDF, the energy of the perturbation from the DNS, and the shape of the perturbation as it evolves in time.}
The energy \ALrevise{in the DNS} matches \ALrevise{the linearized solution} extremely well until $t\approx 125$, \ALrevise{at which point the energy of the DNS starts to drop more rapidly}. At $t=125$, the energy ratio of the perturbation is $||\mathbf{u}'||_E^2/||\mathbf{U}||_E^2\approx 0.11$ due to the reduced energy in the laminar profile and the growth of the perturbation. The sizeable portion of energy contribution from the perturbation suggests that the assumption of linearity at this point likely breaks down. In Fig.~\ref{fig:DNSOptWDFb}, we show the evolution of the \ALrevise{perturbation} through time. Until $t\approx 75$, we see excellent agreement between the optimal perturbation computed via the linearized equations and the DNS. As mentioned, after this time the relative size of the perturbation becomes sufficiently large, and nonlinear effects distort the field, as shown in the final snapshot at $t=125$. 

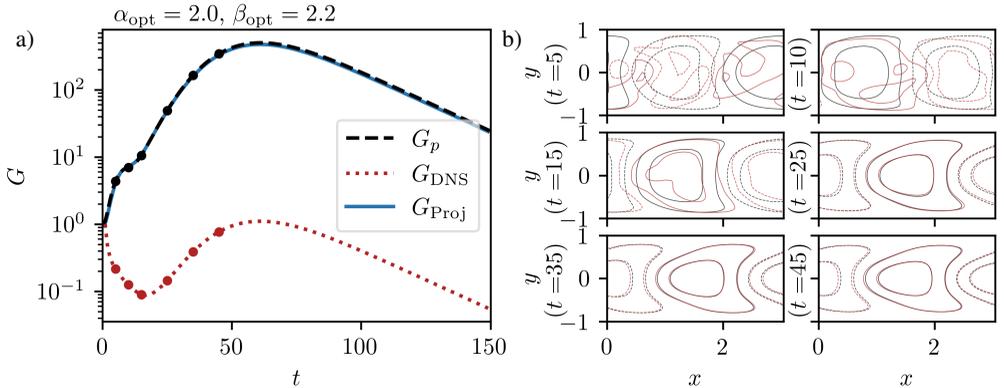
\begin{figure}
    \includegraphics[width=\textwidth]{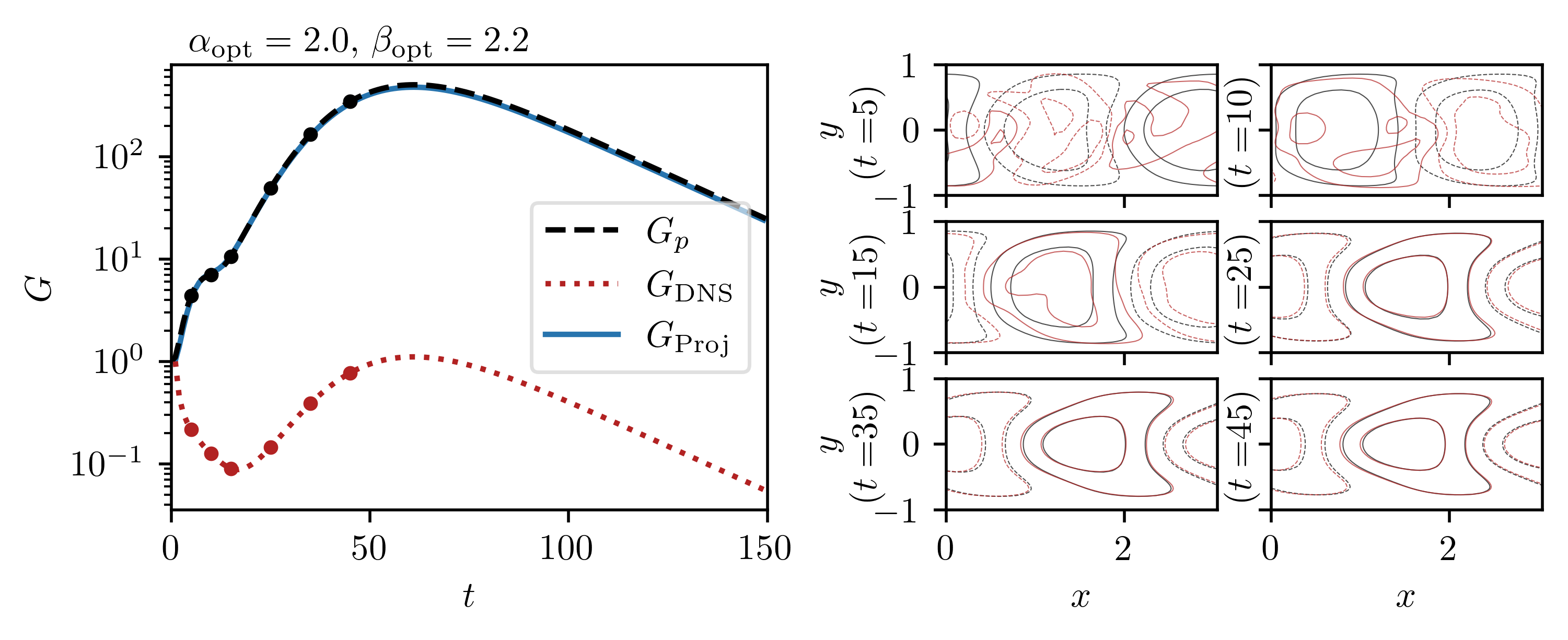}

    \captionsetup[subfigure]{labelformat=empty}
    \begin{picture}(0,0)
    \put(7,145){\contour{white}{ \textcolor{black}{a)}}}
    \put(190,145){\contour{white}{ \textcolor{black}{b)}}}
    \end{picture} 
    \begin{subfigure}[b]{0\textwidth}\caption{}\vspace{-10mm}\label{fig:DNSPertPDFa}\end{subfigure}
    \begin{subfigure}[b]{0\textwidth}\caption{}\vspace{-10mm}\label{fig:DNSPertPDFb}\end{subfigure}
    \vspace{-5mm}
    
    \caption{\ALrevise{(a) Energy of the optimal perturbation for decelerating PDF ($\Rey=500$, $\kappa=0.1$) determined by the linearized equations of motion $G_p$ and energy of a random perturbation in a DNS $G_{\text{DNS}}$, along with the energy of the random perturbation projected onto the optimal perturbation $G_{\text{Proj}}$. (b) Stream function of the perturbation as it evolves in time (black is from the linearized equations and red is from the DNS).}}
\label{fig:DNSPertPDF}
\end{figure}

\ALrevise{In Fig.\ \ref{fig:DNSPertPDF} we show the optimal perturbation in a DNS of PDF. Again, the energy of the DNS agrees well with the energy of the solution in the linearized equations at early times and begins to deviate around $t=50$. At the peak of $t=60$, the energy ratio of the perturbation is $||\mathbf{u}'||_E^2/||\mathbf{U}||_E^2\approx 0.03$. Figure \ref{fig:DNSOptWDF} shows the evolution of the perturbation through time. Here, the perturbation in the DNS remains similar to the evolution in the linearized equations, even when the energy starts to differ at $t=60$.}

\begin{figure}
    \includegraphics[width=\textwidth]{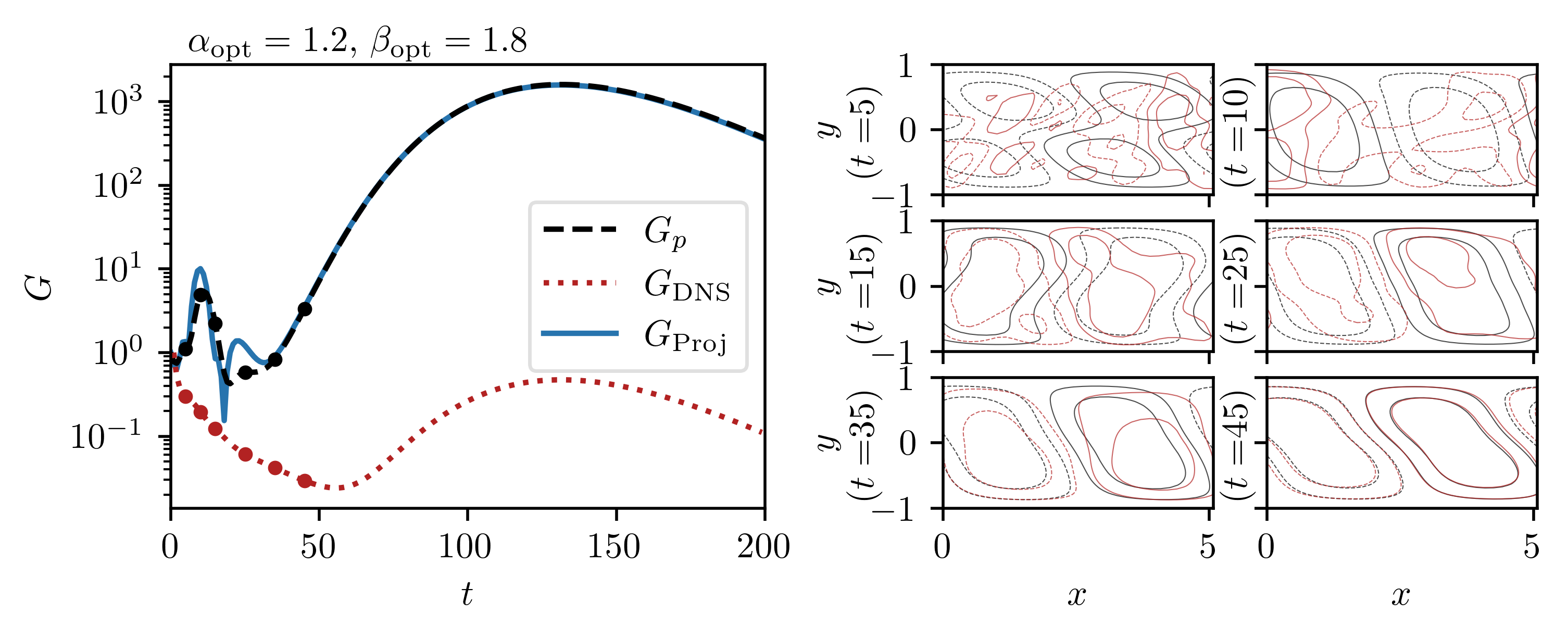}

    \captionsetup[subfigure]{labelformat=empty}
    \begin{picture}(0,0)
    \put(7,145){\contour{white}{ \textcolor{black}{a)}}}
    \put(190,145){\contour{white}{ \textcolor{black}{b)}}}
    \end{picture} 
    \begin{subfigure}[b]{0\textwidth}\caption{}\vspace{-10mm}\label{fig:DNSPertWDFa}\end{subfigure}
    \begin{subfigure}[b]{0\textwidth}\caption{}\vspace{-10mm}\label{fig:DNSPertWDFb}\end{subfigure}
    \vspace{-5mm}
    
    \caption{\ALrevise{(a) Energy of the optimal perturbation for decelerating WDF ($\Rey=500$, $\kappa=0.1$) determined by the linearized equations of motion $G_p$ and energy of a random perturbation in a DNS $G_{\text{DNS}}$, along with the energy of the random perturbation projected onto the optimal perturbation $G_{\text{Proj}}$. (b) Stream function of the perturbation as it evolves in time (black is from the linearized equations and red is from the DNS).}}
\label{fig:DNSPertWDF}
\end{figure}
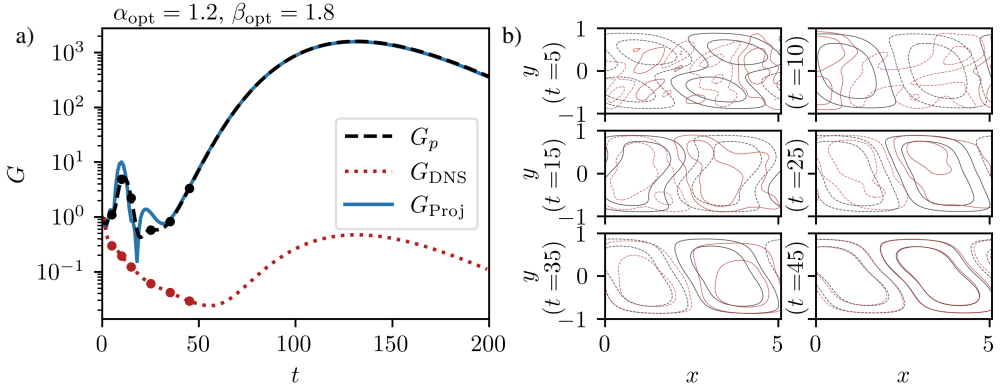

\ALrevise{We have established that the optimal perturbation exhibits large growth even in a DNS. As a next step, we investigate the relevance of this growth in the presence of random perturbations. We perform DNS on a grid of size $[N_x,N_y,N_z]=[32,81,32]$ and a domain size of $[L_x,L_y,L_z]=[2\pi/\alpha_{\text{opt}},2,2 \pi/\beta_{\text{opt}}]$ with a timestep of $\Delta t=0.01$. Here, we chose $\beta_{\text{opt}}$ as the strictly spanwise perturbation that causes the largest amplification for $\kappa=0.1$ and $\Rey=500$. This choice of domain size allows the largest growing spanwise and streamwise perturbations to grow simultaneously. We then perturb this flow with Gaussian noise at every grid point such that $\mathbf{u}'\sim \mathcal{N}(0,\varepsilon^2)$ where $\varepsilon=10^{-7}$. Even though this noise is initially not incompressible, the DNS enforces incompressibility after the first timestep.}

\ALrevise{In Fig.\ \ref{fig:DNSPertWDFa}, we show the energy of the optimal perturbation, the randomly perturbed field, and the projection of the random perturbation onto the optimal perturbation for WDF at $\Rey=500$ and $\kappa=0.1$. The random perturbation exhibits a decay in energy until $t\approx 50$, followed by a growth in energy that peaks when the energy of the optimal perturbation peaks. At early times, this energy decay occurs because most modes decay, despite the energy associated with the optimal perturbation growing. It is not until later in the evolution of this noise that the optimal perturbation can grow sufficiently large to increase the energy of the random perturbation. The projection of the random perturbation onto the optimal perturbation shows that despite the drop in energy of the overall perturbation, the part associated with the optimal perturbation exhibits the expected growth. The slight mismatch between the projection and the energy of the optimal perturbation is conjectured to come from interactions between the modes, as they do not remain orthogonal during their temporal evolution. We also show the evolution of the random perturbation in Fig.\ \ref{fig:DNSPertWDFb}. The random perturbation differs substantially from the optimal perturbation at early times, but begins to tilt against the laminar shear at $t\approx 25$, before exhibiting close agreement with the optimal perturbation at $t\geq 45$. We compute the streamlines for the random perturbation assuming the flow is two-dimensional at $z=0$. Choosing a different $z$-location would influence the field at early times, but the random field becomes a streamwise perturbation, like the optimal perturbation, at later times.}

\ALrevise{Figure \ref{fig:DNSPertPDF} shows the same results for PDF at $\Rey=500$ and $\kappa=0.1$. The energy of the random perturbation decreases until $t\approx 16$ before increasing and peaking at the same time as the optimal perturbation. Due to the faster timescale over which growth happens in this case, the projection of the random perturbation onto the optimal perturbation matches the energy of the optimal perturbation even at early times. Furthermore, the streamlines of the random perturbation agree with those of the optimal perturbation at $t\geq 25$. These figures show that, despite exhibiting far less growth, the random perturbation evolves into the optimal perturbation.}

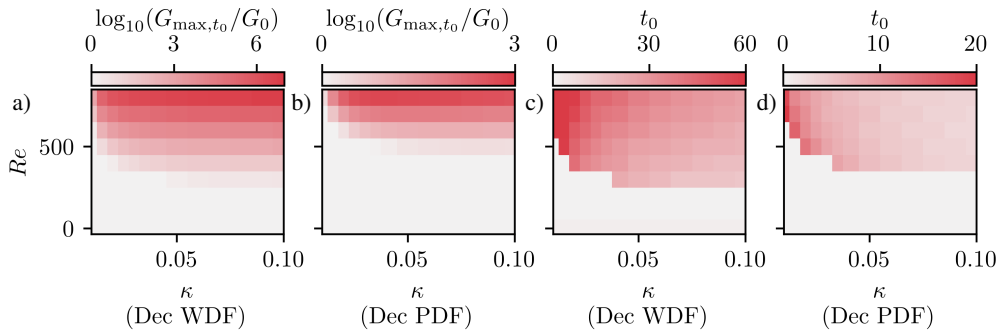
\begin{figure}
    \includegraphics[width=\textwidth]{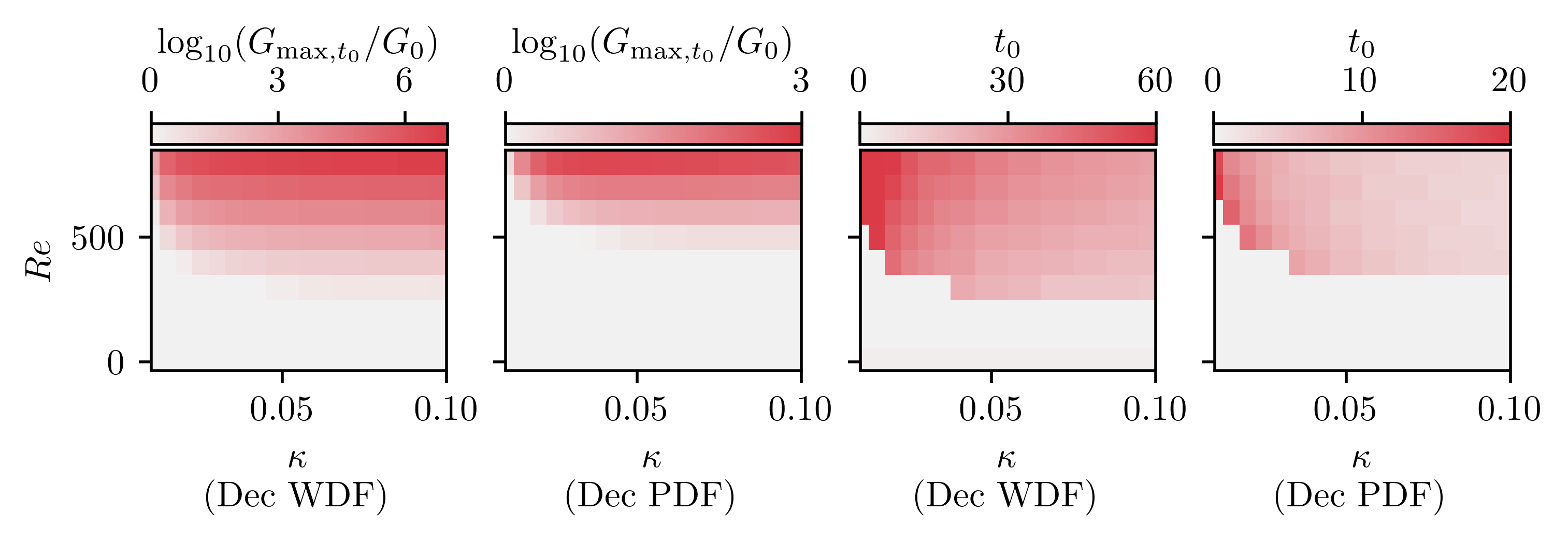}

    \captionsetup[subfigure]{labelformat=empty}
    \begin{picture}(0,0)
    \put(5,98){\contour{white}{ \textcolor{black}{a)}}}
    \put(110,98){\contour{white}{ \textcolor{black}{b)}}}
    \put(198,98){\contour{white}{ \textcolor{black}{c)}}}
    \put(285,98){\contour{white}{ \textcolor{black}{d)}}}
    \end{picture} 
    \begin{subfigure}[b]{0\textwidth}\caption{}\vspace{-10mm}\label{fig:Gmaxt0a}\end{subfigure}
    \begin{subfigure}[b]{0\textwidth}\caption{}\vspace{-10mm}\label{fig:Gmaxt0b}\end{subfigure}
    \begin{subfigure}[b]{0\textwidth}\caption{}\vspace{-10mm}\label{fig:Gmaxt0c}\end{subfigure}
    \begin{subfigure}[b]{0\textwidth}\caption{}\vspace{-10mm}\label{fig:Gmaxt0d}\end{subfigure}
    \vspace{-5mm}
    
    \caption{\ALrevise{(a) and (b) maximum growth normalized by the maximum growth of perturbations in the constant flow at various $\Rey$ and $\kappa$ for decelerating WDF and decelerating PDF. (c) and (d) optimal perturbation timing at various $\Rey$ and $\kappa$ for decelerating WDF and decelerating PDF.}}
\label{fig:Gmaxt0}
\end{figure}

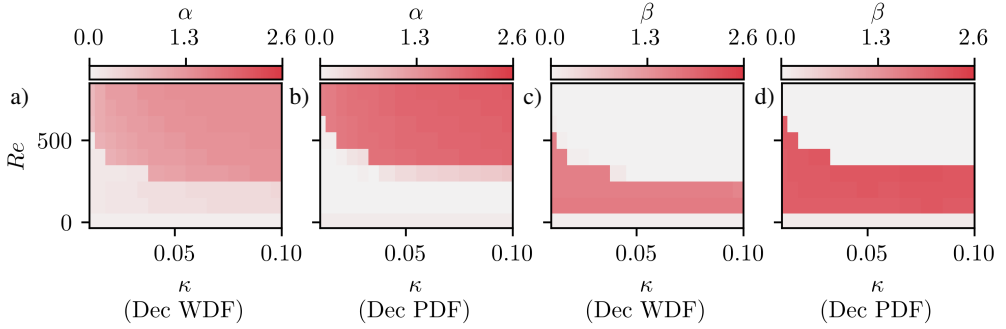
\begin{figure}
    \includegraphics[width=\textwidth]{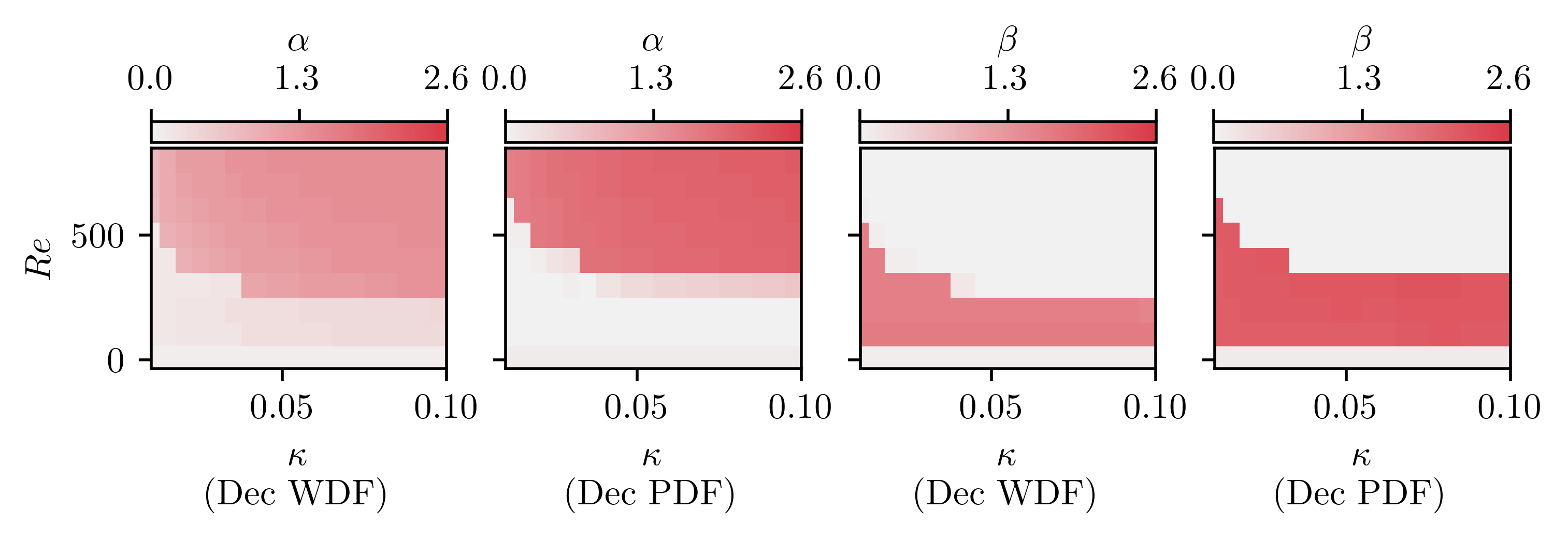}

    \captionsetup[subfigure]{labelformat=empty}
    \begin{picture}(0,0)
    \put(5,98){\contour{white}{ \textcolor{black}{a)}}}
    \put(110,98){\contour{white}{ \textcolor{black}{b)}}}
    \put(198,98){\contour{white}{ \textcolor{black}{c)}}}
    \put(285,98){\contour{white}{ \textcolor{black}{d)}}}
    \end{picture} 
    \begin{subfigure}[b]{0\textwidth}\caption{}\vspace{-10mm}\label{fig:AlphaBetaa}\end{subfigure}
    \begin{subfigure}[b]{0\textwidth}\caption{}\vspace{-10mm}\label{fig:AlphaBetab}\end{subfigure}
    \begin{subfigure}[b]{0\textwidth}\caption{}\vspace{-10mm}\label{fig:AlphaBetac}\end{subfigure}
    \begin{subfigure}[b]{0\textwidth}\caption{}\vspace{-10mm}\label{fig:AlphaBetad}\end{subfigure}
    \vspace{-5mm}
    
    \caption{\ALrevise{(a) and (b) optimal perturbation streamwise wavenumber at various $\Rey$ and $\kappa$ for decelerating WDF and decelerating PDF. (c) and (d) optimal perturbation spanwise wavenumber at  $\Rey$ and $\kappa$ for decelerating WDF and decelerating PDF.}}
\label{fig:AlphaBeta}
\end{figure}

\subsection {Optimal perturbation timing}
\label{sec:Stability5}

\ALrevise{Up to this point we have predominantly considered perturbations applied at the first instant of acceleration or deceleration. However, as noted in Figs.\ \ref{fig:CGrowth} and \ref{fig:PGrowth}, perturbations can grow larger when applied at later times. In this section, we investigate this behavior by sweeping over the time $t_0$ when we apply the perturbation. Figure \ref{fig:Gmaxt0} shows the maximum amplification $G_{\text{max},t_0}=\max_{\alpha,\beta,t,t_0} G(t)$ and the time at which the perturbation is applied at various $\Rey$ and $\kappa$ for decelerating WDF and PDF. Similar to the results in Fig.\ \ref{fig:GkRe}, there is a massive increase in the amplification as $\Rey$ and $\kappa$ increase. For WDF, this increase in amplification now appears at lower values of both $\Rey$ and $\kappa$, in comparison to the $t_0=0$ case. Figures \ref{fig:Gmaxt0c} and \ref{fig:Gmaxt0d} show that this increase coincides with a substantial delay in the application of the perturbation. At low $\kappa$, this delay is largest, and decreases and converges as $\kappa$ increases. When $\kappa$ is small, the timescale associated with wall motion dominates; when $\kappa$ is large, the viscous timescale dominates.}

\ALrevise{In Fig.\ \ref{fig:AlphaBeta}, we show the optimal wavenumbers at which $G_{\text{max},t_0}$ is achieved. Qualitatively, these results agree with the results at $t_0=0$. At low $\kappa$ and $\Rey$, the optimal perturbations are spanwise with $\beta \approx 1.6$ for WDF and $\beta \approx 2.2$ for PDF. At high $\Rey$, the optimal perturbations are streamwise with $\alpha \approx 0.8$, at low $\kappa$, and $\alpha \approx 1.4$, at high $\kappa$, for WDF. In PDF, the optimal perturbations are at $\alpha \approx 1.6$, at low $\kappa$, and $\alpha \approx 2.1$, at high $\kappa$. Again, this highlights that branch switching occurs at sufficiently high $\Rey$ and $\kappa$.}

\begin{figure}
    \includegraphics[width=\textwidth]{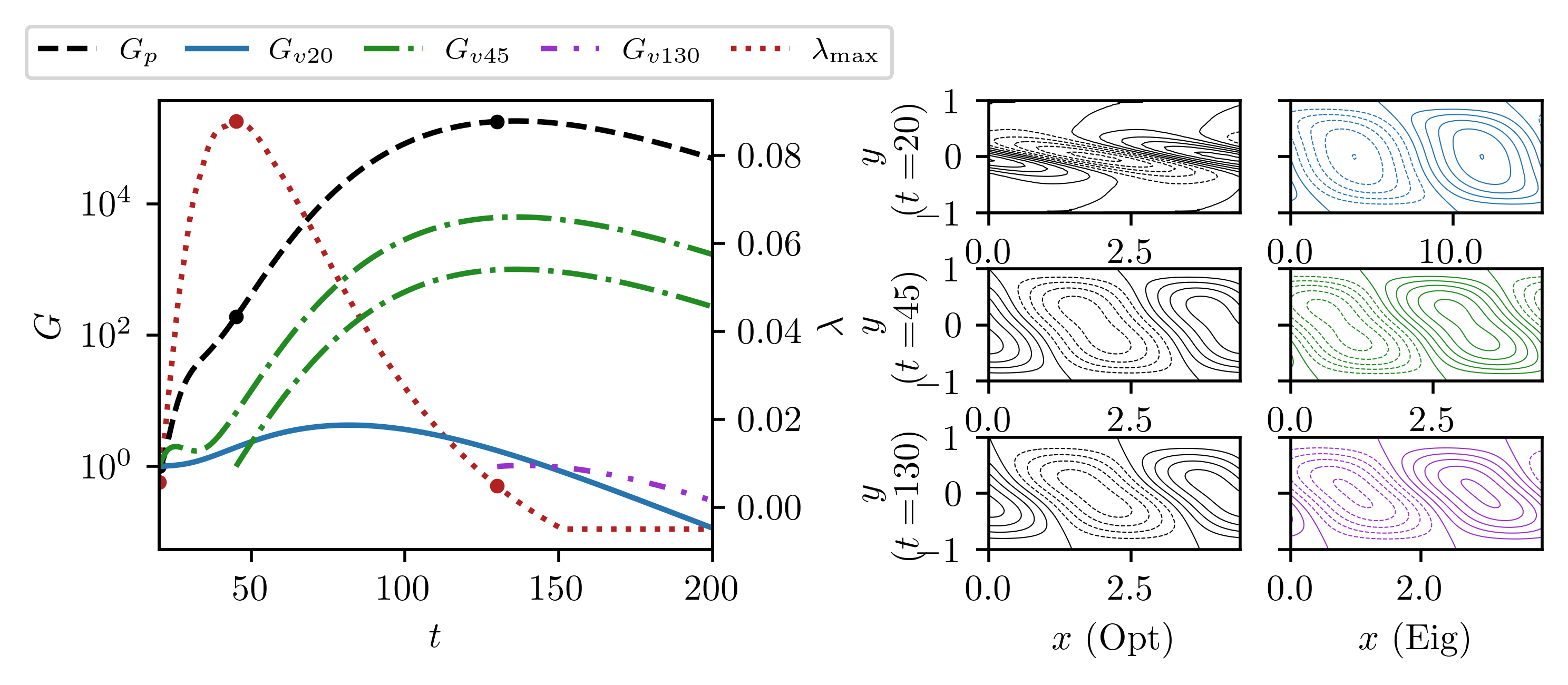}

    \captionsetup[subfigure]{labelformat=empty}
    \begin{picture}(0,0)
    \put(10,145){\contour{white}{ \textcolor{black}{a)}}}
    \put(197,145){\contour{white}{ \textcolor{black}{b)}}}
    \end{picture} 
    \begin{subfigure}[b]{0\textwidth}\caption{}\vspace{-10mm}\label{fig:WDFEigPerta}\end{subfigure}
    \begin{subfigure}[b]{0\textwidth}\caption{}\vspace{-10mm}\label{fig:WDFEigPertb}\end{subfigure}
    \vspace{-5mm}
    
    \caption{\ALrevise{(a) Evolution of the optimal perturbation and of eigenvectors of the instantaneous linear operator for decelerating WDF ($\Rey=500$, $\kappa=0.1$) determined by the linearized equations of motion and the largest eigenvalue of the instantaneous linear operator. (b) Stream function of the perturbation as it evolves in time, and the stream function of the eigenvector of the instantaneous linear operator at that time.}}
\label{fig:WDFEigPert}
\end{figure}

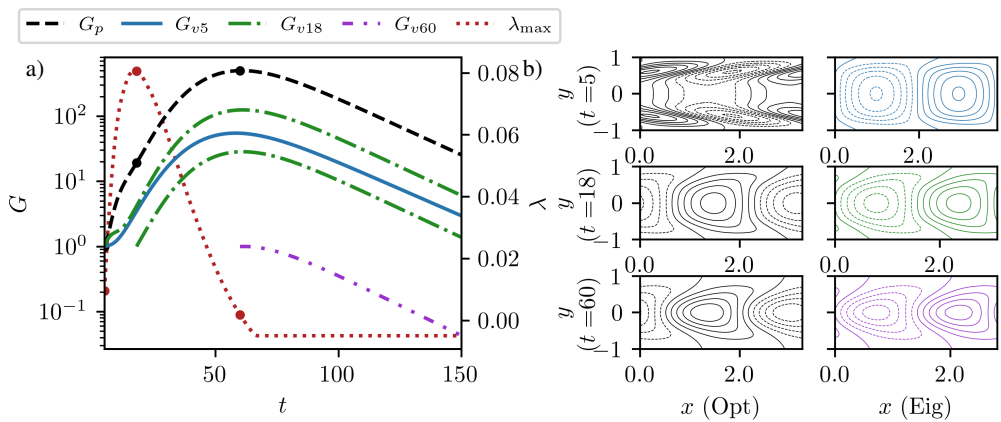
\begin{figure}
    \includegraphics[width=\textwidth]{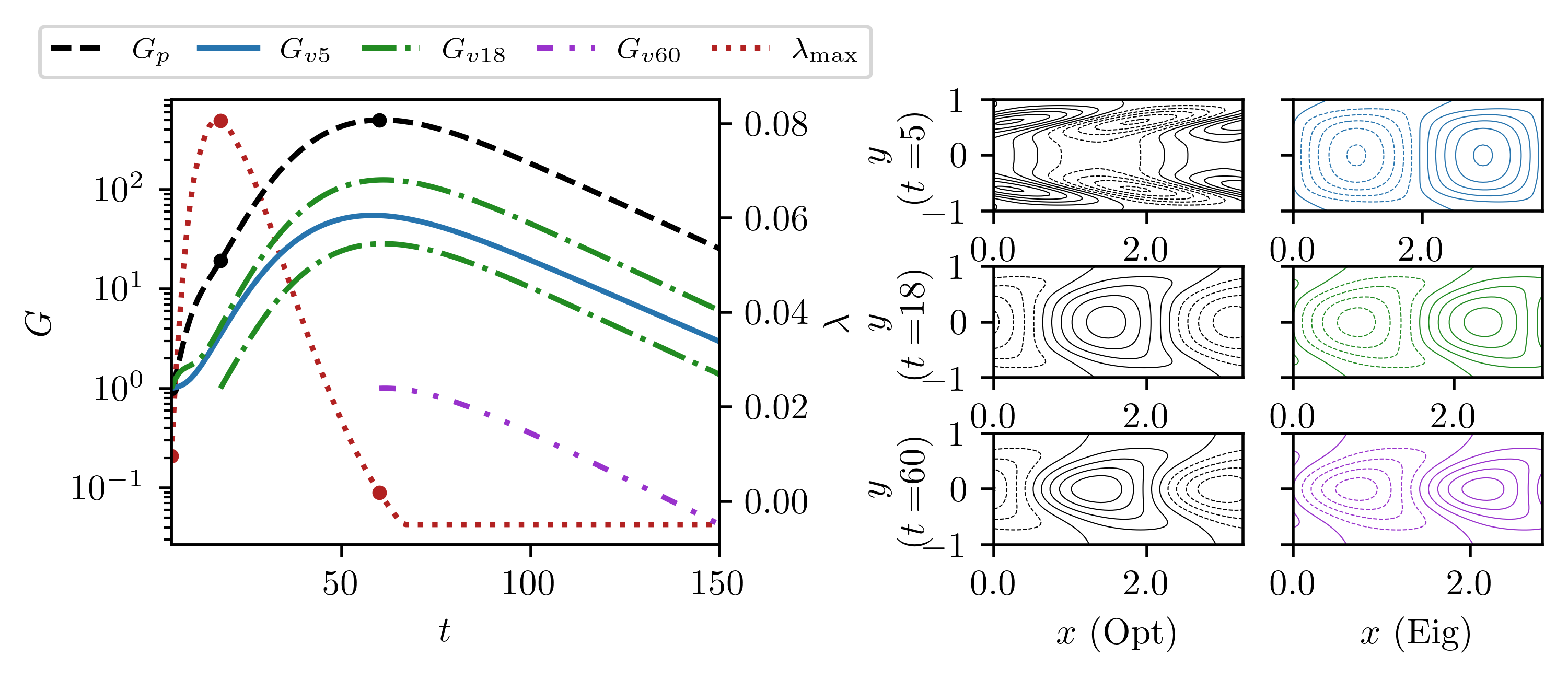}

    \captionsetup[subfigure]{labelformat=empty}
    \begin{picture}(0,0)
    \put(10,145){\contour{white}{ \textcolor{black}{a)}}}
    \put(197,145){\contour{white}{ \textcolor{black}{b)}}}
    \end{picture} 
    \begin{subfigure}[b]{0\textwidth}\caption{}\vspace{-10mm}\label{fig:PDFEigPerta}\end{subfigure}
    \begin{subfigure}[b]{0\textwidth}\caption{}\vspace{-10mm}\label{fig:PDFEigPertb}\end{subfigure}
    \vspace{-5mm}
    
    \caption{\ALrevise{(a) Evolution of the optimal perturbation and of eigenvectors of the instantaneous linear operator for decelerating PDF ($\Rey=500$, $\kappa=0.1$) determined by the linearized equations of motion and the largest eigenvalue of the instantaneous linear operator. (b) Stream function of the perturbation as it evolves in time, and the stream function of the eigenvector of the instantaneous linear operator at that time.}}
\label{fig:PDFEigPert}
\end{figure}

\ALrevise{Next, we investigate the shape of the optimal perturbations at the optimal timing and inform the results with linear stability analysis using the quasi-steady state approximation. Figure \ref{fig:WDFEigPerta} shows the energy of the optimal perturbation, the real part of the largest instantaneous eigenvalue, and the energy of evolving different eigenvectors forward in time for WDF at $\Rey=500$ and $\kappa=0.1$. Figure \ref{fig:WDFEigPertb} shows the evolution of the optimal perturbation, and the eigenvectors of the instantaneous linear operator at various times for WDF. This figure highlights three important times in the evolution of the optimal perturbation. First, the initial phase of the optimal perturbation occurs around when the maximum eigenvalue of the instantaneous linear operator becomes positive. Second, when the maximum eigenvalue peaks, the shape of the optimal perturbation and the instantaneous eigenvector closely resemble one another. Third, when the maximum eigenvalue drops below zero, the energy of the optimal perturbation begins to decrease.}

\ALrevise{Due to this close relationship between the optimal perturbation and the quasi-steady state approximation, we now consider the influence of perturbing the flow with the instantaneous eigenvectors. At the initial perturbation time $t=20$, the optimal perturbation and the instantaneous eigenvector are much different. In particular, the instantaneous eigenvector exhibits a much lower streamwise wavenumber, resulting in a larger structure. When we perturb with this eigenvector $G_{v20}$, we see minimal growth. At the peak of the maximum eigenvalue, the optimal perturbation and the instantaneous eigenvector closely match one another. However, if we perturb with the eigenvector at the peak $G_{v45}$, it results in orders of magnitude less growth. Also, if we perturb with this eigenvector at $t=20$, it too results in far less growth. Although the instantaneous stability of the flow plays an important role in the evolution of the optimal perturbation, the instantaneous eigenvectors do not sufficiently align with the optimal perturbation, needed to experience the maximum growth in energy. The optimal perturbation exhibits larger energy than the instantaneous eigenvectors because it can exhibit significant levels of transient growth before matching the shape of the eigenvector at the peak of the maximum eigenvalue. Past this point, the shape of the eigenvector remains nearly constant, so the optimal perturbation too maintains this shape. This also indicates that the quasi-steady state approximation is reasonable over this time interval, as the eigenvector does not change substantially. Lastly, we show the energy of the eigenvector once the eigenvalue becomes negative $G_{v130}$, which experiences minimal growth, as expected.}

\ALrevise{In Fig.\ \ref{fig:PDFEigPert}, we show the same results as in Fig.\ \ref{fig:WDFEigPert} for PDF at $\Rey=500$ and $\kappa=0.1$. The optimal perturbation timing occurs when the maximum eigenvalue turns positive. The optimal perturbation starts with streamlines that are much more anti-aligned with the flow than the instantaneous eigenvector at this time. Then, the optimal perturbation grows and aligns with the instantaneous eigenvector at the peak of the maximum eigenvalue. This growth continues until the maximum eigenvalue drops below zero. As with WDF, the initial growth of the optimal perturbation plays an important role which is missed when only considering the instantaneous eigenvectors. In both WDF and PDF, as $\Rey$ increases (at a sufficiently high $\kappa$) the peak of the maximum eigenvalue increases to larger times, thus leading to longer times over which this initial growth increases the energy of the perturbation.}

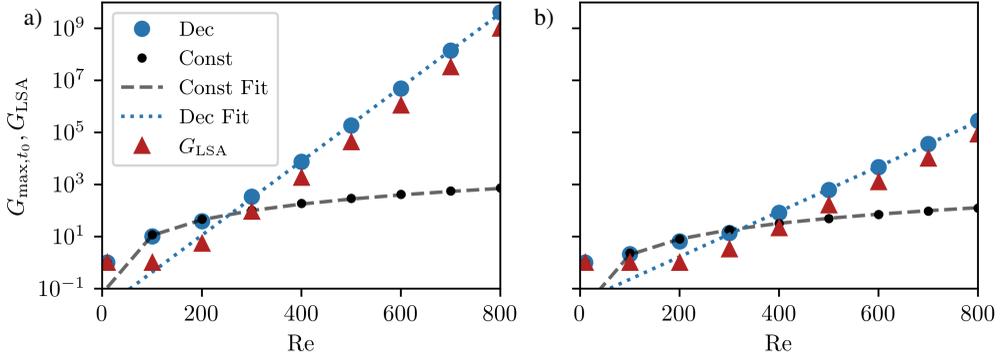
\begin{figure}
    \includegraphics[width=\textwidth]{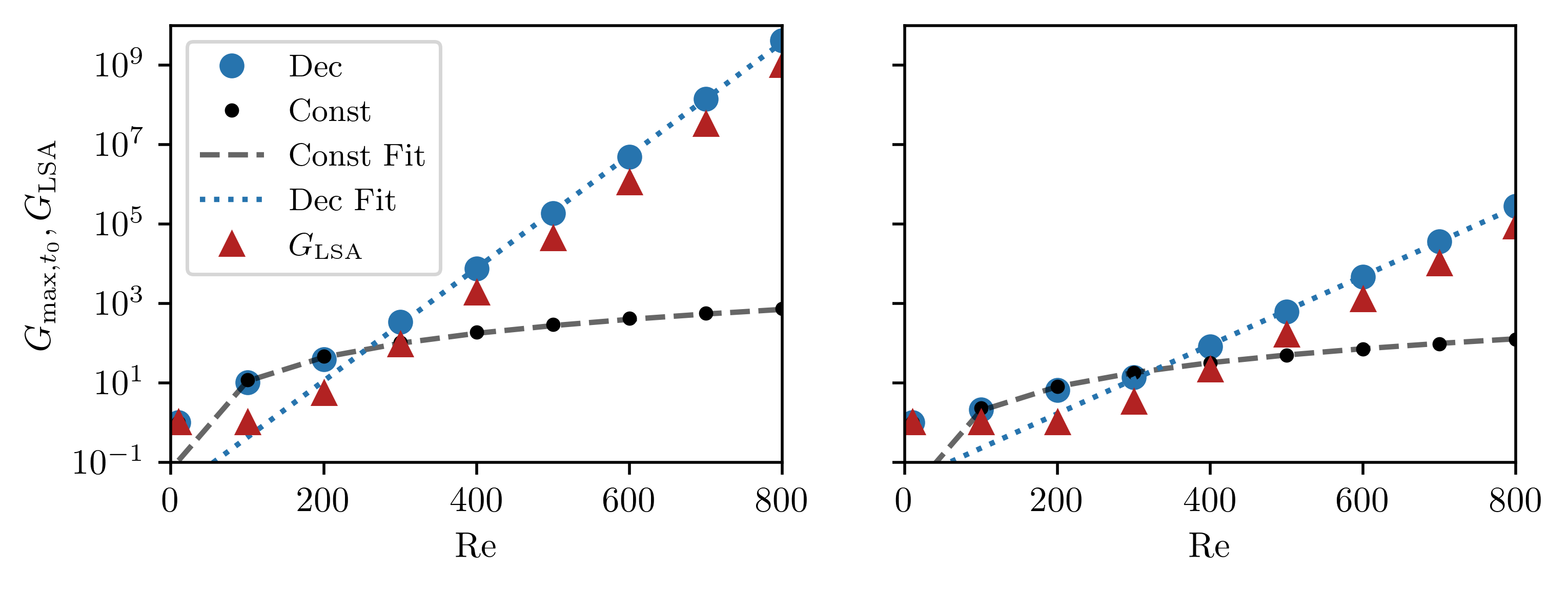}

    \captionsetup[subfigure]{labelformat=empty}
    \begin{picture}(0,0)
    \put(10,140){\contour{white}{ \textcolor{black}{a)}}}
    \put(202,140){\contour{white}{ \textcolor{black}{b)}}}
    \end{picture} 
    \begin{subfigure}[b]{0\textwidth}\caption{}\vspace{-10mm}\label{fig:GScaleOptt0a}\end{subfigure}
    \begin{subfigure}[b]{0\textwidth}\caption{}\vspace{-10mm}\label{fig:GScaleOptt0b}\end{subfigure}
    \vspace{-5mm}
    
    \caption{\ALrevise{$G_{\text{max},t_0}$ and $G_{\text{LSA}}$ as a function of $\Rey$ for decelerating flow at fixed $\kappa=0.1$ 
 (``Dec'') and $G_{\text{max},t_0}$ for constant flow (``Const'') for (a) WDF and (b) PDF. The fitting lines are described in the text.}}
\label{fig:GScaleOptt0}
\end{figure}

\ALrevise{Finally, we end by showing that $G_{\text{max},t_0}$ can be accurately approximated by integrating the instantaneous eigenvalue curve. \citet{Moron2022} showed for pulsatile flows with specific waveforms that $G_{\text{max},t_0}$ can be well approximated by
\begin{equation}
    G_{\text{LSA}}=e^{2\lambda_i},
\end{equation}
where 
\begin{equation}
    \lambda_i=\int_{t_0}^{t_0+\Delta t_u} \lambda_{\text{max}}(t) dt.
\end{equation}
Here, $\Delta t_u$ stands for the time window where $\lambda_{\text{max}}>0$. Note that \citet{Moron2022} include a time period $T$ that cancels out; this is not relevant to our case as our flow is not pulsatile. In Fig.\ \ref{fig:GScaleOptt0}, we show how $G_{\text{max},t_0}$ and $G_{\text{LSA}}$ change as we increase $\Rey$ at $\kappa=0.1$. Similar to $t_0=0$ (Fig.\ \ref{fig:Gscale}), the scaling of $G_{\text{max},t_0}$ changes from $\Rey^2$ to $10^{\Rey}$ as $\Rey$ increases. For WDF, this transition occurs at $\Rey \approx 250$, which is far lower than the $t_0=0$ case, where it takes place at $\Rey \approx 400$. For PDF, transition arises near the same point, because the optimal perturbation timing is closer to zero. Additionally, Fig.\ \ref{fig:GScaleOptt0} shows that $G_{\text{LSA}}$ is a good proxy for the optimal energy growth. It shows the correct scaling at high $\Rey,$ even though it slightly underestimates the growth.}

\section{Conclusions}
\label{sec:Conclusions}

We undertook a systematic investigation of the stability of unsteady accelerating and decelerating flows. An exact analytical solution for laminar flows with arbitrary wall motion and pressure gradients in channels has been derived, from which we selected a specific exponentially decaying profile for wall motion and flow rate that isolated the effects of acceleration and deceleration. With these analytical profiles, we investigated the stability of the flow with respect to these base flows through the linearized equations of motion by computing optimal perturbations. This investigation showed that deceleration can cause perturbations to exhibit massive amplification, while perturbations in the accelerating case never exceeded the maximum amplification of perturbations for a constant profile. Furthermore, the maximum amplification seen in the decelerating case exhibits both the $\Rey^2$ scaling seen in the constant case at low $\Rey$, and the $10^{\Rey}$ scaling seen in some unsteady cases at high $\Rey$. 

This change in scaling arises due to a branch switching of the dominant perturbation in the decelerating case. At low $\Rey$ and $\kappa$, we found that spanwise \ALrevise{perturbations} are most amplified, but upon increasing these values streamwise \ALrevise{perturbations} became most amplified. This was not the case for accelerating flows where spanwise \ALrevise{perturbations} always showed the largest amplification. We then studied the evolution of the optimal perturbations in time. In the case of acceleration, we found that the optimal perturbations stayed as streamwise vortices through time, similar to what was observed by \citet{Butler1993}. However, in the case of deceleration, the perturbations grew due to \ALrevise{the Orr mechanism, or the down-gradient Reynolds stress mechanism \citep{Butler1993}}. When streamlines are aligned with the laminar shear, there is growth. In the case of deceleration, the flow stops advecting the streamlines causing them to align opposite to the laminar shear of the base flow for extended periods of time, leading to growth. This behavior does not happen in constant or accelerating flows since the laminar profile eventually aligns the streamlines with the laminar shear of the base flow. \ALrevise{Furthermore, we found that this large growth of perturbations appears in DNS, both when the perturbation is applied directly or when random noise is imposed. Finally, we discovered that the optimal timing, when the perturbation is applied, is given by the instant when the maximum real eigenvalue of the instantaneous linear operator becomes positive.}


In the future, we intend to further explore how much the flow needs to decelerate for perturbations to exhibit the massive amplification shown here by using other temporal functions for the wall motion and flow rate. Additionally, these results could be extended to inform control strategies to avoid laminar transition. In the case of decelerating flows, this would require breaking up the formation of streamwise perturbations, and, in the case of acceleration, this would require breaking up spanwise perturbations. The fundamental insights from this study \ALrevise{could also} have implications for understanding the emergence of instabilities around accelerating and decelerating bodies of more complex geometries. \ALrevise{In particular, it may be the case that the destabilization around decelerating bodies could also be due to extended growth via the Orr mechanism.}

\section*{Acknowledgments}
This work was supported by the US Department of Defense Vannevar Bush Faculty Fellowship (Grant Number: N00014-22-1-2798), the Army Research Office (Grant Number: W911NF-21-1-0060), and the Air Force Office of Scientific Research (Grant Number: FA9550-21-1-0178). 

\section*{Declaration of Interests}
The authors report no conflict of interest.
\bibliographystyle{jfm}
\bibliography{jfm}

\begin{appendices}

\section{Additional discussion on laminar flow solutions} \label{sec:AppendixA}

\begin{figure}
\centering
    \includegraphics[width=.5\textwidth]{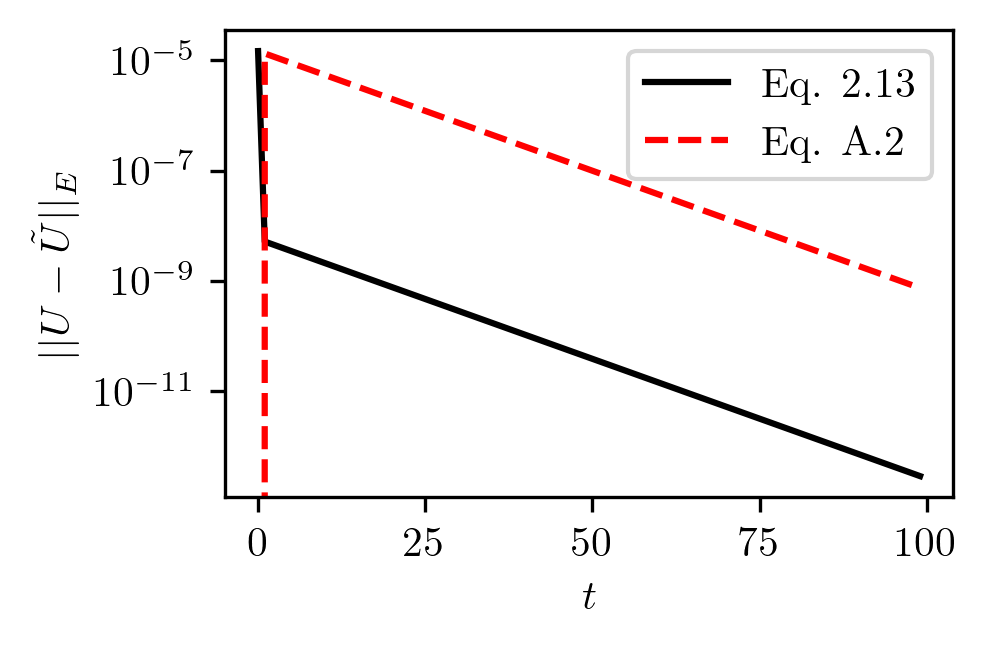}
  \caption{Error in the laminar profile for $K=100$ modes using different forms of the laminar solution.}
\label{fig:LamChoice}
\end{figure}

We further discuss the choice of functions in Eq.\ \ref{eq:NSELamWall}, validate the laminar solutions against other analytical solutions for specific flows, and provide a list of solutions for a family of unsteady flows. As mentioned in Sec.~\ref{sec:Laminar}, we could seek solutions in the wall-motion case of the form 
\begin{equation} \label{eq:NSELamWallSimple}
    U(y,t)=f_w(y,t)+g_w(t)y.
\end{equation}
Following the procedure from Sec.~\ref{sec:Laminar}, this results in the laminar solution
\begin{equation} \label{eq:CouetteLam2}
    U(y,t)= \sum_{n=1}^\infty e^{-a_nt} \left(\dfrac{2 (-1)^n}{\pi n} \int_0^t e^{a_nt'}\dfrac{dg_w(t')}{dt} dt' +C_n\right)\sin(n\pi y) +g_w(t) y.
\end{equation}
Although this equation has fewer terms than Eq.\ \ref{eq:CouetteLam}, the error is larger since it requires approximating $y$ in terms of $\sin(n\pi y)$. In Fig.~\ref{fig:LamChoice}, we show the error between $U$ approximated with $K=10^5$ using Eq.\ \ref{eq:CouetteLam2} ($U$) and $K=100$ modes using both Eq.\ \ref{eq:CouetteLam} and Eq.\ \ref{eq:CouetteLam2} ($\tilde{U}$) for decelerating WDF ($\Rey=500$ and $\kappa=0.1$). In the $K=10^5$ case, there is no discernible error between the two solutions. Figure~\ref{fig:LamChoice} shows that a judicious choice of the form of Eq.\ \ref{eq:CouetteLam} results in multiple orders of magnitude higher accuracy than the solution in Eq.\ \ref{eq:CouetteLam2}. The one exception to this is at $t=0$ where there is a discontinuity in $dg_c/dt$, when evolving from constant simple shear, that does not change the results in Eq.\ \ref{eq:CouetteLam2}.

\begin{figure}
\centering
    \includegraphics[width=\textwidth]{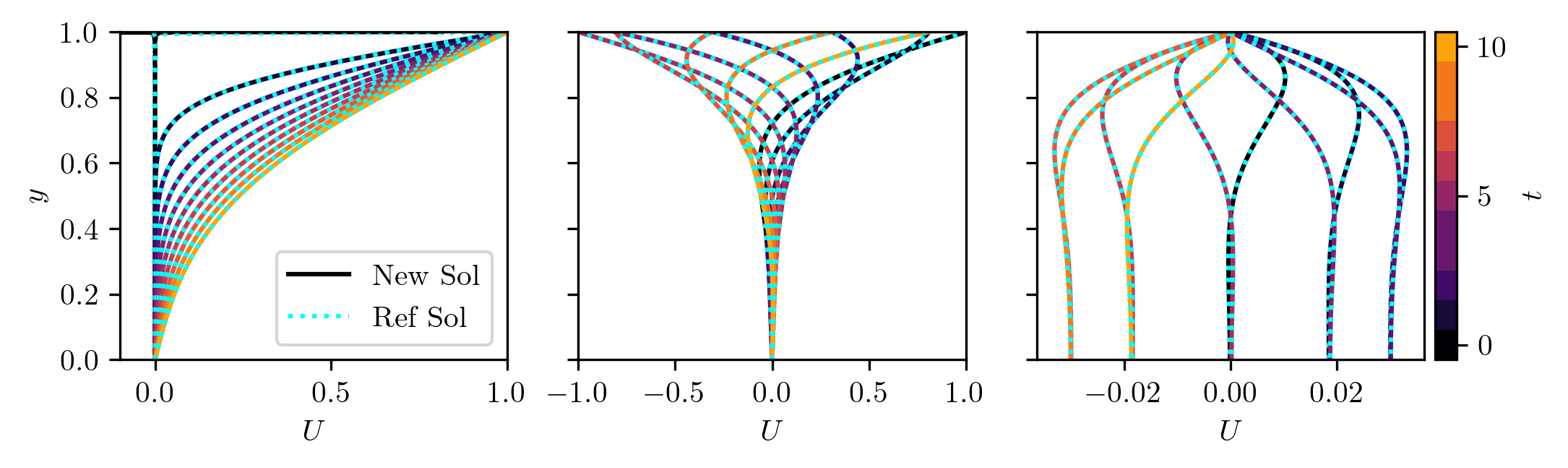}

    \captionsetup[subfigure]{labelformat=empty}
    \begin{picture}(0,0)
    \put(-185,108){\contour{white}{ \textcolor{black}{a)}}}
    \put(-60,108){\contour{white}{ \textcolor{black}{b)}}}
    \put(51,108){\contour{white}{ \textcolor{black}{c)}}}
    \end{picture} 
    \begin{subfigure}[b]{0\textwidth}\caption{}\vspace{-10mm}\label{fig:Validatea}\end{subfigure}
    \begin{subfigure}[b]{0\textwidth}\caption{}\vspace{-10mm}\label{fig:Validateb}\end{subfigure}
    \begin{subfigure}[b]{0\textwidth}\caption{}\vspace{-10mm}\label{fig:Validatec}\end{subfigure}
    \vspace{-5mm}
    
    \caption{Laminar solutions using reference equations and the equations presented here for (a) Stokes' first problem, (b) Stokes' second problem, and (c) Womersley flow.}
\label{fig:Validate}
\end{figure}
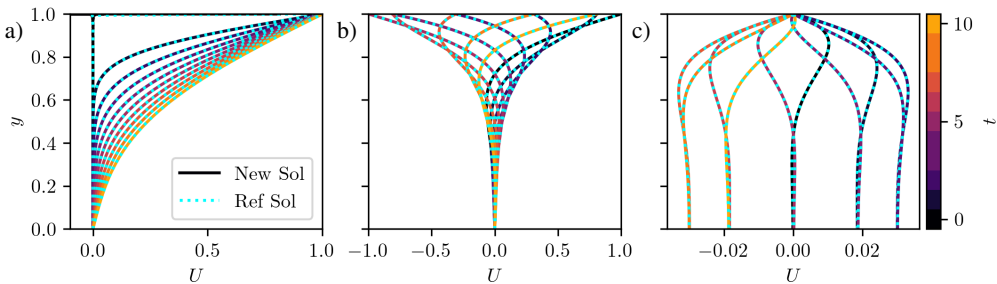

Next, we validate the laminar solution with three canonical problems: Stokes' first problem ($g_w=H(t)$), Stokes' second problem ($g_w=\sin(2\pi t)$), and Womersley flow ($g_p=-(2/\Rey) \sin(2\pi t)$) \citep{Batchelor2000,Majdalani2008}. In Fig.~\ref{fig:Validate} we compare the solutions Eq.\ \ref{eq:CouetteLam2}, Eq.\ \ref{eq:CouetteLam}, and Eq.\ \ref{eq:NSELamPresfin} against these three canonical problems at $\Rey=10$. Notably, we present both Stokes' 2nd problem and the Womersley flow after the transient from the initial condition has approached zero.

We summarize in Table \ref{Table} the integrals in Eq.\ \ref{eq:CouetteLam2}, Eq.\ \ref{eq:CouetteLam}, and Eq.\ \ref{eq:NSELamPresfin} for some interesting boundary conditions, and for generic Womersley flow. 
The cases that we include are Stokes' first problem, a generalization of Stokes' second problem for all periodic functions, the polynomial $t^m$, Laguerre polynomials ($L_m^{(\alpha)}(t)$), which is the natural basis for polynomial functions with $t\in[0,\infty]$, the exponential decaying flow considered above, and a generalization of Womersley flow for all periodic functions. Note that, in the polynomial cases, we are assuming that the polynomial is twice differentiable. For $t^m$ this implies $m\in\mathbb{Z}^+$ and $m>1$. Also, $\Gamma$ represents the ordinary gamma function, if it takes one argument, and the incomplete gamma function, if it takes two input values.

\begin{table}
\caption{Integral expressions in the laminar solutions for specific flows.}
$$
\begin{array}{lcll} \text{Flow} & \text { Eq.\ } & \text { Integral } 
\\ \hline 
&&&\\ g_w(t)=H(t) & \ref{eq:CouetteLam2} & 1
\\ g_w(t)=\sum\limits_{m=-\infty}^\infty \hat{g}_m e^{i\omega_m t} & \ref{eq:CouetteLam} & \sum\limits_{m=-\infty}^\infty \dfrac{-\omega_m^2 \hat{g}_m}{a_n+i\omega_m}\left(e^{(a_n+i\omega_m)t}-1\right)
\\ &&&\\ g_w(t)=t^m \quad & \ref{eq:CouetteLam} & \dfrac{-m(m-1)a_nt^m}{(-a_nt)^m}\left(\Gamma (m-1)-\Gamma(m-1,-a_nt)\right)
\\ 
&&&\\ g_w(t)=\sum\limits_{m=0}^\infty \hat{g}_m L^{(0)}_m(t) & \ref{eq:CouetteLam} & \sum\limits_{m=2}^\infty \sum\limits_{l=0}^\infty \dfrac{(-1)^l \hat{g}_m}{a_n^{l+1}}\left(e^{a_nt}L_{m-2-l}^{(2+l)}(t)-L_{m-2-l}^{(2+l)}(0)\right)
\\ 
&&&\\ g_w(t)=u_ie^{-\kappa t}+u_f\left(1-e^{-\kappa t}\right) & \ref{eq:CouetteLam} & \dfrac{\kappa^2(u_i-u_f)}{a_n-\kappa}\left(e^{(a_n-\kappa)t}-1\right)
\\ 
&&&\\ g_p(t)=\sum\limits_{m=-\infty}^\infty \hat{g}_m e^{i\omega_m t} & \ref{eq:NSELamPresfin} & \sum\limits_{m=-\infty}^\infty \dfrac{i\omega_m \hat{g}_m}{b_n+i\omega_m}\left(e^{(b_n+i\omega_m)t}-1\right)
\\ &&&\\ 
\hline \end{array}
$$
\label{Table}
\end{table}

\section{Converting the energy norm to the $L_2$-norm} \label{sec:AppendixD}

Here we outline the procedure for converting the energy norm to the $L_2$-norm. This procedure depends upon how we discretize $\mathscr{L}$. Specifically, in Eq.\ \ref{eq:LinearEq} we must approximate the two derivative operators $D^2$ and $D^4.$ Early approaches \citep{Farrell1988} used finite-difference methods for approximating these derivatives, but we instead choose Chebyshev differentiation matrices \citep{Reddy1993,Weideman2000} to approximate derivatives. Spectral methods exhibit superior convergence, which reduces the number of collocation points required for our computations. Additionally, Chebyshev collocation points (also known as Gauss-Lobatto points \citep{Peyret2002}) are well suited for our flows of interest since they are more densely clustered near the walls of the domain where we exhibit larger gradients. However, defining $\mathbf{q}$ on these collocation points necessitates accounting for the non-uniform grid when evaluating Eq.\ \ref{eq:q}.

To address this non-uniform spacing, consider the inner product
\begin{equation} \label{eq:norm}
    \left<\hat{\omega}_y,\hat{\omega}_y\right>_L=\int_{-1}^1 \hat{\omega}_y^*\hat{\omega}_y dy \approx \sum_{i=1}^{M-1} \hat{\omega}_y(y_i)^*\hat{\omega}_y(y_i)\Delta y_i,
\end{equation}
where 
\begin{equation}
    y_i=\cos\left(\dfrac{\pi i}{M}\right), \qquad \qquad i=0,...,M,
\end{equation}
with $M$ grid points.
With this definition, we can construct a diagonal weight matrix $\mathbf{W}$ to account for $\Delta y_i$ and transform the inner product on a non-uniform grid into a standard $L_2$-norm with
\begin{equation} \label{eq:wnorm}
    \left<\hat{\omega}_y,\hat{\omega}_y\right>_L\approx ||\mathbf{W}^{1/2}\hat{\omega}_y||^2.
\end{equation}
While this allows us to convert all terms in the integral in Eq.\ \ref{eq:q} to standard $L_2$-norms we are still left with the presence of derivates of $\mathbf{q}$ in the energy norm. 

To proceed, we must convert Eq.\ \ref{eq:q} to one inner product in terms of $\hat{v}$ and one in terms of $\hat{\omega}_y$. For this reason, we consider the inner product
\begin{multline} \label{eq:innerprod}
    \left<(D+\sqrt{k^2})\hat{v},(D+\sqrt{k^2})\hat{v}\right>_L= \\
    \left<D\hat{v},D\hat{v}\right>_L+\left<\sqrt{k^2}\hat{v},\sqrt{k^2}\hat{v}\right>_L+\left<D\hat{v},\sqrt{k^2}\hat{v}\right>_L+\left<\sqrt{k^2}\hat{v},D\hat{v}\right>_L.
\end{multline}
The right-hand-side of Eq.\ \ref{eq:innerprod} includes the two terms in Eq.\ \ref{eq:q}, with two additional cross terms. We can evaluate these cross terms by rewriting the above equations in integral form
\begin{equation} \left<D\hat{v},\sqrt{k^2}\hat{v}\right>_L+\left<\sqrt{k^2}\hat{v},D\hat{v}\right>_L=\int_{-1}^1 \dfrac{\partial \hat{v}}{\partial y}^* \sqrt{k^2}\hat{v} dy + \int_{-1}^1 \sqrt{k^2}\hat{v}^* \dfrac{\partial \hat{v}}{\partial y} dy.
\end{equation}
Integrating by parts, which we demonstrate on the second integral, we see that the resulting terms sum to zero
\begin{multline} 
\left<D\hat{v},\sqrt{k^2}\hat{v}\right>_L+\left<\sqrt{k^2}\hat{v},D\hat{v}\right>_L= \\
    \int_{-1}^1 \dfrac{\partial \hat{v}}{\partial y}^* \sqrt{k^2}\hat{v} dy + \left.\sqrt{k^2}\hat{v}^*\hat{v}\right|_{-1}^1 -\int_{-1}^1 \sqrt{k^2}\hat{v} \dfrac{\partial \hat{v}}{\partial y}^* dy=0.
\end{multline}
This allows us to rewrite Eq.\ \ref{eq:q} as 
\begin{equation} \label{eq:q2}
    ||\mathbf{q}||_E^2=\left<(D+\sqrt{k^2})\hat{v},(D+\sqrt{k^2})\hat{v}\right>_L+\left<\hat{\omega}_y,\hat{\omega}_y\right>_L,
\end{equation}
which we combine with Eq.\ \ref{eq:wnorm} to convert the energy norm into the $L_2$-norm
\begin{equation} \label{eq:Energy2L2}
    ||\mathbf{q}||_E^2=\left\Vert \ 
    \left[\begin{array}{ll}
    \mathbf{W}^{1/2}(D+\sqrt{k^2}) & 0 \\
    0 & \mathbf{W}^{1/2}  \end{array}\right] \mathbf{q} \ \right\Vert^2= || \mathbf{V}\mathbf{q} ||^2.
\end{equation}
We use $\mathbf{V}$ to in \ref{eq:Growth2} to compute the maximum amplification.

\section{Computation of nonnormal growth} \label{sec:AppendixB}

\begin{figure}
\centering
    \includegraphics[width=.5\textwidth]{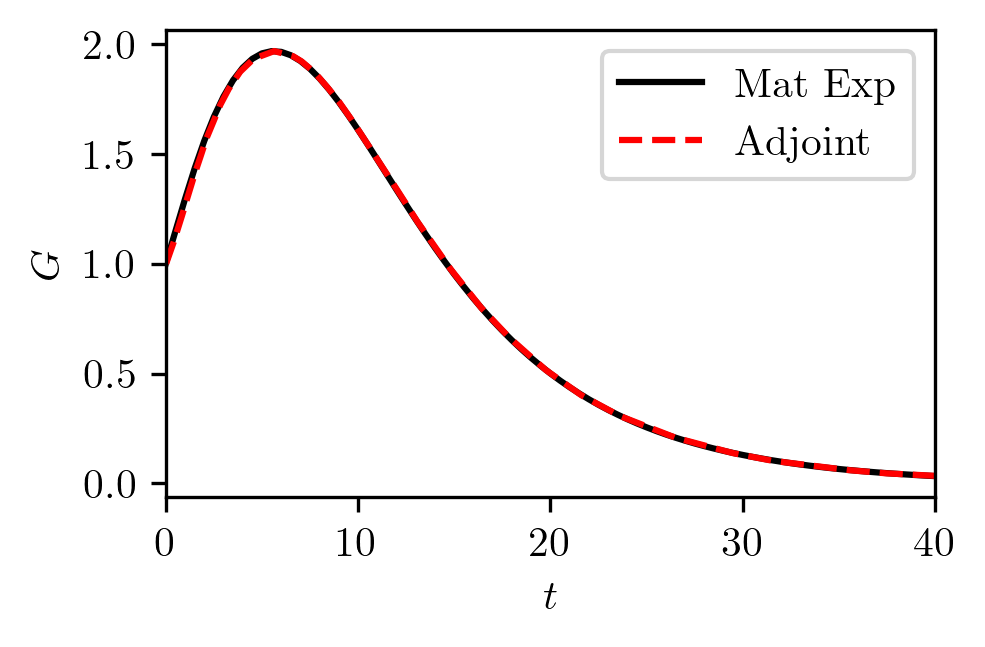}
  \caption{Maximum amplification of energy density $G$ using the adjoint method and the matrix exponential method.}
\label{fig:Adj}
\end{figure}

We compare our use of the matrix exponential method for computing the energy amplification to the adjoint method for computing nonnormal growth. For this comparison, we use the laminar pressure-driven flow that is impulsively stopped to a zero flow rate given in \citet{Nayak2017}. Figure~\ref{fig:Adj} shows the growth for perturbations of the laminar base flow using both the matrix exponential method (Eq.\ \ref{eq:Growth2}) and the adjoint method \citep{Nayak2017}. The two curves are in excellent agreement suggesting both choices give equivalent solutions. The matrix exponential method has the advantage that only matrix multiplications for the forward solution are needed, while the adjoint method requires solving a forward problem and a backward problem iteratively. For the channel flow cases we considered in this work, the linearized equations of motion result in a small matrix, however, this method would not be feasible when a larger matrix is required, in which case the adjoint method would be preferable.

\section{Solving for the pressure gradient when given a flow rate} \label{sec:AppendixC}

Here, we introduce a method for numerically approximating the pressure gradient $g_p$ for a prescribed flow rate assuming no even wall motion (i.e., $g_e(t)=0)$. Even though we could numerically approximate both the integral and the time derivative in Eq.\ \ref{eq:Q}, we instead perform integration by parts to eliminate the time derivative in Eq.\ \ref{eq:Q}, which leaves us with an approximation of the integral only. Using integration by parts we obtain the following equation:
\begin{equation} \label{eq:intrelation}
    e^{-b_nt}\int_0^t e^{b_nt'}\dfrac{dg_p}{dt} dt'=\hat{g}_p(t)- e^{-b_n t} \hat{g}_p(0) + e^{-b_n t} H(t)\hat{g}_0 -b_n e^{-b_n t} \int_0^t e^{b_nt'}\hat{g}_p dt'.
\end{equation}
Combining Eq.\ \ref{eq:Q} with Eq.\ \ref{eq:intrelation}, 
we find
\begin{multline} \label{eq:Q2}
    Q(t)=\sum_{n=0}^\infty \dfrac{32 e^{-b_nt} \Rey}{(2\pi n+\pi)^4} \left(H(t)\hat{g}_0-\hat{g}_p(0)-b_n \int_0^t e^{b_nt'}\hat{g}_p(t') dt'\right) +\dfrac{2 (-1)^n e^{-b_nt} C_{2,n} }{2\pi n+\pi}.
\end{multline}
Using the trapezoidal rule to approximate the integral in Eq.\ \ref{eq:Q2} we arrive at an expression for the temporal evolution of the pressure gradient
\begin{multline} \label{eq:gp}
    \hat{g}_p(l\Delta t)\sum_{n=0}^\infty d_n \approx \hat{g}_p((l-1)\Delta t) \sum_{n=0}^\infty d_n  e^{-b_n \Delta t} \\
    -\sum_{n=0}^\infty \sum_{j=1}^{l-1} d_n (e^{-b_n (l-j)\Delta t} \hat{g}_p(j\Delta t)+e^{-b_n (l-(j-1))\Delta t} \hat{g}_p((j-1)\Delta t) +R (l \Delta t),
\end{multline}
where $d_n=4 \Delta t / (2\pi n+ \pi)^2$ and
\begin{equation} \label{eq:R}
    R(t) =-Q(t) + \sum_{n=0}^\infty \dfrac{32 e^{-b_nt} \Rey}{(2\pi n+\pi)^4} \left( H(t)\hat{g}_0-\hat{g}_p(0)\right) +\dfrac{2 (-1)^n e^{-b_nt} C_{2,n}}{2\pi n+\pi}.
\end{equation}
Note that the factor of $2$ from the trapezoidal rule has been incorporated into $d_n$.
Finally, we can compute the velocity profile corresponding to this pressure gradient by again using integration by parts (Eq.\ \ref{eq:intrelation}) to simplify Eq.\ \ref{eq:NSELamPres} and by approximating the resulting integral with the trapezoidal rule leading to
\begin{multline} \label{eq:uLapprox}
    U(y,l \Delta t) \approx \\
    -\sum_{n=0}^\infty \sum_{j=1}^l \left\{\dfrac{d_n (-1)^n}{2}\left[ e^{-b_n (l-j) \Delta t}\hat{g}_p(j \Delta t) + e^{-b_n (l-(j-1)) \Delta t}\hat{g}_p((j-1) \Delta t) \right]\right\}\cos\left[\left(n+\dfrac{1}{2}\right)\pi y\right]\\
    + \sum_{n=0}^\infty e^{-b_n l \Delta t} \left[\dfrac{16 (-1)^n \Rey}{(2\pi n+\pi)^3} \left( H(l \Delta t)\hat{g}_0-\hat{g}_p(0)\right)+C_{2,n}\right] \cos\left[\left(n+\dfrac{1}{2}\right)\pi y\right].
\end{multline}

\end{appendices}

\end{document}